%% file: Robust2.tex
\documentclass[aps,superscriptaddress,showpacs,letterpaper,twocolumn,reprint,nofootinbib,longbibliography,pra]{revtex4-2}
\usepackage{graphicx} 
\usepackage{orcidlink}
\usepackage{amsmath} 
\usepackage{amssymb}   
\usepackage{color}  

\newcommand{\ket}[1]{|#1\rangle}
\newcommand{\bra}[1]{\langle#1|}
\newcommand \be{\begin{equation}}
\newcommand \ee{\end{equation}}
\newcommand \bea{\begin{eqnarray}}
\newcommand \eea{\end{eqnarray}}

\begin{document}

\title{Numerically optimized amplitude-robust controlled-Z gate for ultracold neutral atoms with individual addressing capability}

\author{K.V. Kozenko~\orcidlink{0009-0008-5652-8281}}
\affiliation{Rzhanov Institute of Semiconductor Physics SB RAS, 630090 Novosibirsk, Russia}
\affiliation{Novosibirsk State University, 630090 Novosibirsk, Russia}

\author{V.V. Gromyko~\orcidlink{0009-0004-5026-3535}}
\affiliation{Rzhanov Institute of Semiconductor Physics SB RAS, 630090 Novosibirsk, Russia}
\affiliation{Institute of Laser Physics SB RAS, 630090 Novosibirsk, Russia}

\author{I. I. Beterov~\orcidlink{0000-0002-6596-6741}}
\affiliation{Rzhanov Institute of Semiconductor Physics SB RAS, 630090 Novosibirsk, Russia}
\affiliation{Novosibirsk State University, 630090 Novosibirsk, Russia}
\affiliation{Novosibirsk State Technical University, 630073 Novosibirsk, Russia}
\affiliation{Institute of Laser Physics SB RAS, 630090 Novosibirsk, Russia}

\author{I.I. Ryabtsev~\orcidlink{0000-0002-5410-2155}}
\affiliation{Rzhanov Institute of Semiconductor Physics SB RAS, 630090 Novosibirsk, Russia}
\affiliation{Novosibirsk State University, 630090 Novosibirsk, Russia}

\date{\today}

\begin{abstract} We numerically optimized a scheme for a symmetric controlled-Z (CZ) gate based on the Rydberg blockade in neutral atoms, with the aim of increasing its robustness to variations in the Rabi frequency. The scheme exploits analytically defined phase profiles of the laser pulse and proves almost an order of magnitude more robust to Rabi-frequency variations than previously proposed protocols. We designed this amplitude-robust protocol with a smooth pulse shape and optimized it for a finite Rydberg blockade strength, which makes it suitable for experimental implementation in both single-photon and two-photon configurations.

We further show that the protocol can be applied to individually addressed Rydberg excitation, accounting for the asymmetry of the Rabi frequencies of two atoms excited by tightly focused laser beams. This allows the effects of residual thermal motion of the trapped atoms to be reduced. Finally, we examined the performance of the protocol for single-photon and two-photon Rydberg excitation schemes at finite blockade strength, and demonstrated its advantages for individual addressing at finite temperatures of the trapped atoms.

\end{abstract}

\maketitle

\section{Introduction}

The notable progress in quantum computing and quantum simulation with ultracold neutral atoms in latest years~\cite{Saffman2025, Saffman2026, Shi2022} relies on the abilities to trap thousands of atoms in arrays of optical dipole traps~\cite{Chiu2025, Manetsch2025, Pichard2024, Norcia2024, Gyger2024,Lin2025, Wang2026} and to implement  entangling gates with  fidelities above 99\%~\cite{Evered2023,Tsai2025,Radnaev2025,Zhang2026}, even reaching 99.7\%~\cite{Tsai2025}. The next step towards scalable quantum computing is a recent implementation of the quantum error correction algorithms~\cite{Zhang2026,Bluvstein2025,Reichardt2024}. Therefore, large-scale atomic arrays have already demonstrated the potential of becoming a leading platform for universal  quantum computing~\cite{Baranes2026}, including the implementation of Shor's factorization algorithm in the near future~\cite{Cain2026}. 

\begin{figure}[!t]
\includegraphics[width=\columnwidth]{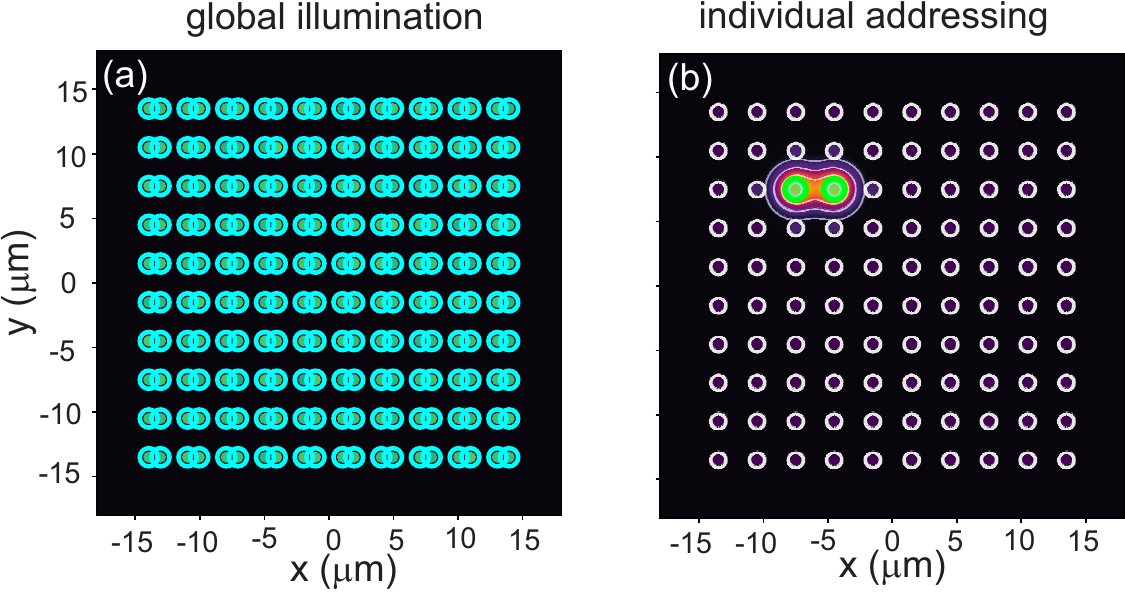}
\caption{ (a) Scheme of atomic array illuminated by global Rydberg beams entangling pairs of nearby atoms. (b) Scheme of atomic array with regularly spaced atoms. An individually addressed pair of atoms is entangled when excited to Rydberg state in the regime of Rydberg blockade. 
}
\label{Arrays}
\end{figure}

High-fidelity entanglement in atomic arrays commonly expoits Rydberg blockade when two nearby atoms cannot be simulutaneously excited into Rydberg state~\cite{Jaksch2000,Lukin2001}. Two architectures of Rydberg excitation in atomic arrays are usually considered nowadays~\cite{Radnaev2025}. In the first case the atoms are excited into Rydberg states by uniform laser beams that illuminate the whole atomic array~\cite{Ebadi2021, Evered2023}, as shown in Fig.\ref{Arrays}(a). The interaction between the atoms is controlled by the interatomic distance, which can be varied in parallel for the whole array~\cite{Bluvstein2022}. This has allowed for the demonstration of phase transitions within the atomic array and has been used for quantum simulations~\cite{Ebadi2021}. 

Although it is possible to entangle individual atoms by shuttling them into the interaction zone~\cite{Bluvstein2024,Dudinets2026,Reichardt2024}, there is an alternative architecture of two-qubit gates based on individual addressing to stationary atoms. Figure~\ref{Arrays}(b) illustrates two-qubit gates performed by individually addressable laser beams~\cite{Graham2022}. This scheme has no restrictions from the limited velocities of the atomic motion in moving optical tweezers~\cite{Radnaev2025}. However, the gate fidelities, obtained experimentally in such arrangements, are typically smaller, compared to experiments with global beams~\cite{Graham2022,Radnaev2025}.

 The use of Rydberg blockade to entangle the atoms was first proposed theoretically~\cite{Jaksch2000} and then experimentally demonstrated more than fifteen years ago~\cite{Saffman2010,Wilk2010}. However, it took almost ten more years to achieve high-fidelity entanglement with a symmetric gate protocol, which was first theoretically proposed~\cite{Shi2019a} and then experimentally implemented in a modified form~\cite{Levine2019}.  Since then, a family of symmetric gate schemes for further increase in fidelity of entanglement has been proposed~\cite{Levine2019, Jandura2022, Jandura2023, Fu2022, Evered2023,Tsai2025, Vybornyi2023} and  experimentally demonstrated with fidelities greater than 99.7\% for strontium-88 atoms~\cite{Tsai2025} and 99.5\% for rubidium atoms~\cite{Evered2023}. Increase of the gate fidelities opens the way to universal quantum computing with error correction, based on arrays of ultracold neutral atoms~\cite{Bluvstein2024,Bluvstein2025,Reichardt2024}.

The universal quantum computing requires entanglement of individual pairs of atoms, which is difficult to achieve in globally illuminated atomic arrays~\cite{Saffman2016, Henriet2020}. 
Therefore, individual addressing during Rydberg excitation of two atoms, as shown in Fig.\ref{Arrays}(b), is of great interest~\cite{Radnaev2025}. The problem of individual entanglement, when global beams are used for Rydberg excitation, can be solved using several approaches. Atoms can be transported to the interaction zone using a movable atomic tweezer~\cite{Chiu2025}. Trapping of atoms using fiber arrays allows for the individual control of the intensity of each dipole trap. It is possible to excite few selected atoms in the array, even  when global laser beams are used for Rydberg excitation~\cite{Li2024}.

Individual addressing to the atoms in stationary atomic arrays requires tightly focused laser beams with waists substantially smaller than interatomic distance, in order to avoid cross-talk with neighboring qubits~\cite{Radnaev2025}. In experiments with two-photon Rydberg excitation and individual addressing, the entanglement fidelity is slightly smaller~\cite{Ma2023, Radnaev2025} compared to schemes based on global illumination of the whole array. The highest fidelity of an experimentally demonstrated CZ gate with individual addressing is 99.35\% \cite{Radnaev2025}, to the best of our knowledge. Similar-type symmetric entangling gate protocols are used in both the experiments with global laser beams and with individual addressing~\cite{Radnaev2025}. One possible reason for the increased gate errors in the latter case is the spatial inhomogeneity of the Rabi frequency in focused laser beams, arising due to residual thermal motion of the atoms and laser beam pointing errors.  A three-photon excitation can be a potential solution~\cite{Bezuglov2025, Spirin2026}, although it can be challenging technically.  An alternative way to reduce the effect of inhomogeneity of the Rabi frequency on entanglement fidelity is the use of amplitude-robust gate protocols~\cite{Jandura2022}.

\begin{figure}[!t]
\includegraphics[width=\columnwidth]{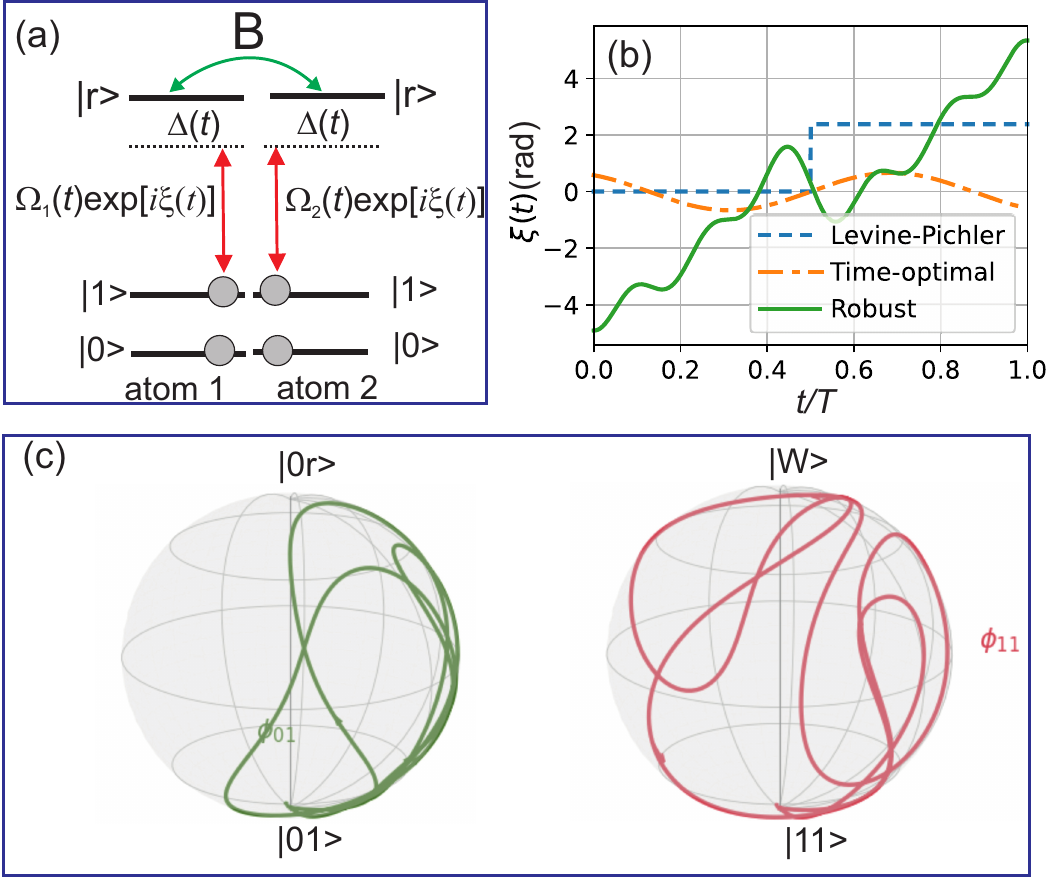}
\caption{ (a)   Energy level structure of two atom qubits 1 and 2 with logical states represented by ground state sublevels $\ket{0},\ket{1}$ and Rydberg state $\ket{r}$ coupled to state $\ket{1}$ by resonant laser fields with Rabi frequencies $\Omega_1\left(t\right)$ and  $\Omega_2\left(t\right)$ and detuning  $\Delta(t)$. $\mathsf{B}$ is the Rydberg blockade strength between atoms in states $\ket{r}$, which prevents simultaneous excitation of two Rydberg atoms. (b) Phase profiles of laser pulses for amplitude-robust (solid line), time-optimal (dash-dotted line) and Levine-Pichler (dashed line) CZ gate protocols. (c) The numerically calculated trajectories of two initial quantum states $\ket{01}$ and $\ket{11}$ on a Bloch sphere for amplitude-robust protocol with perfect blockade.
}
\label{Blockade}
\end{figure}

The schemes of entangling gates with neutral atoms are commonly based on Rydberg blockade. The hyperfine sublevels of the ground state  of alkali-metal atoms (usually rubidium and cesium) are used as qubit logical states $\ket{0}$ and $\ket{1}$, as shown in Fig.~\ref{Blockade}(c)~\cite{Saffman2016}. For strontium-88 atoms the states $\ket{0}$ and $\ket{1}$ are coupled by   optical transitions~\cite{Endres2024, Tsai2025}. Each of two atoms in state $\ket{1}$ is coupled to a Rydberg state $\ket{r}$ by laser fields with Rabi frequencies $\Omega_1$ and $\Omega_2$ which can be identical or different. When the atom is excited to the Rydberg state $\ket{r}$ and returns to the ground state, it accumulates a phase shift $\pi$ which can be used for the CZ gate~\cite{Jaksch2000,Saffman2010}. The pairwise interaction of Rydberg atoms is described by a blockade shift $\mathsf{B}$. In the regime of Rydberg blockade, which is illustrated in Fig.~\ref{Blockade}(a), two atoms are simultaneously illuminated by laser radiation, which is tuned to the resonance with the transition to the  Rydberg state. Due to pairwise Rydberg interactions, the collective energy state $\ket{rr}$ acquires a large energy shift $\mathsf{B}$, which is known as the blockade strength~\cite{Jaksch2000}. Simultaneous laser excitation of two nearby Rydberg atoms thus becomes impossible~\cite{Lukin2001}. 

Most of high-fidelity gate protocols for neutral atoms  are symmetric, and both interacting atoms are illuminated by identical laser pulses~\cite{Levine2019, Tsai2025}, in contrast to the initially implemented so-called $\pi$-$2\pi$-$\pi$ scheme with individually addressed atoms~\cite{Saffman2010}. This scheme was recently revised for long-rangle entanglement~\cite{Cole2025}. The advantage of a symmetric gate protocol is that both atoms interact with the same laser field simultaneously. Apart from reducing the time spent by the atoms in the Rydberg state, the symmetry of laser pulses  allows for the use of global laser beams with uniform intensities. This ensures well defined Rabi frequencies for ground-Rydberg transitions~\cite{Levine2019, Evered2023, Tsai2025}. 

In general, for two-qubit entanglement with individual addressing the symmetry of the laser pulses is not required~\cite{Isenhower2010}. However, at present, no clear advantages for two-qubit gate performance have been identified in the cases when the atoms interact with different laser fields~\cite{Jandura2022}. At the same time, when the atoms are illuminated by tightly focused beams, the asymmetry of laser fields, caused by the thermal motion of the trapped atoms and laser beam pointing instability can be detrimental for the gate fidelities.

Two-photon Rydberg excitation is commonly used in experiments with rubidium~\cite{Levine2019, Ebadi2022, Fu2022, Evered2023} and cesium atoms~\cite{Radnaev2025, Chinnarasu2025}. Entanglement fidelities around 96.7\% were reported in the experiment with single-photon Rydberg excitation of cesium atoms using UV light~\cite{Chow2024}. The highest fidelities of 99.6\% and 99.7\% were reported in experiments with single-photon Rydberg excitation of strontium-88 atoms~\cite{Endres2024,Tsai2025}.  Recently, a three-photon Rydberg excitation of single rubidium atoms in an optical dipole trap was reported~\cite{Beterov2024}. 

In this paper, we consider the application of amplitude-robust gate protocol for single-photon and two-photon excitation schemes. 
The paper is organized as follows. In Sec.~\ref{sec.Optimization} we describe how to reduce the sensitivity of the gate infidelity to Rabi-frequency variations, and analyze the effects of laser-pulse asymmetry, Rabi-frequency gradient, and finite Rydberg blockade strength. In Sec.~\ref{sec.TwoPhoton} we consider application of our gate protocol to commonly used schemes of two-photon Rydberg excitation. In Sec.~\ref{sec.MonteCarlo} we perform a simple numerical simulation of the performance of our gate protocol at finite temperatures of trapped atoms. We discuss the applicability of our gate protocols in the experiment in Sec.~\ref{sec.Discussion}. The results are summarized in Sec.~\ref{sec.Conclusion}.

\section{Numerical optimization}
\label{sec.Optimization}

For a two-atom system with two logical states $\ket{0}$, $\ket{1}$ and Rydberg state $\ket{r}$ the Hamiltonian has the form

\be
\label{Hamiltonian}
{\mathcal H}={\mathcal H}_{\rm 1}\otimes I + I\otimes {\mathcal H}_{\rm 2} + {\sf B}\ket{rr}\bra{rr},
\ee
\noindent where $\rm 1,2$ label each of interacting atoms, and
$$
{\mathcal H}_{1,2}=\frac{\Omega_{1,2}(t)\mathrm{exp}\left[i\xi(t)\right]}{2}\ket{r}_{1,2}\bra{1} +\Delta(t)\ket{r}_{1,2}\bra{r} + \rm H.c.~.
$$
Here $\Omega_{1,2}(t)$ is a time-dependent Rabi frequency, $\xi(t)$ is a phase of the laser pulse, $\Delta(t)$ is detuning from the resonance and {\sf B} is a blockade strength. In the idealized case of perfect blockade it is possible simply to remove the states $\ket{rr}$ with double Rydberg excitation from the Hamiltonian. We used this approach as a starting point for our optimization.
\begin{figure}[!t]
\includegraphics[width=\columnwidth]{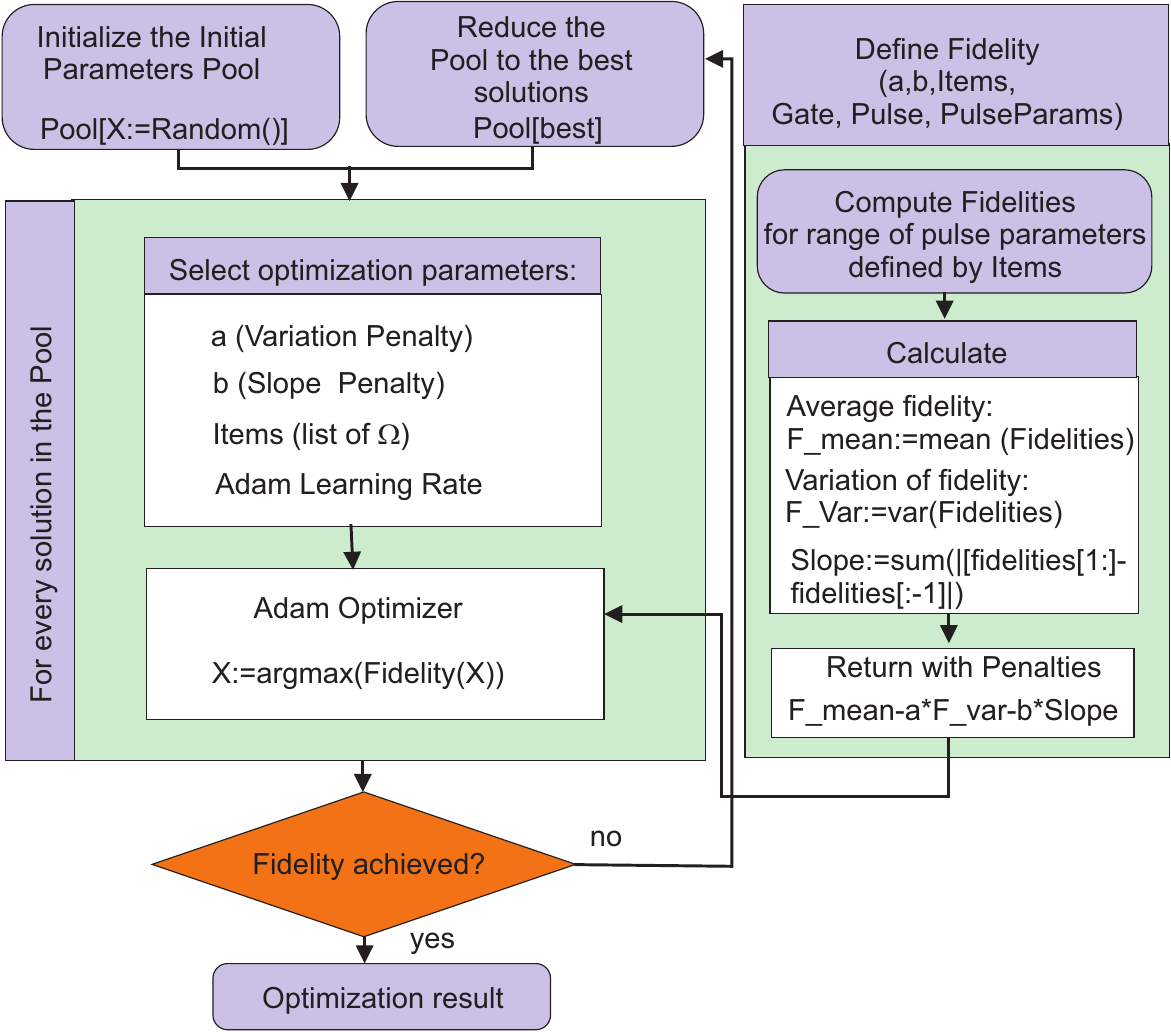}
\caption{Algorithm of numerical optimization of amplitude-robust gates. We start optimization by a random pool of the initial parameters. Then we set the optimization parameters and run the optimization using the customly designed fidelity, averaged over a set of Rabi frequencies. The optimization is run in several steps, with the modified optimization parameters after each step.
}
\label{Algorithm}
\end{figure}

We utilized the customized RydOpt~\cite{Rydopt2025} Python package for the analysis and optimization of gate performance. This package is designed for the quantum optimization of the Hamiltonian dynamics in an arbitrary quantum system with analytically defined controls leveraging the JAX library for Python and implementing the Adam optimization algorithm. Our customization of the RydOpt package primarily involved redefining the target gate fidelity to enable simultaneous optimization across a range of Rabi frequencies. A detailed description of our optimization algorithm is provided in Fig.~\ref{Algorithm}. 

Based on our previous work~\cite{Beterov2026}, we considered a laser pulse with constant Rabi frequency and detuning, along with a phase profile parametrized by chopped random basis (CRAB) initial phase profiles augmented by an additional linear shift.

We began with the initialization of a random set of initial parameters defining the phase profile, gate duration, and detuning. We used uniform distributions: gate duration \(T \in [10.0, 25.0]\), detuning \(\Delta \in [-1.0, 1.0]\), and Rabi frequency \(\Omega_0 \in [0.0, 1.0]\). For the CRAB phase profile, the linear coefficient was initialized uniformly from \([-1.0, 1.0]\), while the sine and cosine amplitude coefficients for the frequency components were chosen independently from \([-0.5, 0.5]\). The number of harmonics was varied from 4 to 16, since it was empirically established that further complication of the spectrum did not yield an increase in operation accuracy. The optimization procedure itself consisted of three consecutive steps; throughout the entire process, we assumed identical Rabi frequency values for each of the two atoms.

To ensure the robustness of the protocol, we minimized the following error function:
\begin{equation}
\mathcal{L} = 1 - \left(F_{\mathrm{mean}} - \alpha F_{\mathrm{var}} - \beta F_{\mathrm{slope}}\right).
\tag{2}
\end{equation}

The mean fidelity value,
\begin{equation}
F_{\mathrm{mean}} = \frac{1}{5}\sum_{k=1}^{5} F_k,
\tag{3}
\end{equation}
was calculated on a discrete set of shifted Rabi frequencies. For this, the protocol fidelity \(F_k = F(\Omega + \epsilon_k)\) was evaluated at five points, where \(\epsilon_k\) was chosen depending on the optimization step. The fidelity itself was determined by the expression~\cite{Rydopt2025} 
\begin{equation}
F = \frac{\mathrm{Tr}\left(U_{\mathrm{targ}}^{\dagger} U(T)\right)^2}{d^2},
\tag{5}
\end{equation}
where \(U_{\mathrm{targ}}\) is the target unitary operator of the required quantum gate, and \(U(T)\) is the actual evolution operator of the system by the end of the control pulse \(T\). The dimension of the Hilbert space is denoted by \(d\).

Optimization of the protocol based only on maximizing \(F_{\mathrm{mean}}\) does not guarantee its robustness, since the algorithm may converge to a solution with high accuracy at the center \((\epsilon_k = 0)\) and sharp drops at the edges. To prevent such edge effects, penalties are introduced into the objective function. The component \(F_{\mathrm{var}}\) represents the variance of the fidelity values on a given set of shifts:
\begin{equation}
F_{\mathrm{var}} = \frac{1}{5}\sum_{k=1}^{5}\left(F_k - F_{\mathrm{mean}}\right)^2.
\tag{6}
\end{equation}
It limits the spread of fidelity within the selected range, preventing individual values from deviating too strongly from the center. However, this metric is not sufficiently sensitive to the appearance of sharp jumps between fidelities with close \(\epsilon_k\). To exclude such local deviations, an additional penalty \(F_{\mathrm{slope}}\) is introduced:
\begin{equation}
F_{\mathrm{slope}} = \sum_{k=1}^{4}\left|F_{k+1} - F_k\right|.
\tag{7}
\end{equation}

The combination of these penalties makes it possible to find protocols with lower sensitivity to fluctuations of the control field.

For the initial selection of the most successful starting point, after initialization, 10 independent runs were performed, and for each of them sequential optimization was carried out with different values of the penalty coefficients. At the first stage, \(\alpha = 0.15\), \(\beta = 0.3\) were used, while the learning rate was varied in the range \([0.05, 0.01, 0.001]\), and for the calculation of \(F_{\mathrm{mean}}\), the values \(\epsilon_k \in [-0.01, -0.005, 0, 0.005, 0.01]\) were used. Based on the comparison results, the run with the flattest error profile was selected, which was then used as the starting point for further optimization.

At the next stage, the contribution of the penalties was increased to \(\alpha = 0.5\), \(\beta = 0.5\), while the remaining parameters were kept unchanged. After that, the range of \(\epsilon_k\) was successively expanded first to \([-0.02, -0.01, 0, 0.01, 0.02]\), and then to \([-0.05, -0.03, 0, 0.03, 0.05]\). Such a sequential change of parameters made it possible to gradually increase the requirements for the robustness of the protocol. We have found that the increase of the number of optimization stages results in slight variations of the shape of the dependence of gate infidelities on variation of Rabi frequencies without general improvement.

\begin{figure}[!t]
\includegraphics[width= \columnwidth]{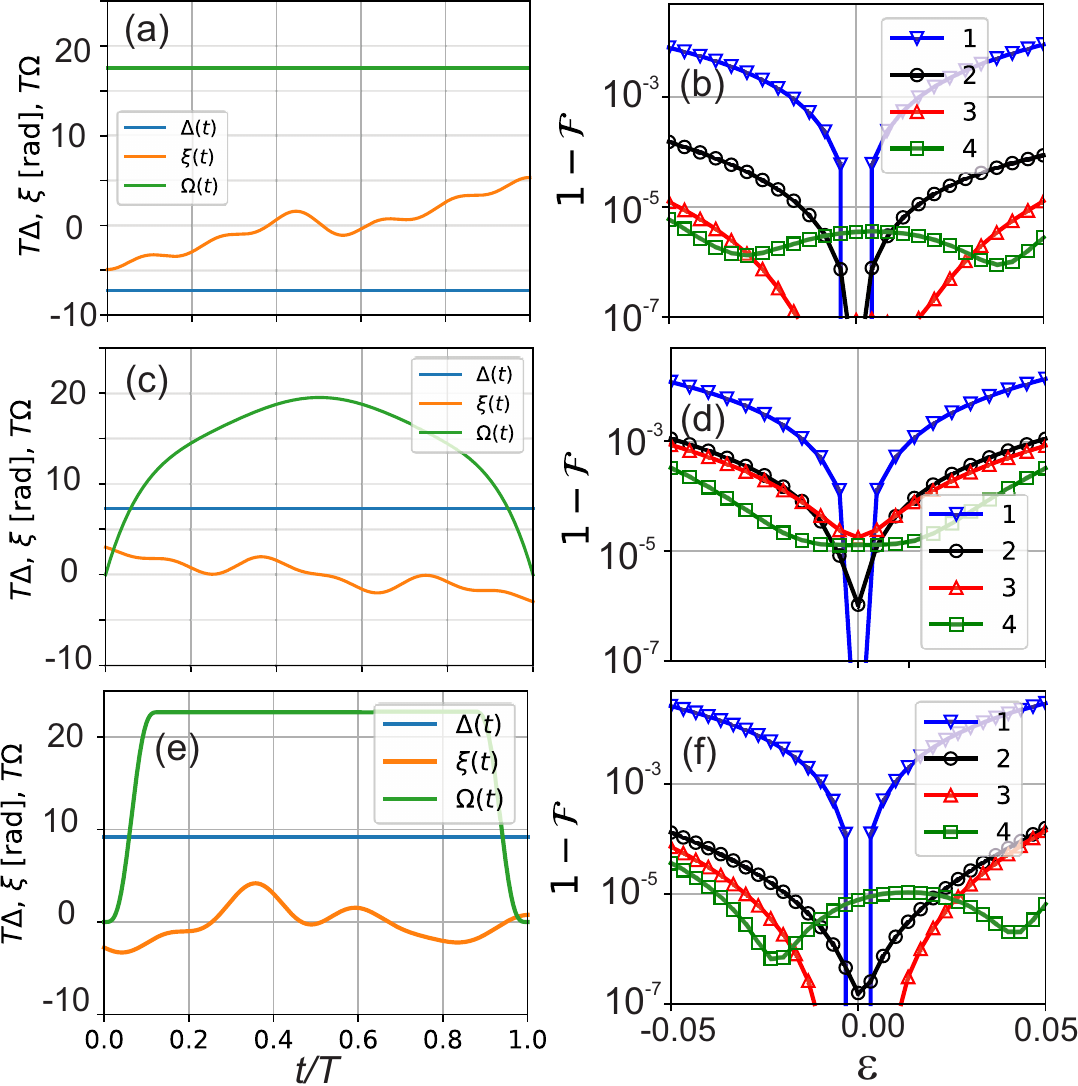}
\caption{Left-hand panel: Numerically optimized pulse profiles for the amplitude-robust CZ gate with   phase profile $\xi\left(t\right)$(solid lines),  time-dependent Rabi frequency $\Omega\left(t\right)$ (dashed lines) and Rydberg detuning $\Delta$ (dash-dotted lines). All profiles are presented for the final step of numeric optimization. Right-hand panel: Dependence of the gate infidelity on the variation of the Rabi frequency $\epsilon$ at four consequent steps of numerical optimization: default single-point optimization (1, downward triangles), first run of optimization for a range of Rabi frequencies (2, circles), second run of optimization with increased penalties (3, upward triangles),  final optimization round for a wider range of Rabi frequencies (4, squares). Results for a rectangular pulse are shown in plots (a) and (b). Results for a smooth shape of Rabi frequency profile defined by  Bernstein basis polynomials are shown in plots (c) and (d). Results for the smooth pulse profile with constant Rabi frequency and smooth edges are shown in plots (e) and (f).
\label{PulseShapes}
}

\end{figure}

We optimized the phase profiles using different amplitude shapes of laser pulse. In addition to rectangular pulses with constant Rabi frequencies we considered smooth pulse shapes, which are more suitable for experimental implementation. Several optimization steps are illustrated in Fig.~\ref{PulseShapes}. Figure~\ref{PulseShapes}(a) shows the simplest case, where both the Rabi frequency and detuning remain constant during the laser pulse. The dependence of the gate infidelity on the variation in Rabi frequency $\epsilon$ is shown in  Fig.~\ref{PulseShapes}(b) for four sequential optimization steps. Initially, we optimized the parameters of randomly defined pulse profile using the standard procedure from the RydOpt library. The result of this initial optimization is illustrated by the curve marked with downward triangles in Fig.~\ref{PulseShapes}(b). The infidelity exhibits a sharp minimum near the optimal value of Rabi frequency and increases   rapidly with variations in the Rabi frequency. Similar behavior is observed for Levine-Pichler~\cite{Levine2019} and time-optimal~\cite{Jandura2022,Evered2023} gate protocols, as demonstrated in our previous work~\cite{Beterov2026}. The second step of optimization is shown by circles in Fig.~\ref{PulseShapes}(b). Here, the randomly initialized pulse profile was optimized for maximum fidelity averaged over five different  Rabi frequency values, incorporating penalties for both fidelity variation  and the slope of the fidelity dependence  on Rabi frequency. The slope was defined as the difference between fidelities at the smallest and largest Rabi frequency  values. Reducing this slope makes the gate performance insensitive to the sign of  the Rabi frequency variation. The next optimization step is represented by upward triangles. At this stage, we started from the pulse parameters obtained in the previous step and increased the penalties. Finally, we expanded the range of Rabi frequencies used for optimization, the  result is shown by squares in Fig.~\ref{PulseShapes}(b). This final stage yielded slightly higher infidelities in the center, but lower infidelities at the edges. Similar dependence of gate infidelity on variation in Rabi frequency was shown in Fig.~2 of Ref.~\cite{Jandura2023}. For the variation of  Rabi frequency by 5\% our gate protocol gives intrinsic gate infidelities below $10^{-5}$, while the authors of Ref.~\cite{Jandura2023} claimed the infidelities below $10^{-4}$. Therefore, our gate protocol provides the amplitude robustness, that is almost an order of magnitude greater than in the previous proposal~\cite{Jandura2023}.

In experiments implementation of laser pulses with extremely short rise and fall times, resembling the constant Rabi frequency  profile   in Fig.~\ref{PulseShapes}(a), is technically challenging. Therefore, we also considered pulse profiles with smoother shapes. An example is the profile defined by Bernstein basis, shown in Fig.~\ref{PulseShapes}(c). Such profiles were used in experiments for high-fidelity entanglement of Rb atoms~\cite{Fu2022}. The corresponding dependence of gate infidelity on Rabi frequency variation is shown in  Fig.~\ref{PulseShapes}(d). In contrast to the  constant Rabi frequency case, this approach yielded  substantial robustness improvement at the final optimization step. A smooth pulse profile with constant Rabi frequency but finite rise and fall times is shown in Fig.~\ref{PulseShapes}(e). The robustness of this protocol to variation of Rabi frequency is shown in Fig.~\ref{PulseShapes}(f). Although the protocol's robustness is inferior to that in Fig.\ref{PulseShapes}(b), its smooth pulse shape offers advantages for experimental implementation. Moreover, our calculations indicate that, for multi-photon excitation schemes, smooth gate profiles provide higher fidelities than those with constant Rabi frequencies~\cite{Beterov2026}. For individual addressing, the gate protocol should be robust not only to variations in the global Rabi frequency $\Omega_{1,2} =\Omega_0\to \Omega_0(1\pm\epsilon)$, but also to variations in the individual Rabi frequencies for each of the two atoms, which can be introduced by the replacment $\Omega_2 \to \Omega_1(1\pm\alpha)$.

We compared the performance of the amplitude-robust gate with other modern entangling gate protocols assuming perfect blockade. The pulse profile with constant Rabi frequency and detuning, and a single step in the phase of the laser pulse, is the first high-fidelity symmetric entangling gate protocol based on Rydberg blockade~\cite{Levine2019} and is known as the Levine-Pichler gate. In our numerical simulations we obtained alternative dimensionless parameters of Levine-Pichler gate for $\Omega_0=1$ with the same total duration $T=8.58531$, detuning $\Delta/\Omega_0=-0.3773671$ and phase shift $\xi=2.38074$~\cite{Beterov2026}. This phase profile is shown  in Fig.~\ref{Blockade}(b). The gate performance with these parameters is identical to the original proposal~\cite{Levine2019}.  It is also possible to design a gate scheme with smoothly varied phase of laser pulse. For the same gate duration, it requires slightly smaller Rabi frequencies~\cite{Jandura2022} and is known as the time-optimal gate. The phase profile for the time-optimal gate obtained using the RydOpt package is also shown in Fig.~\ref{Blockade}(b). 

\begin{figure}[!t]
\includegraphics[width= \columnwidth]{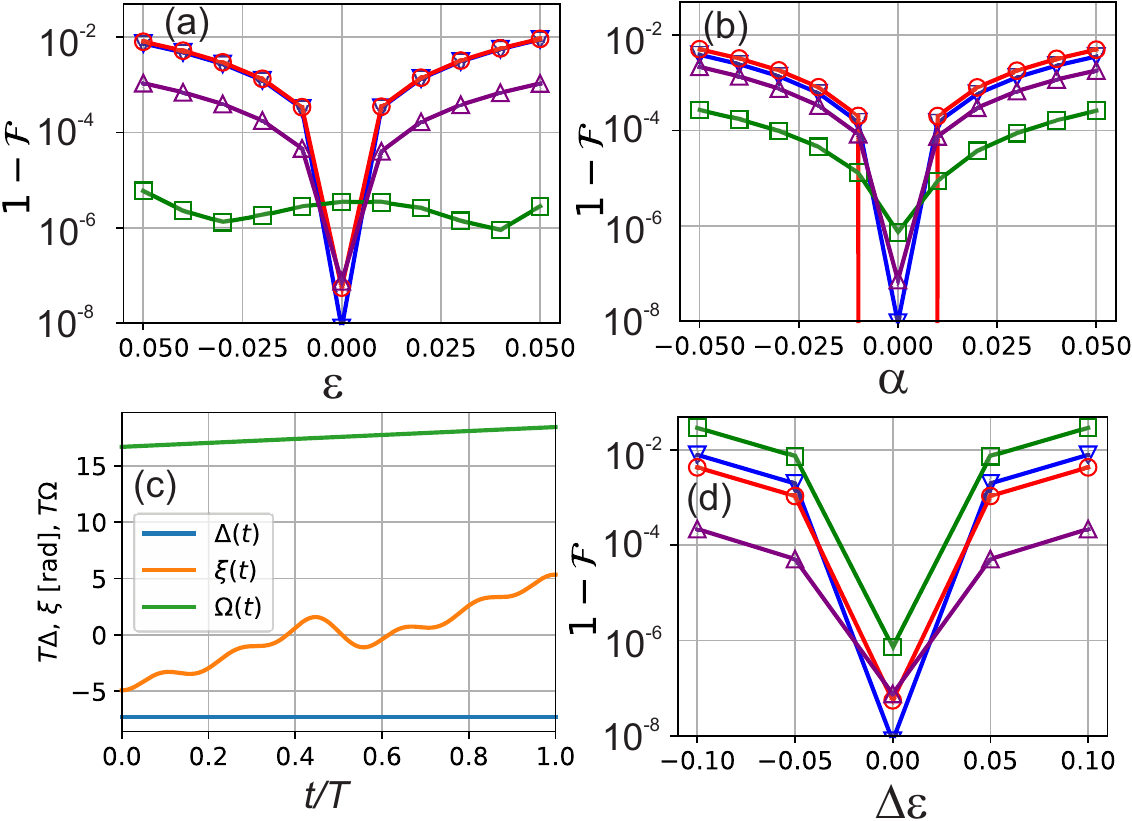}
\caption{(a) Dependence of the gate infidelity on the variation in the Rabi frequency $\epsilon$ for the Levine-Pichler gate (down-pointing triangles), the time-optimal gate (circles), counterdiabatic (up-pointing triangles) and the amplitude-robust gate (squares). 
(b) Dependence of the gate infidelity on the variation in the ratio of Rabi frequencies $\alpha$ for the Levine-Pichler gate (down-pointing triangles), the time-optimal gate (circles), counterdiabatic (up-pointing triangles)  and the amplitude-robust gate (squares).
(c) Time profile of laser pulse for the laser pulse with the gradient of the Rabi frequency $\Delta \epsilon=0.1$. (d) Dependence of the gate infidelity on the gradient of the Rabi frequency $\Delta\epsilon$ for the Levine-Pichler gate (down-pointing triangles), the time-optimal gate (circles), counterdiabatic (up-pointing triangles), and the amplitude-robust gate (squares).}
\label{RabiVariation}
\end{figure}

The dependence of gate infidelities on variations  in the Rabi frequency $\epsilon$  is shown in Fig.~\ref{RabiVariation}(a) in comparison with the time-optimal and  Levine-Pichler gates. The robustness of our gate scheme  outperforms that of the amplitude-robust gate, reported in Ref.~\cite{Jandura2023}. The dependence of gate infidelities on variations $\alpha$ in the ratio of the Rabi frequencies  $\Omega_2$ and $\Omega_1$ for two atoms is shown in Fig.~\ref{RabiVariation}(b), also compared with the time-optimal, Levine-Pichler  and counterdiabatic gate of our previous work~\cite{Beterov2026}. It is clear that the proposed gate protocol is robust to both variations of the global Rabi frequency and the ratio between the Rabi frequencies for each of the two atoms. Similarly to the previous work~\cite{Beterov2026}, we    analyzed the sensitivity of the fidelity to variations in Rabi frequency during the pulse. For this purpose, we used a linear time profile of the Rabi frequency $\Omega_0[1+\Delta\epsilon(t-T/2)/T]$, as shown in Fig.~\ref{RabiVariation}(c). The dependence of the gate infidelities on $\Delta\epsilon$ is shown in Fig.~\ref{RabiVariation}(d). We have found that the amplitude-robust gate is more sensitive to the gradient of Rabi frequency compared to both the time-optimal and Levine-Pichler gates, while the counterdiabatic gate is the least sensitive. This confirms the findings from our previous paper~\cite{Beterov2026}.  However, precise control of the shape of the laser pulse is a technical issue. The variation of the Rabi frequency in tightly focused laser beams due to the displacement of atoms between laser pulses is typically much larger than the effect of variation of atomic position during the laser pulse. However, for infidelities below 0.1\% this factor should be carefully considered. For extremely short gate times on the order of 100~ns it can be neglected in most cases.
\begin{figure}[!t]
\includegraphics[width= \columnwidth]{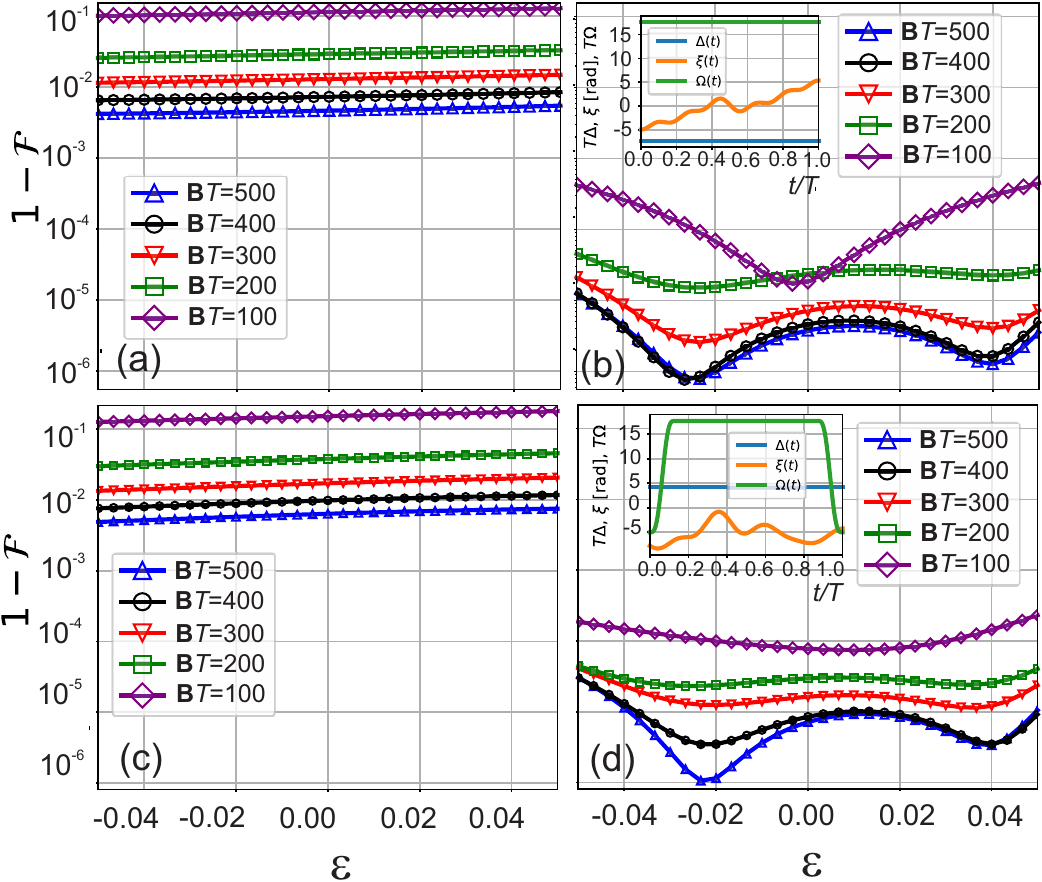}
\caption{(a,b) Dependence of the gate infidelity on the variation in the Rabi frequency $\epsilon$ at finite blockade strengths $\mathsf {B} T$ for amplitude-robust gate with rectangular pulse profile (shown in inset) before (a) and after (b) optimization for blockade energy.
(c,d) Dependence of the gate infidelity on the variation in the Rabi frequency $\epsilon$ at finite blockade strengths $ \mathsf {B} T$ for amplitude-robust gate with smooth pulse profile (shown in inset) before (c) and after (d) optimization for blockade energy.}
\label{FiniteBlockade}
\end{figure}

Finite blockade strength \textbf{B} is one of the major sources of gate infidelities. Figure~\ref{FiniteBlockade}(a) shows the increase of the infidelity  of the amplitude-robust gate with rectangular intensity profile when the blockade strength $\mathsf{B}T$ is decreased. Our optimization procedure allows finding the gate parameters for finite values of blockade strength, if the full Hamiltonian from Eq.~\eqref{Hamiltonian} is used. Optimization results for a range of $\mathsf{B}T$ are illustrated in Fig.~\ref{FiniteBlockade}(b). Even in the worst case $\mathsf{B}T=100$ it is possible to keep the infidelity below 0.1\%.   Optimization at finite blockade strengths preserves robustness of the gate performance to variation of Rabi frequency. Figures~\ref{FiniteBlockade}(c) and \ref{FiniteBlockade}(d) illustrate similar dependences for amplitude-robust gates with smooth shapes of laser pulse, which are most suitable for experimental implementation. We listed the dimensionless pulse parameters in the Appendix~\ref{appendixA}.

\section{Two-photon CZ gate protocol}
\label{sec.TwoPhoton}

\begin{figure}[!t]
\includegraphics[width=\columnwidth]{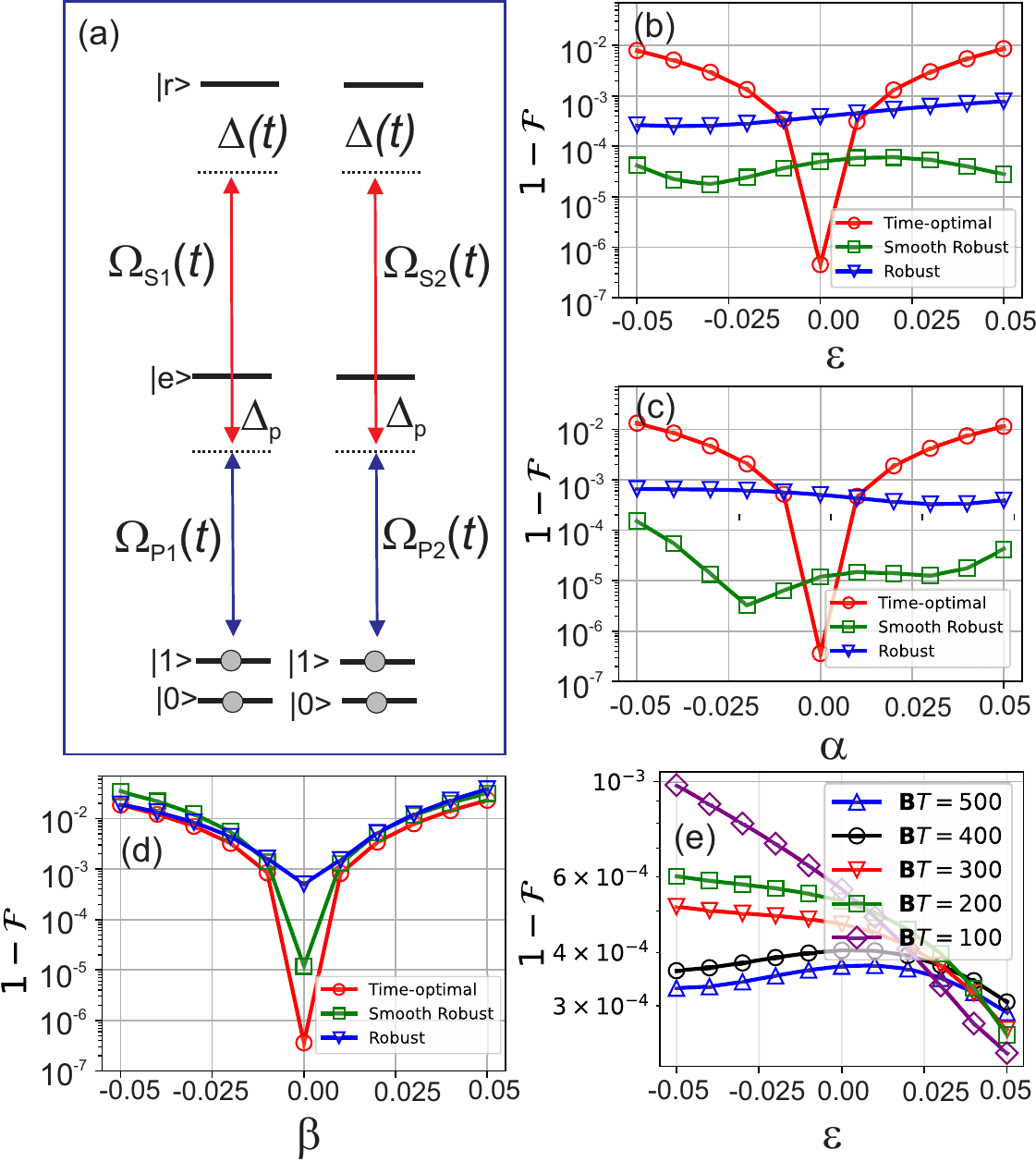}
\caption{(a) Schematic of the two-photon Rydberg excitation  with Rabi frequencies $\Omega_P(t), \Omega_S(t)$ via the intermediate state $\ket{e}$, with a detuning $\Delta_P$ from the intermediate state. (b)-(d) Dependence of the gate infidelity on the (b) variation $\epsilon$ in the global Rabi frequencies, (c) variation $\alpha$ in the ratio  of Rabi frequencies between two atoms and (d) variation $\beta$ in the Rabi frequencies between first and second excitation steps for the smooth time-optimal gate (circles),  amplitude-robust gate with smooth pulse shape (squares) and  amplitude-robust gate with rectangular pulse shape (triangles) in the regime of perfect blockade. The blue detuning from the intermediate state is $T\Delta_P/(2\pi)=-5000$. Finite lifetimes of Rydberg states and intermediate excited state here are not taken into account. (e) Dependence on the gate infidelity of the smooth amplitude-robust gate on the variation in the Rabi frequencies $\epsilon$ for different values of blockade strength $\mathrm{B}$. The blue detuning from the intermediate state is $T\Delta_P/(2\pi)=-5000$. We used the pulse profiles, optimized for finite blockade strength in a single-photon configuration.}
\label{TwoPhoton}
\end{figure}

Two-photon schemes of Rydberg excitation, shown in Fig.~\ref{TwoPhoton}(a), are the most common in experiments with high-fidelity gates for ultracold neutral atoms~\cite{Levine2019,Evered2023,Maller2015}. For rubidium atoms, the ground-state 5S\textsubscript{1/2} atoms are excited through the intermediate 6P\textsubscript{3/2} state to Rydberg \textit{nS} or \textit{nD} states using laser fields with 420~nm and 1013~nm wavelengths at the first and second excitation steps, respectively~\cite{Levine2019}. For cesium atoms with the 6S\textsubscript{1/2} ground state, the excitation path goes through the intermediate  7P\textsubscript{1/2} state using laser radiation at 459~nm and 1038~nm at the first and second excitation steps, respectively~\cite{Maller2015}. The lifetimes of the intermediate excited states for rubidium and cesium atoms for these configurations are 118~ns for rubidium 6P\textsubscript{3/2} state and 155~ns for cesium 7P\textsubscript{3/2} state~\cite{Vsibalic2017ARC}. These lifetimes are close to each other, which ensures similar gate performance. 

Another scheme for two-photon laser excitation in rubidium through the intermediate  5P\textsubscript{3/2}  state uses laser fields with 780~nm and 480~nm wavelengths~\cite{Saffman2010}. The disadvantage of this scheme is a shorter 26~ns lifetime of the intermediate 5P\textsubscript{3/2} state. However, below we will show that this scheme is of interest for individual addressing due to the smaller difference in laser excitation wavelengths, comparing to the excitation through 6P\textsubscript{3/2}  state. Another advantage of this scheme is a smaller residual Doppler shift in counterpropagating beams geometry.

The Hamiltonian describing two-photon excitation for a single atom labeled as 1 or 2 is written similarly to our previous works, taking into account a time-dependent phase of the laser pulse $\xi(t)$~\cite{Saffman2020,Beterov2026}:
 
\begin{eqnarray}
\label{Htwophoton}
{\mathcal H}_{\rm 1,2}&=&\frac{\Omega_{P\rm 1/2}(t)}{2}\ket{p}_{\rm 1,2}\bra{1}+\frac{\Omega_{S\rm 1,2}(t)\mathrm{exp}\left[i\xi(t)\right]}{2}\ket{r}_{\rm 1,2}\bra{p}\nonumber\\
&+&\Delta_P\ket{p}_{\rm 1,2}\bra{p} +\Delta(t)\ket{r}_{\rm 1,2}\bra{r} + \rm H.c.. 
\end{eqnarray}

\noindent In order to suppress spontaneous decay from the intermediate excited states, large detuning $\Delta_P$ on the order of several GHz is required~\cite{Evered2023}. In this case, the intermediate excited state can be eliminated adiabatically, and the three-level system is reduced to a two-level system with an effective two-photon Rabi frequency $\Omega_{\mathrm {two-photon}}=\Omega_P \Omega_S /2 |\Delta_P|$. Here $\Omega_P$ and $\Omega_S$ are the Rabi frequencies at the first and second excitation steps, respectively, and $\Delta_P$ is the detuning from the intermediate excited state, as shown in Fig.~\ref{TwoPhoton}(a). The difference between Rabi frequencies at the first and second steps of laser excitation $\Omega_P  \neq \Omega_S$, results in an additional light shift of the two-photon resonance $(\Omega_S^2-\Omega_P^2)/(4\Delta_P)$~\cite{Saffman2010}. 

In our simulations we applied the phase shift  only to the second step of laser excitation, as it can be seen from the Eq.~\eqref{Htwophoton}. We used two identically shaped pulse profiles 
$\Omega_S\left(t\right)=\Omega_P\left(t\right)=\sqrt{2|\Delta_P| \Omega\left(t\right)}$, where $\Omega\left(t\right)$ is a Rabi frequency taken from the pulse profile of the optimized  single-photon gate protocol. We found that blue detuning from the intermediate state $\Delta_P <0$ provides slightly better  performance for amplitude-robust gate protocols at moderate values of $|\Delta_P| T \sim 2\pi\cdot 5000$.

At first, we assumed perfect blockade and sufficiently large detuning from the intermediate state $T\Delta_P/(2\pi)=-5000$ which ensures applicability of adiabatic elimination of the intermediate state. Figure~\ref{TwoPhoton}(b) shows the calculated dependence of the gate infidelity on the variation $\epsilon$ of the global two-photon Rabi frequency. In order to avoid the undesired light shift we have taken identical Rabi frequencies  $\Omega_P(t)=\Omega_{\mathrm{two-photon}}(t)\sqrt{(1+\epsilon)}$ and $\Omega_S(t)=\Omega_{\mathrm{two-photon}}(t)\sqrt{(1+\epsilon)}$ for both steps of laser excitation. For an effective two-photon Rabi frequency we used the pulse profiles, shown in Fig.~\ref{PulseShapes}(a) and Fig.~\ref{PulseShapes}(e).

In contrast to a single-photon case, shown in Fig.~\ref{PulseShapes}, at two-photon configuration the smooth laser pulse profiles outperform gate protocols with rectangular pulses, for both amplitude-robust and time-optimal gate protocols.  From Fig.~\ref{TwoPhoton}(b) it is clear that the robustness to variation in Rabi frequency is preserved in  two-photon configuration, when we adopt the pulse parameters, optimized for a single-photon case. 

In the second case, the intensities of laser fields acting on atoms are different, but the Rabi frequencies  remain identical at the first and second excitation steps   for each of the atoms. Therefore, light shifts are avoided. This means  $\Omega_{S2}(t)=\Omega_{S1}(t)(1+\alpha)$ and $\Omega_{P2}(t)=\Omega_{P1}(t)(1+\alpha)$ with  $\Omega_{S1}(t)=\Omega_{P1}(t)$ and $\Omega_{S2}(t)=\Omega_{P2}(t)$. The illumination of the atomic array by weakly focused laser beams with intensities slightly dropping  near the array edges is an example of such a configuration.  The gate infidelities shown in Fig.~\ref{TwoPhoton}(c) illustrate robustness of the gate protocol to such asymmetry in laser intensities experienced by the atoms.

Now we consider the effect of the imbalance between laser intensities of the first and second steps of laser excitation with $\Omega_{S1}(t)=\Omega_{P1}(t)(1+\beta)$ and $\Omega_{S2}(t)=\Omega_{P2}(t)(1+\beta)$ with $\Omega_{P1}(t)=\Omega_{P2}(t)$.  In this case, the two-photon Rabi frequency is identical for both atoms. The calculated dependences of gate infidelities on $\beta$  are shown in  Fig.~\ref{TwoPhoton}(d). All gate protocols demonstrate poor performance, since they are not robust to the light shifts, which are caused by the imbalance of light intensities at the fist and second step of laser excitation. Such imbalance may result from the independent drifts of intensities of two different lasers used at the first and second steps of laser excitation or from different spatial intensity profiles of the laser beams used at first and second steps of laser excitation. This problem can be overcome by a single-photon or three-photon laser excitation~\cite{Bezuglov2025, Spirin2026}.

Finally, we studied the gate performance at finite values of blockade strength $\sf B$. We used the pulse profiles optimized for the finite blockade strengths in a single-photon configuration, which are previously illustrated in Fig.~\ref{FiniteBlockade}(d). The calculated dependences of gate infidelities on variation $\epsilon$ of the global two-photon Rabi frequency are shown in Fig.~\ref{TwoPhoton}(e). Both amplitude robustness and small infidelities are illustrated even for relatively weak blockade strengths around $\mathsf {B}T=100$.

\section{Monte-Carlo simulation of atomic motion}
\label{sec.MonteCarlo}

The thermal motion of atoms in the optical dipole trap is an important source of uncertainty in the Rabi frequency for tightly focused laser beams, which results in gate errors.  Near the trap minimum, the potential is harmonic and can be described as $U(r,z)=U_0+\frac{1}{2}M\omega_r r^2+\frac{1}{2}M\omega_z z^2$ where $M$ is the mass of the atom, $\omega_r $  and $\omega_z $ are radial and axial trap frequencies and $U_0$ is the potential depth at the trap center~\cite{Grimm2000}. For red-detuned laser beam with waist $w_0$ and Rayleigh length $z_R=\pi w_0^2/\lambda$ the trap frequencies are approximated as following: 
\be 
\label{trap_frequencies}
\omega_r\approx\sqrt{\frac{4|U_0|}{M w_0^2}}, \ \ \omega_z\sim\sqrt{\frac{2|U_0|}{M z_R^2}}.
\ee
\noindent
For trap depth $U_0/k_B=100\,\mu$K, where $k_B$ is a Boltzmann constant, wavelength $\lambda=850$~nm  and beam waist $w_0=1,\mu$m  we obtain $\omega_r/(2\pi)$=31~kHz and $\omega_z/(2\pi)$=6~kHz for \textsuperscript{87}Rb atoms.

The trajectories of the thermal motion of the atom in an optical dipole trap can be treated classically. We consider a sample of the initial positions of atoms $x_0$, $y_0$ and $z_0$. For a finite temperature $T_a$ the uncertainties of radial and axial positions are described by Gaussian distributions with standard deviation
\be 
\label{trap_variances}
\sigma_r\approx\sqrt{\frac{k_B T_a}{M \omega_r^2}}, \ \ \sigma_z\approx\sqrt{\frac{k_B T_a}{M \omega_z^2}}.
\ee
\noindent The initial atomic velocities $v_{x0}$, $v_{y0}$, $v_{z0}$ are described by the standard deviation of the single-dimensional velocity distribution $\sigma_v\approx\sqrt{\frac{k_B T_a}{M}}$. The analytic solution for the classical motion in a three-dimensional harmonic trap is
\begin{eqnarray}
\label{trajectories}
x&=&x_0\textrm{cos}(\omega_r t)+\frac{v_{x0}}{\omega_r}\textrm{sin}(\omega_r t),\nonumber \\
y&=&y_0\textrm{cos}(\omega_r t)+\frac{v_{y0}}{\omega_r}\textrm{sin}(\omega_r t), \nonumber \\
z&=&z_0\textrm{cos}(\omega_z t)+\frac{v_{z0}}{\omega_z}\textrm{sin}(\omega_z t).  
\end{eqnarray}

\begin{figure}[!t]
\includegraphics[width=\columnwidth]{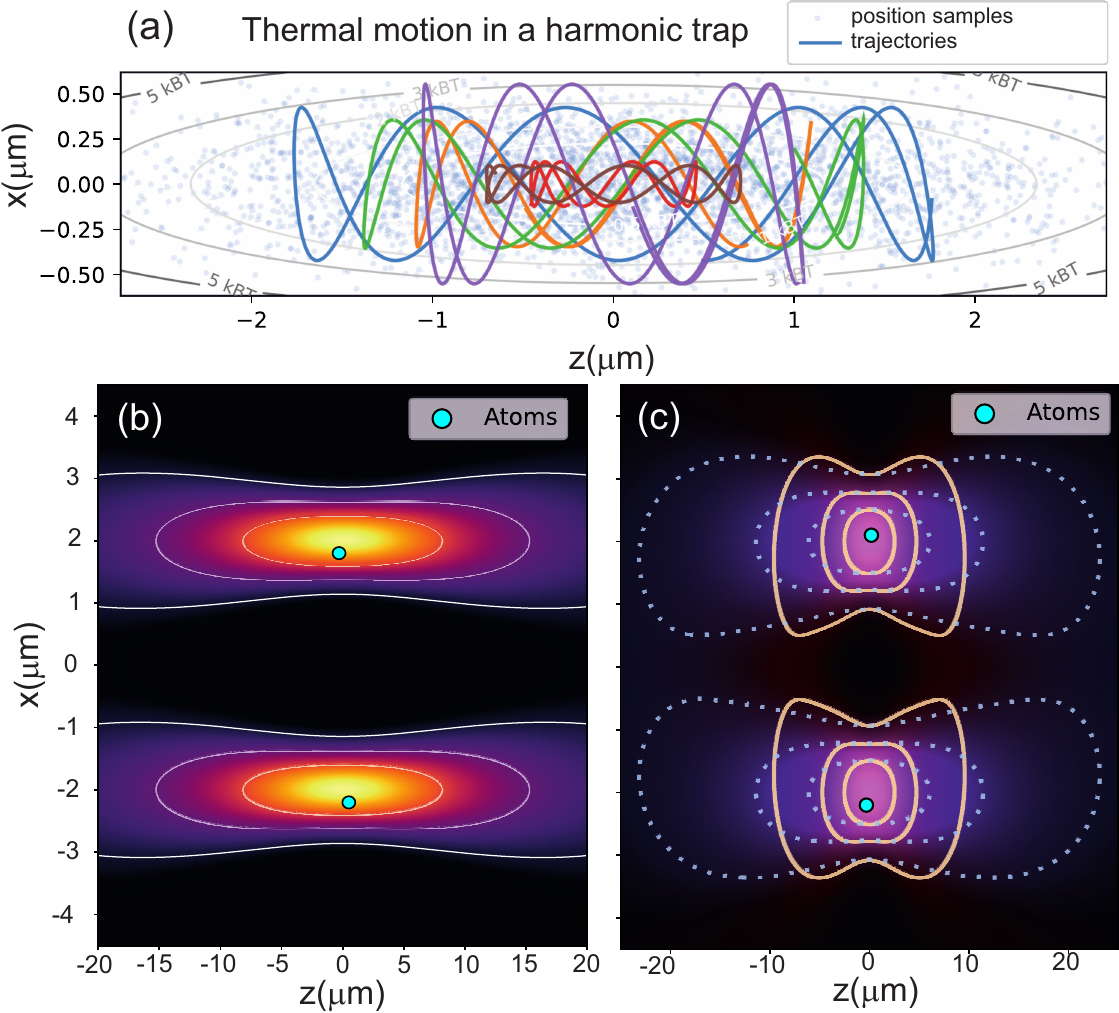}
\caption{ (a) Example of classical trajectories of the atomic motion in the optical dipole trap at temperature $T_a=10\,\mu$K. (b) Spatial distribution of the intensity of two trapping laser beams with 850~nm wavelength, 1~$\mu$m waist and  4~$\mu$m distance between the traps. (c)  Contour plot of the spatial distribution of two laser beams with wavelengths 420~nm and 1013~nm and 1~$\mu$m waist, overlapped with the trapping beams.}
\label{AtomicTrap}
\end{figure}

\noindent  From the Eq.~\eqref{trajectories} the atomic velocity along the Z axis is written as $v_z=-z_0 \omega_z  \textrm{sin}(\omega_z t)+ v_{z0}\textrm{cos}(\omega_z t)$. The thermal distribution of the atomic positions and several trajectories of the atomic motion are illustrated in Fig.~\ref{AtomicTrap}(a) at temperature $T_a=10\,\mu$K.

We performed a simple simulation of the dependence of gate performance on the temperature $T_a$ of trapped \textsuperscript{87}Rb atoms in two spatially separated optical dipole traps, as shown in Fig.~\ref{AtomicTrap}(b) for two traps with 1~$\mu$m waist and  4~$\mu$m distance between the traps, assuming the same 850~nm wavelength of the trapping light. At first, we considered perfect blockade and used the laser pulse profiles optimized for this case. At this stage we neglected finite lifetimes of excited states. The reason is to clearly show the temperature-dependent effects of displacement and motion of atoms, excited by spatially inhomogeneous Rydberg beams. In contrast, the effects of finite lifetimes depend on the choice of the particular Rydberg state, and on the detuning from the intermediate excited states, which can be arbitrarily selected.

The spatial distribution of two laser beams with wavelengths 420~nm and 1013~nm, which are used for two-photon Rydberg excitation through the intermediate 6P\textsubscript{3/2} state in   modern experiments on entanglement of rubidium atoms,  is shown in Fig.~\ref{AtomicTrap}(c). It illustrates the difference between the Rayleigh lengths of two tightly focused beams with 1~$\mu$m waist, overlapped with the trapping beams. Due to the difference in wavelengths, the first step laser beam  with 420~nm linewidth has much longer Rayleigh length $z_{420}=7.5~\mu$m, compared to the Rayleigh length of the second step beam with 1013~nm linewidth $z_{1013}=3.1~\mu$m. Therefore,  it is not possible to  balance the Rabi frequencies for both steps of laser excitation in axial and radial directions simultaneously.

\begin{figure}[!t]
\includegraphics[width=\columnwidth]{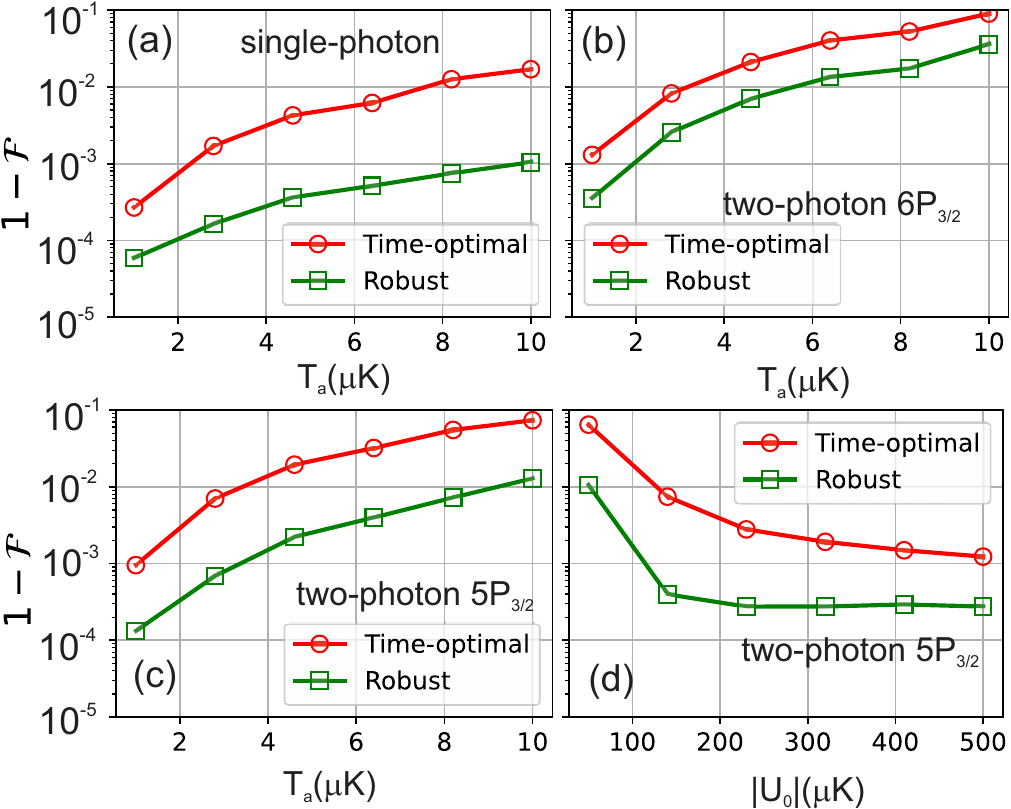}
\caption{ (a)-(d) Monte-Carlo simulation of the dependence of the gate infidelities on the temperature of trapped atoms  for time-optimal (circles) and amplitude-robust (squares) gate protocols with potential depth $|U_0|=100\,\mu$K for the the cases of (a) single-photon laser excitation around 300~nm wavelength (297~nm for rubidium and 302~nm for strontium-88 atoms),  and gate duration $T=100$~ns;  (b) two-photon laser excitation of rubidium atoms with 420~nm and 1013~nm wavelengths and  $T=1\,\mu$s; (c) two-photon laser excitation of rubidium atoms with 780~nm and 480~nm wavelengths and  $T=1\,\mu$s. (d) dependence of the gate infidelities on the potenital depth at temperature $T=5\,\mu$K for two-photon laser excitation of rubidium atoms with 780~nm and 480~nm wavelengths and  $T=1\,\mu$s.}
\label{MonteCarlo}
\end{figure}

To clearly illustrate the difference between amplitude-robust and time-optimal gate protocols, we first considered a simple case of a single-photon laser excitation of two atoms by two focused laser beams with 297~nm wavelength and 1~$\mu$m waist, overlapped with the trapping beams. We sampled the atomic trajectories using Eqs.~\eqref{trap_variances}-\eqref{trajectories} with  $U_0/k_B=100\,\mu$K and temperatures in the range from 1~$\mu$K to 10~$\mu$K. From each trajectory we selected one atomic position and velocity at random time during one period of classical motion along Z axis. Then we calculated the variation of the Rabi frequency in a focused laser beam depending on the atomic position as $\Omega(r,z)=\Omega(r=0,z=0)\mathrm{exp}(-r^2/w_0^2)\mathrm{exp}(-z^2/(2 z_R^2))$ and the Doppler shift $kv_z$ of the resonance where $k$ is the wavenumber. 

Single-photon laser excitation results in much larger residual Doppler shifts, compared to two-photon and three-photon schemes with counterpropagating beam geometry. The phase error due to the Doppler effect is linear in time. Therefore, in the simulations of the single-photon scheme we assumed short gate duration $T=100$~ns. Short gate times are much easier to implement for a single-photon excitation, as it does not require large detunings from the intermediate excited states, resulting in large laser intensities.
 
As  here we were interested in the intrinsic limits on gate infidelity due to thermal motion, Rydberg lifetimes and other noise sources were not taken into account. Results of the simulation of the dependence of gate infidelities on temperature are shown in Fig.~\ref{MonteCarlo}(a) for time-optimal  (circles) and amplitude-robust (squares) gate protocols with constant Rabi frequencies. From the simulation it is clear that amplitude-robust protocol substantially outperforms the time-optimal scheme. Single-photon excitation by UV light is  used in experiments with strontium-88 atoms.  For alkali-metal atoms,  two-photon excitation schemes are more common, but the entanglement with single-photon excitation has been already reported for cesium~\cite{Chow2024}. 

The simulation results for the case of two-photon excitation by laser beams with wavelengths of 420~nm and 1013~nm, as shown in Fig.~\ref{AtomicTrap}(c), are presented in Fig.~\ref{MonteCarlo}(b). We used $\Delta_P/(2\pi)=-5$~GHz detuning from the intermediate state, and a potential depth $U_0=100\,\mu$K. The gate duration was $T=1\, \mu$s. We neglected the finite lifetimes of the Rydberg and intermediate state to study the intrinsic gate infidelity due to atomic motion. However, results of our simulations are not fully scalable, as the influence of the residual Doppler shifts depends on the gate duration. In general, shorter gate durations require increases in both the blockade strength $\mathsf{B}$ and the detuning from the intermediate state $\Delta_P$. This reduces the effects of finite lifetimes and Doppler shifts of the resonance, thereby improving gate fidelity. Perfect compensation of the Doppler effect is possible when three-photon laser excitation is used~\cite{Demtroder2015,Ryabtsev2011}. 

Although at two-photon excitation the amplitude-robust protocol still demonstrates reduced infidelities, compared with the time-optimal scheme, the difference between them is smaller, compared to the single-photon case. The main source of infidelity is the imbalance in Rabi frequencies for the first and second steps of laser excitation, when the atoms are moving in the axial direction. This imbalance results in light shifts that strongly worsen the gate performance. In this case, the gate fidelity of both schemes can be improved by using Rydberg beams with waists slightly larger than the trap waist, or by increasing the potential depth (which provides tighter localization of atoms), and by further cooling of atoms. All these approaches were successfully employed in a recent experiment on a high-fidelity  gate with individual addressing~\cite{Radnaev2025}. However, increasing the waist of the Rydberg excitation beams increases cross-talk between neighboring atoms, and higher potential depths require more laser power for large-scale atomic arrays. Although the overall gate fidelity in our simulations in such conditions increases, the difference in performance of two gate protocols remains unchanged.

Surprisingly, an alternative scheme of two-photon laser excitation through the loweer intermediate 5P\textsubscript{3/2} state, using wavelengths of 780~nm and 480~nm, provides substantially better results, as shown in Fig.~\ref{MonteCarlo}(e). This is due to the  reduced differences in the Rayleigh lengths for the 780~nm and 480~nm laser beams, which are 4.0~$\mu$m and 6.5~$\mu$m, respectively. Another advantage of this scheme is the reduced Doppler shift in the counterpropagating beam geometry.

We also consider the dependence of the gate infidelities on the potential depth $U_0$, which determines the tightness of the localization of trapped atoms. For this analysis, we used two-photon excitation with wavelengths of 780~nm and 480~nm. As $|U_0|$ increases, the gate infidelities decrease for both gate protocols, as shown in Fig.~\ref{MonteCarlo}(f). The advantages of amplitude-robust gate are more pronounced for shallow traps, which are of interest for large-scale atomic arrays.

\begin{figure}[!t]
\includegraphics[width=\columnwidth]{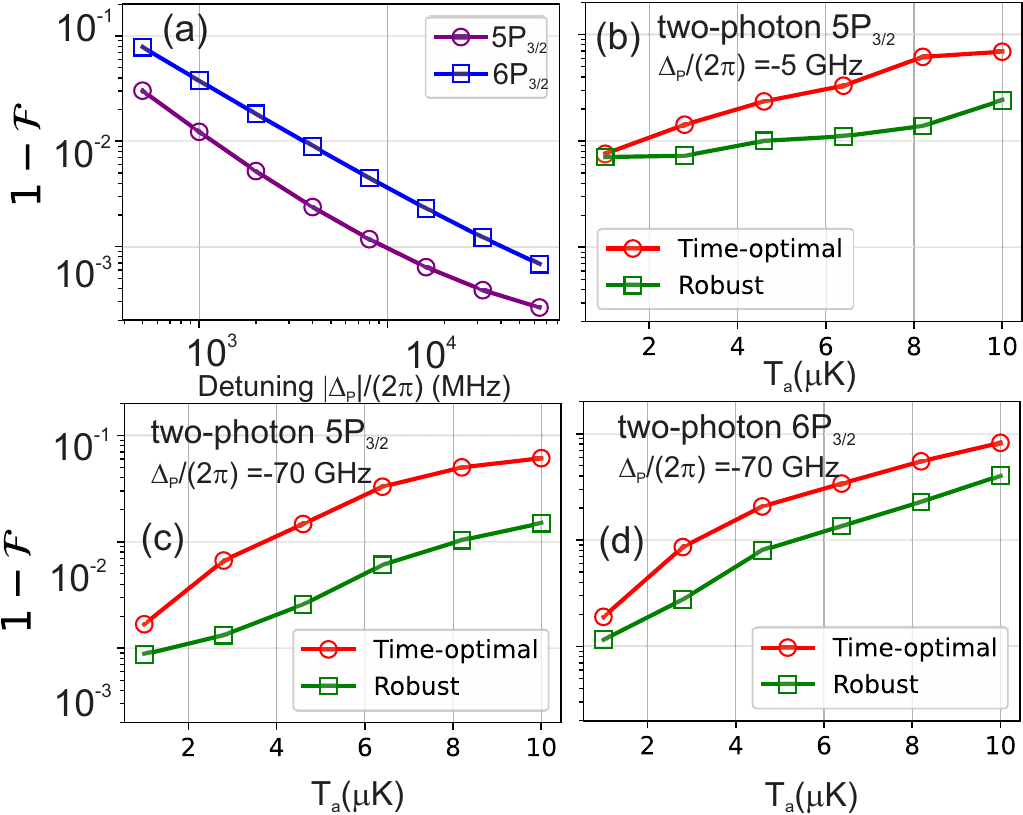}
\caption{ (a) Dependence of  the gate infidelity on detuning from the intermediate state $\Delta_P$ for the schemes of two-photon laser excitation of Rb 76S state through the 5P\textsubscript{3/2} state (squares) and  6P\textsubscript{3/2} state (circles) for gate duration $T=100$~ns taking into account atomic lifetimes. (b)-(d) Monte-Carlo simulation of the dependence of the gate infidelities on the temperature of trapped atoms  for time-optimal (circles) and amplitude-robust (squares) gate protocols with potential depth $|U_0|=100\,\mu$K taking into account atomic lifetimes and finite blockade strength for the the cases of  (b) two-photon laser excitation of rubidium atoms with 780~nm and 480~nm wavelengths and  $T=0.1\,\mu$s and blue detuning  $\Delta_P/(2\pi)=-70$~GHz from the intermediate 5P\textsubscript{3/2} state; (c) two-photon laser excitation of rubidium atoms with 780~nm and 480~nm wavelengths and  $T=0.1\,\mu$s and  detuning  $\Delta_P/(2\pi)=-5$~GHz from the intermediate 5P\textsubscript{3/2} state; (d) (b) two-photon laser excitation of rubidium atoms with 420~nm and 1013~nm wavelengths and  $T=0.1\,\mu$s and  blue detuning  $\Delta_P/(2\pi)=-70$~GHz from the intermediate 6P\textsubscript{3/2} state.
}
\label{Lifetimes}
\end{figure}

This laser excitation   limited by relatively short lifetimes of 26~ns of the 5P\textsubscript{3/2} state in Rb, in contrast with longer lifetime  of 118~ns of the 6P\textsubscript{3/2} state~\cite{Vsibalic2017ARC}. We compare the dependence of the infidelities of amplitude-robust gates on the detuning $\Delta_P$ from the intermediate state for these two excitation schemes, as shown in Fig.~\ref{Lifetimes}(a). Here we consider excitation to a Rydberg 76S state with 180~$\mu$s room-temperature lifetime~\cite{Beterov2009} and a gate duration of $T=100$~ns. Although  excitation through the 6P\textsubscript{3/2} state is clearly advantageous for the same values of  $\Delta_P$, it is possible to  achieve reasonably high fidelities for both excitation schemes at higher values of detuning. In this case the better amplitude robustness of the laser excitation scheme through the intermediate 5P\textsubscript{3/2} state can be employed to obtain reasonably high gate fidelities with focused laser beams.  Increasing the detuning from the intermediate state requires large values of Rabi frequencies at both excitation steps. However,  this is much easier to achieve with tightly focused laser beams, compared to global illumination of the whole atomic array.

We performed Monte-Carlo simulation of gate performance for two two-photon laser excitation schemes taking into account finite lifetimes of intermediate and Rydberg states and finite blockade strength. The energy of the repulsive van der Waals interaction of two Rb 76S atoms at the spacing of around 4~$\mu$m is approximately described by the  $C_6=-2058$~GHz$\cdot \mu\mathrm{m}^6$ coefficient~\cite{Vsibalic2017ARC}. For the interatomic distance of 4~$\mu$m and 100~ns gate duration it gives the dimensionless blockade strength $\sf BT=640$. Therefore, we used the laser pulse profiles optimized for $\sf BT=500$ both for amplitude-robust and time-optimal gate. In the simulation, the interaction energies were calculated for all random spatial positions of trapped atoms to take into account shot-to-shot variation of the blockade strength.

The simulated gate infidelities for moderate detuning from the intermediate  5P\textsubscript{3/2} state $\Delta_P/(2\pi)=-5$~GHz for laser excitation scheme with 780~nm and 480~nm wavelengths demonstrates relatively large infidelities for both gate schemes. With increase of the detuning from the intermediate state   $\Delta_P/(2\pi)=-70$~GHz it becomes possible to reach the infidelities below 0.1\%, as shown in Fig.~\ref{Lifetimes}(c). Amplitude-robust scheme is clearly advantageous over the time-optimal gate. For laser excitation with the same detuning $\Delta_P/(2\pi)=-70$~GHz from the intermediate  6P\textsubscript{3/2} state the performance of the amplitude-robust gate is slightly worse, while it still has some advantage over the time-optimal gate.

\section{Discussion}
\label{sec.Discussion}

In our gate protocols we used chopped random basis (CRAB) initial phase profiles with an additional linear shift, which allowed us to identify analytically defined, scalable phase profiles for various amplitude shapes of laser pulses. In particular, we considered an idealized rectangular pulse with constant Rabi frequency, and a smooth time profile with finite rise and fall times. 

Our gate profiles are scalable. The rise and fall times of the smooth profile are approximately 10\% of the pulse duration. This allows for its experimental implementation using acousto-optical modulators with typical rise times of tens of nanoseconds, which can be reduced to about 5~ns using tightly focused laser beams. The analytically defined time dependences of the phase of the laser pulse can be obtained using electrooptical modulators (EOMs). The free-space EOMs typically have bandwidths less than 100~MHz, which makes it challenging to obtain the required phase profiles for gate times less than 1~$\mu$s. However, the bandwidths of  modern commercial fiber-coupled EOMs reach 40~GHz which substantially simplifies this task even for extremely short gate times.

Our pulse profiles can be applied in experiments on high-fidelity entanglement of neutral atoms in globally illuminated atomic arrays,  reducing the impact of laser intensity variations on gate fidelities. Although for two-photon configuration our protocol is not robust to independent drifts of laser intensities, which result in the undesired light shifts of the two-photon resonance, it still has advantages when the atomic array is illuminated by weakly focused laser beams, with identical spatial intensity distributions, and with a slight drop of the laser intensity at the edges of the array. In this case the effect of spatial inhomogeneity of Rabi frequencies within the atomic array can be substantially reduced.

We studied the robustness of our gate protocol in the case of two individually addressed atoms for single-photon and two-photon Rydberg excitation.   Our scheme strongly reduces gate infidelities caused by non-uniformity in the Rabi frequencies experienced by the two atoms during individual addressing with tightly focused laser beams in a single-photon excitation scheme, which is commonly used in experiments with strontium-88 atoms. This is also confirmed by Monte-Carlo simulations for single-photon excitation of atoms at finite temperatures in an optical dipole trap. In our previous work~\cite{Beterov2026} we have shown that the gate infidelities due to finite Rydberg lifetimes in single-photon excitation case can be reduced below $10^{-4}$ for gate durations around 100~ns, if sufficiently large principal quantum number is chosen. When Rydberg \textit{nP} states are excited, special attention should be paid to precise control of stray electric fields due to higher polarizabilities of  \textit{nP} states compared to \textit{nS} states~\cite{Spirin2026}. 

Two-photon Rydberg excitation driven by tightly focused multi-wavelengths laser beams suffers from the mismatched Rabi frequencies, when the atoms move in both radial and axial directions. This leads to undesirable light shifts, which substatially reduce the fidelities of symmetric gate protocols~\cite{Saffman2025}. We performed Monte-Carlo simulations for two schemes of two-photon laser excitation,  taking into account finite lifetimes of Rydberg and intermediate excited states and several temperature-dependent error sources, arising from random displacement of the atoms from the center of the dipole trap, including spatial inhomogeneity of Rabi frequencies at two steps of laser excitation, Doppler effect and variation of the interatomic interaction energy. The variation of the Rabi frequency resulting from the atomic motion during the 100~ns gate time at temperatures around 10~$\mu$K can be safely neglected. 

In the experiments there are also other error sources, including laser phase noises, laser beam pointing instabilities and time-dependent misalignment between the trapping beams, created using spatial light modulators, and Rydberg excitation beams. These problems are common for all neutral-atom platform experiments. We expect that the Rydberg beam pointing noise will have limited effect for a single-photon excitation, but may further deteriorate the gate performance for two-photon excitation due to light shifts, caused by imbalances between the intensities at the first and second steps of laser excitation. At the same time, our gate protocol will reduce the influence of the mismatch between the trapping and Rydberg laser beams, caused by the thermal drifts in spatial light modulators, since the resulting slow fluctuation of the atomic positions  only weakly affect the light shifts, experienced by the atoms. New approaches based on metasurfaces~\cite{Wang2026} and fiber arrays~\cite{Li2024} are promising to solve the latter problem.

We have shown that the mismatch in the spatial intensity profiles of laser beams at first and second excitation steps has a detrimental effect on the gate fidelity. This problem can be partially addressed by employing a laser excitation scheme through the intermediate 5P\textsubscript{3/2} state for rubidium atoms, using the wavelengths that are closer to each other. This also helps compensate for the residual Doppler effect in counterpropagating beam geometry. Another approach can be based on advanced beam-shaping techniques to keep the spatial distribution of laser beams closer to each other~\cite{Gillen-Christandl2016}. The effect of shorter lifetimes of low-excited 5P\textsubscript{3/2} state compared to 6P\textsubscript{3/2} can be suppressed by the increase of the detuning from the intermediate state. 

We expect that our gate protocol will have potential advantages for three-photon excitation scheme. It has been shown theoretically that three-photon excitation does not suffer from light shifts and allows for substantial reduction of the inhomogeneity of the effective three-photon Rabi frequency in focused laser beams~\cite{Bezuglov2025}. These findings were partly confirmed in our recent experiment~\cite{Spirin2026}, but more research is required. Our gate protocols may also improve gate fidelity by overcoming imperfect matching of the beam profiles in the radial and axial directions.

\section{Conclusion}
\label{sec.Conclusion}

In summary, we have numerically optimized a scheme for a symmetric CZ gate gate to enhance its robustness against variations in the Rabi frequency. The intrinsic robustness of the  gate is almost an order of magnitude increased compared to the previous proposal~\cite{Jandura2023}. We optimized our analytically defined pulse profiles for finite interaction strengths, making them ready for the experimental implementation.

Our scheme is perfectly suited for single-photon Rydberg excitation, which is commonly used in experiments with strontium-88 atoms. For  two-photon Rydberg excitation, most widely used with rudbidium and cesium atoms,  we also demonstrated  advantage compared to commonly used symmetric gate protocols, despite the detrimental effect of the unbalanced light shifts on gate fidelity. This opens the way to high-fidelity entanglement via individual addressing in Rydberg excitation within atomic arrays.

\appendix

\section{Pulse parameters of the amplitude-robust gate}
\label{appendixA}

\begin{table*}
\caption{\label{Table1} Numerically optimized pulse parameters of the amplitude-robust pulse  for $T=1$}
\begin{ruledtabular}
\begin{tabular}{cccc|cccc}
\hline
\multicolumn{4}{c|}{Rectangular profile}&\multicolumn{4}{c}{Smooth profile}\\\hline 
$\sf BT$&$\Omega$ & $\Delta$ &  $c_1$&$\Omega$ & $\Delta$ &  $c_1$ & $\tau$ \\ \hline
Perfect blockade& 17.5508 & -7.290384 &   8.389602 & 22.71428 & 9.1613407 &   0.4423347 & 0.258303\\ \hline
500& 17.4543 & -7.3527088 &   8.327069 &22.42809 & 9.121456 &   0.402433 &  0.248960\\ \hline
400& 17.4363 & -7.3592486 &   8.320766 &22.28584 & 9.121495 &   0.402469&   0.244479\\ \hline
300& 17.3624 & -7.3631701 &   8.316801& 22.40534 & 9.115935 &   0.396924&   0.244022\\ \hline
200& 17.2627 & -7.3957925 &   8.284195&  22.26218 & 9.100121 &   0.381134&   0.239234\\ \hline
100& 17.2910 & -7.521524 &   8.158326 & 21.94527 & 8.882542 &   0.163544&   0.234823\\ 
\end{tabular}
\end{ruledtabular}
\end{table*}

Using a customized version RydOpt~\cite{Rydopt2025} package we optimized  numerically the parameters of the amplitude-robust $C_Z$ gate for rectangular  pulse profile at several different values of blockade strength $\sf B$. The Rabi frequency $\Omega$ and detuning $\Delta$ are listed in Table~\ref{Table1}. The pulse profile is defined by the following ansatz:

\begin{eqnarray}
\label{phase_profile}
\xi\left(t\right)&=&c_1(t-T/2)+\\
&+&\sum_{n=1}^{N}\alpha_n \textrm{sin}\left[\frac{2\pi}{T}n\left(1+\frac{1}{2}\textrm{tanh}(A_n)\left(t-T/2\right)\right)\right]+\nonumber\\
&+&\sum_{n=1}^{N}\beta_n \textrm{cos}\left[\frac{2\pi}{T}n\left(1+\frac{1}{2}\textrm{tanh}(B_n)\left(t-T/2\right)\right)\right].\nonumber
\end{eqnarray}
\noindent Here $N$ can be arbitrary. The obtained linear chirp  $c_1$ of the phase profile is also listed in Table~\ref{Table1}(left-hand panel). The obtained values of the parameters A,B, $\alpha$, $\beta$ in Eq.~\ref{phase_profile} are listed in the Table~\ref{Table2} (left-hand panel).

\begin{table*}
\caption{\label{Table2} Numerically optimized parameters of the rectangular amplitude-robust phase profile $\xi\left(t\right)$}
\begin{ruledtabular}
\begin{tabular}{ccccc|cccc}
\hline
\multicolumn{5}{c|}{Rectangular profile}&\multicolumn{4}{c}{Smooth profile}\\\hline 
\multicolumn{9}{c}{Perfect blockade}\\\hline 
$n$ & $A_n$ & $\alpha_n$ &  $B_n$  & $\beta_n $& $A_n$ & $\alpha_n$ &  $B_n$  & $\beta_n $\\ \hline
1& 0.13288006 & $-0.52012503$ &   1.200519085  & $0.04195050 $& 0.80691385 & $-1.29707873$ &   0.24585137  & $1.74745190 $\\ \hline 
2& 0.04394247 & $-0.903271675$ &   0.17269573 & $0.13556024$&  0.06547802 & $ -0.146930933$ &   0.101617105  & $-0.18819915$\\ \hline 
3& 0.52735281, & $-0.45413667$ &   -1.04072738  & $0.12601295 $& -0.14740044 & $0.228303045$ &   -0.02552445  & $-0.61288804 $\\ \hline
4& 0.88835865 & $-0.62064689$ &     & & 0.82184994 & $0.10687456$ &  -0.3464416  & $ -0.57720572 $\\ \hline
5&&&&& -0.11884555 & $0.5073966$ &   $-0.22318837$  & $ -0.58181292 $\\ \hline
\multicolumn{9}{c}{ $\mathsf{B}T = 500$}\\\hline 
$n$ & $A_n$ & $\alpha_n$ &  $B_n$  & $\beta_n $& $A_n$ & $\alpha_n$ &  $B_n$  & $\beta_n $\\ \hline
1& 0.10451478 & $-0.56598020$ &   $-1.21442020$  & $0.02697618 $& $0.75895250$ & $-1.27770734$ &   0.26043773  & $1.78785527 $\\ \hline 
2&  0.05053879 & $ -0.88657206$ &   0.14298120  & $0.11660236$ &  0.15000384 & $ -0.13677143$ &   0.14020199  & $-0.18637520$\\ \hline 
3& 0.52308065 & $-0.45910349$ &   -1.06399202  & $0.11116937 $& -0.14489804 & $0.19244094$ &   -0.01859949  & $-0.61028683 $\\ \hline
4& 0.86408448 & $-0.61285377$ &     & $  $& 0.87585139 & $0.13807388$ &  -0.32945049  & $ -0.57951516 $\\ \hline 
5&&&&&-0.12694196 & $0.53169680$ &   -0.22010612  & $ -0.60583931 $\\ \hline
\multicolumn{9}{c}{ $\mathsf{B}T= 400$}\\\hline 
$n$ & $A_n$ & $\alpha_n$ &  $B_n$  & $\beta_n $& $A_n$ & $\alpha_n$ &  $B_n$  & $\beta_n $\\ \hline
1& 0.09853586 & $-0.57872456$ &   -1.22338474  & $0.02165976 $ & 0.75801110 & $-1.27571332$ &   0.26541358  & $1.79636443 $\\ \hline 
2&  0.05146921 & $ -0.88231885$ &   0.13746244  & $0.10754482$& 0.16515854 & $ -0.12615040$ &   0.15352967  & $-0.17691115$\\ \hline 
3& 0.51889557 & $-0.46075284$ &   -1.06758142  & $0.10419219 $ &$-0.14604917$ & $0.17576924$ &   -0.01878377  & $-0.60897416 $\\ \hline 
4& 0.85457778 & $-0.61114478$ &    & $  $&0.87964725 & $0.15374154$ &   -0.32277408  & $-0.57946712 $\\ \hline 
5&&&&&$-0.13396932$ & $0.53694409$ &   -0.21844232  & $-0.60951108 $\\ \hline
\multicolumn{9}{c}{ $\mathsf{B}T= 300$}\\\hline 
$n$ & $A_n$ & $\alpha_n$ &  $B_n$  & $\beta_n $& $A_n$ & $\alpha_n$ &  $B_n$  & $\beta_n $\\ \hline
1& 0.09115436 & $-0.59483081$ &   -1.24235809  & $0.00433925 $& 0.74274337 & $-1.27944970$ &   0.27672303  & $1.81143713 $\\ \hline  
2&  0.05579052 & $ -0.87120986$ &   0.13284303  & $0.08623948$&  0.14704898 & $ -0.10822806$ &   0.16748385  & $-0.17052917$\\ \hline 
3& 0.51417094 & $-0.46318254$ &   -1.06639659  & $0.09603453 $& $-0.15532199$ & $0.17144492$ &   -0.01223383  & $-0.59681112 $\\ \hline 
4& 0.83635360 & $-0.59742230$ &     & & 0.89111209 & $0.16720228$ &   -0.32199672  & $-0.57676941 $\\ \hline 
5&&&&&$-0.14029095$ & $0.53982943$ &   -0.21630090  & $-0.61818427 $\\ \hline 
\multicolumn{9}{c}{$\mathsf{B}T = 200$}\\\hline 
$n$ & $A_n$ & $\alpha_n$ &  $B_n$  & $\beta_n $& $A_n$ & $\alpha_n$ &  $B_n$  & $\beta_n $\\ \hline
1& 0.07436273 & $-0.62908751$ &   -1.26135755  & $-0.01013446 $& 0.70239061 & $-1.25187969$ &   0.29171562  & $1.85601401 $\\ \hline  
2&  0.06035638 & $ -0.85429633$ &   0.12038650  & $0.06323481$ &  0.18675236 & $ -0.11221703$ &   0.19709191  & $-0.15800372$\\ \hline 
3& 0.50522369 & $-0.46819031$ &   -1.08155096  & $0.08202378 $ & $-0.16092126$ & $0.15488584$ &   -0.00489378  & $-0.57796860 $\\ \hline 
4& 0.80744433 & $-0.58449715$ &    &  & 0.92344409 & $0.20760311$ &   -0.31229177  & $-0.57940817 $\\ \hline 
5&&&&&$-0.14894663$ & $0.56168497$ &   -0.21158326  & $-0.63603771 $\\ \hline

\multicolumn{9}{c}{ $\mathsf{B}T = 100$}\\\hline 
$n$ & $A_n$ & $\alpha_n$ &  $B_n$  & $\beta_n $& $A_n$ & $\alpha_n$ &  $B_n$  & $\beta_n $\\ \hline
1& $-0.01600492$ & $-0.70317549$ &   -1.26835823  & $0.03499363 $ & 0.43482846 & $-1.03078890$ &   0.50933057  & $1.94475269 $\\ \hline 
2&  $-0.08726589$ & $ -0.79445046$ &   0.14256614  & $0.04927920$&  -0.44078767 & $ 0.40130651$ &   0.09522384  & $0.12691177$\\ \hline 
3& 0.28525007 & $-0.51158571$ &   -0.91836143  & $0.14508431 $ & -0.09147110 & $0.31068280$ &   0.15187274  & $-0.32660344 $\\ \hline
4& 0.70260870 & $-0.71596605$ &     & & 1.01056635 & $0.48520571$ &  -0.15379775  & $ -0.51212704 $\\ \hline 
5&&&&&-0.18432784 & $0.67398065$ &   -0.09459613  & $ -0.55446166 $\\ \hline 

\end{tabular}
\end{ruledtabular}
\end{table*}

For smooth profile we used a  7th-order smoothstep amplitude profile of the laser pulse from RydOpt library:

\be
f(t)=
\begin{cases}
0, & t<0 \ \text{or}\ t>T,\\[6pt]
\Omega\,w\!\left(\dfrac{t}{\tau T/2}\right), & 0\le t<\tau T/2,\\[10pt]
\Omega, & \tau T/2 \le t \le T-\tau T/2,\\[10pt]
\Omega\,w\!\left(\dfrac{T-t}{\tau T/2}\right), & T-\tau T/2 < t \le T.
\end{cases}
\ee

\noindent with
\be
w(x)=-20x^7+70x^6-84x^5+35x^4
\ee
\noindent The Rabi frequency $\Omega$, detuning $\Delta$, linear chirp rate $c_1$ and the characteristic rise time $\tau$ are listed in Table~\ref{Table1} (right-hand panel).
The parameters $A_n$, $B_n$, $\alpha_n$, $\beta_n$ are listed in Table~\ref{Table2} (right-hand panel).

\begin{acknowledgments}
This work was supported by the Foundation for the Advancement of Theoretical Physics and Mathematics "BASIS". 
\end{acknowledgments}

\section*{data availability}
The data that support the findings of this article are openly available \url{https://github.com/beterov/RobustCZ}
\bibliographystyle{apsrev4-2}
\input{Robust2.bbl}

\end{document}

%% file: Robust2.bbl
%

%% file: Robust2.bbl
\begin{thebibliography}{56}%
\makeatletter
\providecommand \@ifxundefined [1]{%
 \@ifx{#1\undefined}
}%
\providecommand \@ifnum [1]{%
 \ifnum #1\expandafter \@firstoftwo
 \else \expandafter \@secondoftwo
 \fi
}%
\providecommand \@ifx [1]{%
 \ifx #1\expandafter \@firstoftwo
 \else \expandafter \@secondoftwo
 \fi
}%
\providecommand \natexlab [1]{#1}%
\providecommand \enquote  [1]{``#1''}%
\providecommand \bibnamefont  [1]{#1}%
\providecommand \bibfnamefont [1]{#1}%
\providecommand \citenamefont [1]{#1}%
\providecommand \href@noop [0]{\@secondoftwo}%
\providecommand \href [0]{\begingroup \@sanitize@url \@href}%
\providecommand \@href[1]{\@@startlink{#1}\@@href}%
\providecommand \@@href[1]{\endgroup#1\@@endlink}%
\providecommand \@sanitize@url [0]{\catcode `\\12\catcode `\$12\catcode
  `\&12\catcode `\#12\catcode `\^12\catcode `\_12\catcode `\%12\relax}%
\providecommand \@@startlink[1]{}%
\providecommand \@@endlink[0]{}%
\providecommand \url  [0]{\begingroup\@sanitize@url \@url }%
\providecommand \@url [1]{\endgroup\@href {#1}{\urlprefix }}%
\providecommand \urlprefix  [0]{URL }%
\providecommand \Eprint [0]{\href }%
\providecommand \doibase [0]{https://doi.org/}%
\providecommand \selectlanguage [0]{\@gobble}%
\providecommand \bibinfo  [0]{\@secondoftwo}%
\providecommand \bibfield  [0]{\@secondoftwo}%
\providecommand \translation [1]{[#1]}%
\providecommand \BibitemOpen [0]{}%
\providecommand \bibitemStop [0]{}%
\providecommand \bibitemNoStop [0]{.\EOS\space}%
\providecommand \EOS [0]{\spacefactor3000\relax}%
\providecommand \BibitemShut  [1]{\csname bibitem#1\endcsname}%
\let\auto@bib@innerbib\@empty
\bibitem [{\citenamefont {Saffman}(2025)}]{Saffman2025}%
  \BibitemOpen
  \bibfield  {author} {\bibinfo {author} {\bibfnamefont {M.}~\bibnamefont
  {Saffman}}\ }\href {https://doi.org/10.48550/arXiv.2505.11218}
  {10.48550/arXiv.2505.11218} (\bibinfo {year} {2025})\BibitemShut {NoStop}%
\bibitem [{\citenamefont {Saffman}(2026)}]{Saffman2026}%
  \BibitemOpen
  \bibfield  {author} {\bibinfo {author} {\bibfnamefont {M.}~\bibnamefont
  {Saffman}}\ }\href {https://doi.org/10.48550/arXiv.2608.30783}
  {10.48550/arXiv.2608.30783} (\bibinfo {year} {2026})\BibitemShut {NoStop}%
\bibitem [{\citenamefont {Shi}(2022)}]{Shi2022}%
  \BibitemOpen
  \bibfield  {author} {\bibinfo {author} {\bibfnamefont {X.-F.}\ \bibnamefont
  {Shi}},\ }\href {https://doi.org/10.1088/2058-9565/ac18b8} {\bibfield
  {journal} {\bibinfo  {journal} {Quantum Science and Technology}\ }\textbf
  {\bibinfo {volume} {7}},\ \bibinfo {pages} {023002} (\bibinfo {year}
  {2022})}\BibitemShut {NoStop}%
\bibitem [{\citenamefont {Chiu}\ \emph {et~al.}(2025)\citenamefont {Chiu},
  \citenamefont {Trapp}, \citenamefont {Guo}, \citenamefont {Abobeih},
  \citenamefont {Stewart}, \citenamefont {Hollerith}, \citenamefont
  {Stroganov}, \citenamefont {Kalinowski}, \citenamefont {Geim}, \citenamefont
  {Evered}, \citenamefont {Li}, \citenamefont {Peters}, \citenamefont
  {Bluvstein}, \citenamefont {Wang}, \citenamefont {Greiner}, \citenamefont
  {Vuletic},\ and\ \citenamefont {Lukin}}]{Chiu2025}%
  \BibitemOpen
  \bibfield  {author} {\bibinfo {author} {\bibfnamefont {N.-C.}\ \bibnamefont
  {Chiu}}, \bibinfo {author} {\bibfnamefont {E.~C.}\ \bibnamefont {Trapp}},
  \bibinfo {author} {\bibfnamefont {J.}~\bibnamefont {Guo}}, \bibinfo {author}
  {\bibfnamefont {M.~H.}\ \bibnamefont {Abobeih}}, \bibinfo {author}
  {\bibfnamefont {L.~M.}\ \bibnamefont {Stewart}}, \bibinfo {author}
  {\bibfnamefont {S.}~\bibnamefont {Hollerith}}, \bibinfo {author}
  {\bibfnamefont {P.}~\bibnamefont {Stroganov}}, \bibinfo {author}
  {\bibfnamefont {M.}~\bibnamefont {Kalinowski}}, \bibinfo {author}
  {\bibfnamefont {A.~A.}\ \bibnamefont {Geim}}, \bibinfo {author}
  {\bibfnamefont {S.~J.}\ \bibnamefont {Evered}}, \bibinfo {author}
  {\bibfnamefont {S.~H.}\ \bibnamefont {Li}}, \bibinfo {author} {\bibfnamefont
  {L.~M.}\ \bibnamefont {Peters}}, \bibinfo {author} {\bibfnamefont
  {D.}~\bibnamefont {Bluvstein}}, \bibinfo {author} {\bibfnamefont {T.~T.}\
  \bibnamefont {Wang}}, \bibinfo {author} {\bibfnamefont {M.}~\bibnamefont
  {Greiner}}, \bibinfo {author} {\bibfnamefont {V.}~\bibnamefont {Vuletic}},\
  and\ \bibinfo {author} {\bibfnamefont {M.~D.}\ \bibnamefont {Lukin}},\ }\href
  {https://doi.org/10.1038/s41586-025-09596-6} {\bibfield  {journal} {\bibinfo
  {journal} {Nature}\ } (\bibinfo {year} {2025})}\BibitemShut {NoStop}%
\bibitem [{\citenamefont {Manetsch}\ \emph {et~al.}(2025)\citenamefont
  {Manetsch}, \citenamefont {Nomura}, \citenamefont {Bataille}, \citenamefont
  {Leung}, \citenamefont {Lv},\ and\ \citenamefont {Endres}}]{Manetsch2025}%
  \BibitemOpen
  \bibfield  {author} {\bibinfo {author} {\bibfnamefont {H.~J.}\ \bibnamefont
  {Manetsch}}, \bibinfo {author} {\bibfnamefont {G.}~\bibnamefont {Nomura}},
  \bibinfo {author} {\bibfnamefont {E.}~\bibnamefont {Bataille}}, \bibinfo
  {author} {\bibfnamefont {K.~H.}\ \bibnamefont {Leung}}, \bibinfo {author}
  {\bibfnamefont {X.}~\bibnamefont {Lv}},\ and\ \bibinfo {author}
  {\bibfnamefont {M.}~\bibnamefont {Endres}},\ }\href
  {https://doi.org/10.1038/s41586-025-09641-4} {\bibfield  {journal} {\bibinfo
  {journal} {Nature}\ }\textbf {\bibinfo {volume} {647}},\ \bibinfo {pages}
  {60} (\bibinfo {year} {2025})}\BibitemShut {NoStop}%
\bibitem [{\citenamefont {Pichard}\ \emph {et~al.}(2024)\citenamefont
  {Pichard}, \citenamefont {Lim}, \citenamefont {Bloch}, \citenamefont
  {Vaneecloo}, \citenamefont {Bourachot}, \citenamefont {Both}, \citenamefont
  {M\'eriaux}, \citenamefont {Dutartre}, \citenamefont {Hostein}, \citenamefont
  {Paris}, \citenamefont {Ximenez}, \citenamefont {Signoles}, \citenamefont
  {Browaeys}, \citenamefont {Lahaye},\ and\ \citenamefont
  {Dreon}}]{Pichard2024}%
  \BibitemOpen
  \bibfield  {author} {\bibinfo {author} {\bibfnamefont {G.}~\bibnamefont
  {Pichard}}, \bibinfo {author} {\bibfnamefont {D.}~\bibnamefont {Lim}},
  \bibinfo {author} {\bibfnamefont {E.}~\bibnamefont {Bloch}}, \bibinfo
  {author} {\bibfnamefont {J.}~\bibnamefont {Vaneecloo}}, \bibinfo {author}
  {\bibfnamefont {L.}~\bibnamefont {Bourachot}}, \bibinfo {author}
  {\bibfnamefont {G.-J.}\ \bibnamefont {Both}}, \bibinfo {author}
  {\bibfnamefont {G.}~\bibnamefont {M\'eriaux}}, \bibinfo {author}
  {\bibfnamefont {S.}~\bibnamefont {Dutartre}}, \bibinfo {author}
  {\bibfnamefont {R.}~\bibnamefont {Hostein}}, \bibinfo {author} {\bibfnamefont
  {J.}~\bibnamefont {Paris}}, \bibinfo {author} {\bibfnamefont
  {B.}~\bibnamefont {Ximenez}}, \bibinfo {author} {\bibfnamefont
  {A.}~\bibnamefont {Signoles}}, \bibinfo {author} {\bibfnamefont
  {A.}~\bibnamefont {Browaeys}}, \bibinfo {author} {\bibfnamefont
  {T.}~\bibnamefont {Lahaye}},\ and\ \bibinfo {author} {\bibfnamefont
  {D.}~\bibnamefont {Dreon}},\ }\href
  {https://doi.org/10.1103/PhysRevApplied.22.024073} {\bibfield  {journal}
  {\bibinfo  {journal} {Phys. Rev. Appl.}\ }\textbf {\bibinfo {volume} {22}},\
  \bibinfo {pages} {024073} (\bibinfo {year} {2024})}\BibitemShut {NoStop}%
\bibitem [{\citenamefont {Norcia}\ \emph {et~al.}(2024)\citenamefont {Norcia},
  \citenamefont {Kim}, \citenamefont {Cairncross}, \citenamefont {Stone},
  \citenamefont {Ryou}, \citenamefont {Jaffe}, \citenamefont {Brown},
  \citenamefont {Barnes}, \citenamefont {Battaglino}, \citenamefont
  {Bohdanowicz}, \citenamefont {Brown}, \citenamefont {Cassella}, \citenamefont
  {Chen}, \citenamefont {Coxe}, \citenamefont {Crow}, \citenamefont {Epstein},
  \citenamefont {Griger}, \citenamefont {Halperin}, \citenamefont {Hummel},
  \citenamefont {Jones}, \citenamefont {Kindem}, \citenamefont {King},
  \citenamefont {Kotru}, \citenamefont {Lauigan}, \citenamefont {Li},
  \citenamefont {Lu}, \citenamefont {Megidish}, \citenamefont {Marjanovic},
  \citenamefont {McDonald}, \citenamefont {Mittiga}, \citenamefont {Muniz},
  \citenamefont {Narayanaswami}, \citenamefont {Nishiguchi}, \citenamefont
  {Paule}, \citenamefont {Pawlak}, \citenamefont {Peng}, \citenamefont
  {Pudenz}, \citenamefont {Rodr\'{\i}guez~P\'erez}, \citenamefont {Smull},
  \citenamefont {Stack}, \citenamefont {Urbanek}, \citenamefont {van~de
  Veerdonk}, \citenamefont {Vendeiro}, \citenamefont {Wadleigh}, \citenamefont
  {Wilkason}, \citenamefont {Wu}, \citenamefont {Xie}, \citenamefont
  {Zalys-Geller}, \citenamefont {Zhang},\ and\ \citenamefont
  {Bloom}}]{Norcia2024}%
  \BibitemOpen
  \bibfield  {author} {\bibinfo {author} {\bibfnamefont {M.~A.}\ \bibnamefont
  {Norcia}}, \bibinfo {author} {\bibfnamefont {H.}~\bibnamefont {Kim}},
  \bibinfo {author} {\bibfnamefont {W.~B.}\ \bibnamefont {Cairncross}},
  \bibinfo {author} {\bibfnamefont {M.}~\bibnamefont {Stone}}, \bibinfo
  {author} {\bibfnamefont {A.}~\bibnamefont {Ryou}}, \bibinfo {author}
  {\bibfnamefont {M.}~\bibnamefont {Jaffe}}, \bibinfo {author} {\bibfnamefont
  {M.~O.}\ \bibnamefont {Brown}}, \bibinfo {author} {\bibfnamefont
  {K.}~\bibnamefont {Barnes}}, \bibinfo {author} {\bibfnamefont
  {P.}~\bibnamefont {Battaglino}}, \bibinfo {author} {\bibfnamefont {T.~C.}\
  \bibnamefont {Bohdanowicz}}, \bibinfo {author} {\bibfnamefont
  {A.}~\bibnamefont {Brown}}, \bibinfo {author} {\bibfnamefont
  {K.}~\bibnamefont {Cassella}}, \bibinfo {author} {\bibfnamefont {C.-A.}\
  \bibnamefont {Chen}}, \bibinfo {author} {\bibfnamefont {R.}~\bibnamefont
  {Coxe}}, \bibinfo {author} {\bibfnamefont {D.}~\bibnamefont {Crow}}, \bibinfo
  {author} {\bibfnamefont {J.}~\bibnamefont {Epstein}}, \bibinfo {author}
  {\bibfnamefont {C.}~\bibnamefont {Griger}}, \bibinfo {author} {\bibfnamefont
  {E.}~\bibnamefont {Halperin}}, \bibinfo {author} {\bibfnamefont
  {F.}~\bibnamefont {Hummel}}, \bibinfo {author} {\bibfnamefont {A.~M.~W.}\
  \bibnamefont {Jones}}, \bibinfo {author} {\bibfnamefont {J.~M.}\ \bibnamefont
  {Kindem}}, \bibinfo {author} {\bibfnamefont {J.}~\bibnamefont {King}},
  \bibinfo {author} {\bibfnamefont {K.}~\bibnamefont {Kotru}}, \bibinfo
  {author} {\bibfnamefont {J.}~\bibnamefont {Lauigan}}, \bibinfo {author}
  {\bibfnamefont {M.}~\bibnamefont {Li}}, \bibinfo {author} {\bibfnamefont
  {M.}~\bibnamefont {Lu}}, \bibinfo {author} {\bibfnamefont {E.}~\bibnamefont
  {Megidish}}, \bibinfo {author} {\bibfnamefont {J.}~\bibnamefont
  {Marjanovic}}, \bibinfo {author} {\bibfnamefont {M.}~\bibnamefont
  {McDonald}}, \bibinfo {author} {\bibfnamefont {T.}~\bibnamefont {Mittiga}},
  \bibinfo {author} {\bibfnamefont {J.~A.}\ \bibnamefont {Muniz}}, \bibinfo
  {author} {\bibfnamefont {S.}~\bibnamefont {Narayanaswami}}, \bibinfo {author}
  {\bibfnamefont {C.}~\bibnamefont {Nishiguchi}}, \bibinfo {author}
  {\bibfnamefont {T.}~\bibnamefont {Paule}}, \bibinfo {author} {\bibfnamefont
  {K.~A.}\ \bibnamefont {Pawlak}}, \bibinfo {author} {\bibfnamefont {L.~S.}\
  \bibnamefont {Peng}}, \bibinfo {author} {\bibfnamefont {K.~L.}\ \bibnamefont
  {Pudenz}}, \bibinfo {author} {\bibfnamefont {D.}~\bibnamefont
  {Rodr\'{\i}guez~P\'erez}}, \bibinfo {author} {\bibfnamefont {A.}~\bibnamefont
  {Smull}}, \bibinfo {author} {\bibfnamefont {D.}~\bibnamefont {Stack}},
  \bibinfo {author} {\bibfnamefont {M.}~\bibnamefont {Urbanek}}, \bibinfo
  {author} {\bibfnamefont {R.~J.~M.}\ \bibnamefont {van~de Veerdonk}}, \bibinfo
  {author} {\bibfnamefont {Z.}~\bibnamefont {Vendeiro}}, \bibinfo {author}
  {\bibfnamefont {L.}~\bibnamefont {Wadleigh}}, \bibinfo {author}
  {\bibfnamefont {T.}~\bibnamefont {Wilkason}}, \bibinfo {author}
  {\bibfnamefont {T.-Y.}\ \bibnamefont {Wu}}, \bibinfo {author} {\bibfnamefont
  {X.}~\bibnamefont {Xie}}, \bibinfo {author} {\bibfnamefont {E.}~\bibnamefont
  {Zalys-Geller}}, \bibinfo {author} {\bibfnamefont {X.}~\bibnamefont
  {Zhang}},\ and\ \bibinfo {author} {\bibfnamefont {B.~J.}\ \bibnamefont
  {Bloom}},\ }\href {https://doi.org/10.1103/PRXQuantum.5.030316} {\bibfield
  {journal} {\bibinfo  {journal} {PRX Quantum}\ }\textbf {\bibinfo {volume}
  {5}},\ \bibinfo {pages} {030316} (\bibinfo {year} {2024})}\BibitemShut
  {NoStop}%
\bibitem [{\citenamefont {Gyger}\ \emph {et~al.}(2024)\citenamefont {Gyger},
  \citenamefont {Ammenwerth}, \citenamefont {Tao}, \citenamefont {Timme},
  \citenamefont {Snigirev}, \citenamefont {Bloch},\ and\ \citenamefont
  {Zeiher}}]{Gyger2024}%
  \BibitemOpen
  \bibfield  {author} {\bibinfo {author} {\bibfnamefont {F.}~\bibnamefont
  {Gyger}}, \bibinfo {author} {\bibfnamefont {M.}~\bibnamefont {Ammenwerth}},
  \bibinfo {author} {\bibfnamefont {R.}~\bibnamefont {Tao}}, \bibinfo {author}
  {\bibfnamefont {H.}~\bibnamefont {Timme}}, \bibinfo {author} {\bibfnamefont
  {S.}~\bibnamefont {Snigirev}}, \bibinfo {author} {\bibfnamefont
  {I.}~\bibnamefont {Bloch}},\ and\ \bibinfo {author} {\bibfnamefont
  {J.}~\bibnamefont {Zeiher}},\ }\href
  {https://doi.org/10.1103/PhysRevResearch.6.033104} {\bibfield  {journal}
  {\bibinfo  {journal} {Phys. Rev. Res.}\ }\textbf {\bibinfo {volume} {6}},\
  \bibinfo {pages} {033104} (\bibinfo {year} {2024})}\BibitemShut {NoStop}%
\bibitem [{\citenamefont {Lin}\ \emph {et~al.}(2025)\citenamefont {Lin},
  \citenamefont {Zhong}, \citenamefont {Li}, \citenamefont {Zhao},
  \citenamefont {Zheng}, \citenamefont {Hu}, \citenamefont {Wu}, \citenamefont
  {Wu}, \citenamefont {Ma}, \citenamefont {Gao}, \citenamefont {Zhu},
  \citenamefont {Su}, \citenamefont {Ouyang}, \citenamefont {Zhang},
  \citenamefont {Rui}, \citenamefont {Chen}, \citenamefont {Lu},\ and\
  \citenamefont {Pan}}]{Lin2025}%
  \BibitemOpen
  \bibfield  {author} {\bibinfo {author} {\bibfnamefont {R.}~\bibnamefont
  {Lin}}, \bibinfo {author} {\bibfnamefont {H.-S.}\ \bibnamefont {Zhong}},
  \bibinfo {author} {\bibfnamefont {Y.}~\bibnamefont {Li}}, \bibinfo {author}
  {\bibfnamefont {Z.-R.}\ \bibnamefont {Zhao}}, \bibinfo {author}
  {\bibfnamefont {L.-T.}\ \bibnamefont {Zheng}}, \bibinfo {author}
  {\bibfnamefont {T.-R.}\ \bibnamefont {Hu}}, \bibinfo {author} {\bibfnamefont
  {H.-M.}\ \bibnamefont {Wu}}, \bibinfo {author} {\bibfnamefont
  {Z.}~\bibnamefont {Wu}}, \bibinfo {author} {\bibfnamefont {W.-J.}\
  \bibnamefont {Ma}}, \bibinfo {author} {\bibfnamefont {Y.}~\bibnamefont
  {Gao}}, \bibinfo {author} {\bibfnamefont {Y.-K.}\ \bibnamefont {Zhu}},
  \bibinfo {author} {\bibfnamefont {Z.-F.}\ \bibnamefont {Su}}, \bibinfo
  {author} {\bibfnamefont {W.-L.}\ \bibnamefont {Ouyang}}, \bibinfo {author}
  {\bibfnamefont {Y.-C.}\ \bibnamefont {Zhang}}, \bibinfo {author}
  {\bibfnamefont {J.}~\bibnamefont {Rui}}, \bibinfo {author} {\bibfnamefont
  {M.-C.}\ \bibnamefont {Chen}}, \bibinfo {author} {\bibfnamefont {C.-Y.}\
  \bibnamefont {Lu}},\ and\ \bibinfo {author} {\bibfnamefont {J.-W.}\
  \bibnamefont {Pan}},\ }\href {https://doi.org/10.1103/2ym8-vs82} {\bibfield
  {journal} {\bibinfo  {journal} {Phys. Rev. Lett.}\ }\textbf {\bibinfo
  {volume} {135}},\ \bibinfo {pages} {060602} (\bibinfo {year}
  {2025})}\BibitemShut {NoStop}%
\bibitem [{\citenamefont {Wang}\ \emph {et~al.}(2026)\citenamefont {Wang},
  \citenamefont {Zhang}, \citenamefont {Zhang}, \citenamefont {Liao},
  \citenamefont {Wang}, \citenamefont {Tian}, \citenamefont {Ji}, \citenamefont
  {Wu}, \citenamefont {Ma}, \citenamefont {Qing}, \citenamefont {Li},
  \citenamefont {Zhang}, \citenamefont {Huang}, \citenamefont {Zhang},
  \citenamefont {Feng}, \citenamefont {Chen},\ and\ \citenamefont
  {Zhai}}]{Wang2026}%
  \BibitemOpen
  \bibfield  {author} {\bibinfo {author} {\bibfnamefont {Y.}~\bibnamefont
  {Wang}}, \bibinfo {author} {\bibfnamefont {Z.}~\bibnamefont {Zhang}},
  \bibinfo {author} {\bibfnamefont {T.}~\bibnamefont {Zhang}}, \bibinfo
  {author} {\bibfnamefont {Y.}~\bibnamefont {Liao}}, \bibinfo {author}
  {\bibfnamefont {H.}~\bibnamefont {Wang}}, \bibinfo {author} {\bibfnamefont
  {Y.}~\bibnamefont {Tian}}, \bibinfo {author} {\bibfnamefont {B.}~\bibnamefont
  {Ji}}, \bibinfo {author} {\bibfnamefont {Y.}~\bibnamefont {Wu}}, \bibinfo
  {author} {\bibfnamefont {L.}~\bibnamefont {Ma}}, \bibinfo {author}
  {\bibfnamefont {C.}~\bibnamefont {Qing}}, \bibinfo {author} {\bibfnamefont
  {C.}~\bibnamefont {Li}}, \bibinfo {author} {\bibfnamefont {W.}~\bibnamefont
  {Zhang}}, \bibinfo {author} {\bibfnamefont {Y.}~\bibnamefont {Huang}},
  \bibinfo {author} {\bibfnamefont {W.}~\bibnamefont {Zhang}}, \bibinfo
  {author} {\bibfnamefont {X.}~\bibnamefont {Feng}}, \bibinfo {author}
  {\bibfnamefont {W.}~\bibnamefont {Chen}},\ and\ \bibinfo {author}
  {\bibfnamefont {H.}~\bibnamefont {Zhai}}\ }\href
  {https://doi.org/10.48550/arXiv.2606.02715} {10.48550/arXiv.2606.02715}
  (\bibinfo {year} {2026})\BibitemShut {NoStop}%
\bibitem [{\citenamefont {Evered}\ \emph {et~al.}(2023)\citenamefont {Evered},
  \citenamefont {Bluvstein}, \citenamefont {Kalinowski}, \citenamefont {Ebadi},
  \citenamefont {Manovitz}, \citenamefont {Zhou}, \citenamefont {Li},
  \citenamefont {Geim}, \citenamefont {Wang}, \citenamefont {Maskara},
  \citenamefont {Levine}, \citenamefont {Semeghini}, \citenamefont {Greiner},
  \citenamefont {Vuletic},\ and\ \citenamefont {Lukin}}]{Evered2023}%
  \BibitemOpen
  \bibfield  {author} {\bibinfo {author} {\bibfnamefont {S.~J.}\ \bibnamefont
  {Evered}}, \bibinfo {author} {\bibfnamefont {D.}~\bibnamefont {Bluvstein}},
  \bibinfo {author} {\bibfnamefont {M.}~\bibnamefont {Kalinowski}}, \bibinfo
  {author} {\bibfnamefont {S.}~\bibnamefont {Ebadi}}, \bibinfo {author}
  {\bibfnamefont {T.}~\bibnamefont {Manovitz}}, \bibinfo {author}
  {\bibfnamefont {H.}~\bibnamefont {Zhou}}, \bibinfo {author} {\bibfnamefont
  {S.~H.}\ \bibnamefont {Li}}, \bibinfo {author} {\bibfnamefont {A.~A.}\
  \bibnamefont {Geim}}, \bibinfo {author} {\bibfnamefont {T.~T.}\ \bibnamefont
  {Wang}}, \bibinfo {author} {\bibfnamefont {N.}~\bibnamefont {Maskara}},
  \bibinfo {author} {\bibfnamefont {H.}~\bibnamefont {Levine}}, \bibinfo
  {author} {\bibfnamefont {G.}~\bibnamefont {Semeghini}}, \bibinfo {author}
  {\bibfnamefont {M.}~\bibnamefont {Greiner}}, \bibinfo {author} {\bibfnamefont
  {V.}~\bibnamefont {Vuletic}},\ and\ \bibinfo {author} {\bibfnamefont {M.~D.}\
  \bibnamefont {Lukin}},\ }\href {https://doi.org/10.1038/s41586-023-06481-y}
  {\bibfield  {journal} {\bibinfo  {journal} {Nature}\ }\textbf {\bibinfo
  {volume} {622}},\ \bibinfo {pages} {268} (\bibinfo {year}
  {2023})}\BibitemShut {NoStop}%
\bibitem [{\citenamefont {Tsai}\ \emph {et~al.}(2025)\citenamefont {Tsai},
  \citenamefont {Sun}, \citenamefont {Shaw}, \citenamefont {Finkelstein},\ and\
  \citenamefont {Endres}}]{Tsai2025}%
  \BibitemOpen
  \bibfield  {author} {\bibinfo {author} {\bibfnamefont {R.~B.-S.}\
  \bibnamefont {Tsai}}, \bibinfo {author} {\bibfnamefont {X.}~\bibnamefont
  {Sun}}, \bibinfo {author} {\bibfnamefont {A.~L.}\ \bibnamefont {Shaw}},
  \bibinfo {author} {\bibfnamefont {R.}~\bibnamefont {Finkelstein}},\ and\
  \bibinfo {author} {\bibfnamefont {M.}~\bibnamefont {Endres}},\ }\href
  {https://doi.org/10.1103/PRXQuantum.6.010331} {\bibfield  {journal} {\bibinfo
   {journal} {Phys Rev X Quantum}\ }\textbf {\bibinfo {volume} {6}},\ \bibinfo
  {pages} {010331} (\bibinfo {year} {2025})}\BibitemShut {NoStop}%
\bibitem [{\citenamefont {Radnaev}\ \emph {et~al.}(2025)\citenamefont
  {Radnaev}, \citenamefont {Chung}, \citenamefont {Cole}, \citenamefont
  {Mason}, \citenamefont {Ballance}, \citenamefont {Bedalov}, \citenamefont
  {Belknap}, \citenamefont {Berman}, \citenamefont {Blakely}, \citenamefont
  {Bloomfield}, \citenamefont {Buttler}, \citenamefont {Campbell},
  \citenamefont {Chopinaud}, \citenamefont {Copenhaver}, \citenamefont {Dawes},
  \citenamefont {Eubanks}, \citenamefont {Friss}, \citenamefont {Garcia},
  \citenamefont {Gilbert}, \citenamefont {Gillette}, \citenamefont {Goiporia},
  \citenamefont {Gokhale}, \citenamefont {Goldwin}, \citenamefont {Goodwin},
  \citenamefont {Graham}, \citenamefont {Guttormsson}, \citenamefont {Hickman},
  \citenamefont {Hurtley}, \citenamefont {Iliev}, \citenamefont {Jones},
  \citenamefont {Jones}, \citenamefont {Kuper}, \citenamefont {Lewis},
  \citenamefont {Lichtman}, \citenamefont {Majdeteimouri}, \citenamefont
  {Mason}, \citenamefont {McMaster}, \citenamefont {Miles}, \citenamefont
  {Mitchell}, \citenamefont {Murphree}, \citenamefont {Neff-Mallon},
  \citenamefont {Oh}, \citenamefont {Omole}, \citenamefont {Parlo~Simon},
  \citenamefont {Pederson}, \citenamefont {Perlin}, \citenamefont {Reiter},
  \citenamefont {Rines}, \citenamefont {Romlow}, \citenamefont {Scott},
  \citenamefont {Stiefvater}, \citenamefont {Tanner}, \citenamefont {Tucker},
  \citenamefont {Vinogradov}, \citenamefont {Warter}, \citenamefont {Yeo},
  \citenamefont {Saffman},\ and\ \citenamefont {Noel}}]{Radnaev2025}%
  \BibitemOpen
  \bibfield  {author} {\bibinfo {author} {\bibfnamefont {A.}~\bibnamefont
  {Radnaev}}, \bibinfo {author} {\bibfnamefont {W.}~\bibnamefont {Chung}},
  \bibinfo {author} {\bibfnamefont {D.}~\bibnamefont {Cole}}, \bibinfo {author}
  {\bibfnamefont {D.}~\bibnamefont {Mason}}, \bibinfo {author} {\bibfnamefont
  {T.}~\bibnamefont {Ballance}}, \bibinfo {author} {\bibfnamefont
  {M.}~\bibnamefont {Bedalov}}, \bibinfo {author} {\bibfnamefont
  {D.}~\bibnamefont {Belknap}}, \bibinfo {author} {\bibfnamefont
  {M.}~\bibnamefont {Berman}}, \bibinfo {author} {\bibfnamefont
  {M.}~\bibnamefont {Blakely}}, \bibinfo {author} {\bibfnamefont
  {I.}~\bibnamefont {Bloomfield}}, \bibinfo {author} {\bibfnamefont
  {P.}~\bibnamefont {Buttler}}, \bibinfo {author} {\bibfnamefont
  {C.}~\bibnamefont {Campbell}}, \bibinfo {author} {\bibfnamefont
  {A.}~\bibnamefont {Chopinaud}}, \bibinfo {author} {\bibfnamefont
  {E.}~\bibnamefont {Copenhaver}}, \bibinfo {author} {\bibfnamefont
  {M.}~\bibnamefont {Dawes}}, \bibinfo {author} {\bibfnamefont
  {S.}~\bibnamefont {Eubanks}}, \bibinfo {author} {\bibfnamefont
  {A.}~\bibnamefont {Friss}}, \bibinfo {author} {\bibfnamefont
  {D.}~\bibnamefont {Garcia}}, \bibinfo {author} {\bibfnamefont
  {J.}~\bibnamefont {Gilbert}}, \bibinfo {author} {\bibfnamefont
  {M.}~\bibnamefont {Gillette}}, \bibinfo {author} {\bibfnamefont
  {P.}~\bibnamefont {Goiporia}}, \bibinfo {author} {\bibfnamefont
  {P.}~\bibnamefont {Gokhale}}, \bibinfo {author} {\bibfnamefont
  {J.}~\bibnamefont {Goldwin}}, \bibinfo {author} {\bibfnamefont
  {D.}~\bibnamefont {Goodwin}}, \bibinfo {author} {\bibfnamefont
  {T.}~\bibnamefont {Graham}}, \bibinfo {author} {\bibfnamefont
  {C.}~\bibnamefont {Guttormsson}}, \bibinfo {author} {\bibfnamefont
  {G.}~\bibnamefont {Hickman}}, \bibinfo {author} {\bibfnamefont
  {L.}~\bibnamefont {Hurtley}}, \bibinfo {author} {\bibfnamefont
  {M.}~\bibnamefont {Iliev}}, \bibinfo {author} {\bibfnamefont
  {E.}~\bibnamefont {Jones}}, \bibinfo {author} {\bibfnamefont
  {R.}~\bibnamefont {Jones}}, \bibinfo {author} {\bibfnamefont
  {K.}~\bibnamefont {Kuper}}, \bibinfo {author} {\bibfnamefont
  {T.}~\bibnamefont {Lewis}}, \bibinfo {author} {\bibfnamefont
  {M.}~\bibnamefont {Lichtman}}, \bibinfo {author} {\bibfnamefont
  {F.}~\bibnamefont {Majdeteimouri}}, \bibinfo {author} {\bibfnamefont
  {J.}~\bibnamefont {Mason}}, \bibinfo {author} {\bibfnamefont
  {J.}~\bibnamefont {McMaster}}, \bibinfo {author} {\bibfnamefont
  {J.}~\bibnamefont {Miles}}, \bibinfo {author} {\bibfnamefont
  {P.}~\bibnamefont {Mitchell}}, \bibinfo {author} {\bibfnamefont
  {J.}~\bibnamefont {Murphree}}, \bibinfo {author} {\bibfnamefont
  {N.}~\bibnamefont {Neff-Mallon}}, \bibinfo {author} {\bibfnamefont
  {T.}~\bibnamefont {Oh}}, \bibinfo {author} {\bibfnamefont {V.}~\bibnamefont
  {Omole}}, \bibinfo {author} {\bibfnamefont {C.}~\bibnamefont {Parlo~Simon}},
  \bibinfo {author} {\bibfnamefont {N.}~\bibnamefont {Pederson}}, \bibinfo
  {author} {\bibfnamefont {M.}~\bibnamefont {Perlin}}, \bibinfo {author}
  {\bibfnamefont {A.}~\bibnamefont {Reiter}}, \bibinfo {author} {\bibfnamefont
  {R.}~\bibnamefont {Rines}}, \bibinfo {author} {\bibfnamefont
  {P.}~\bibnamefont {Romlow}}, \bibinfo {author} {\bibfnamefont
  {A.}~\bibnamefont {Scott}}, \bibinfo {author} {\bibfnamefont
  {D.}~\bibnamefont {Stiefvater}}, \bibinfo {author} {\bibfnamefont
  {J.}~\bibnamefont {Tanner}}, \bibinfo {author} {\bibfnamefont
  {A.}~\bibnamefont {Tucker}}, \bibinfo {author} {\bibfnamefont
  {I.}~\bibnamefont {Vinogradov}}, \bibinfo {author} {\bibfnamefont
  {M.}~\bibnamefont {Warter}}, \bibinfo {author} {\bibfnamefont
  {M.}~\bibnamefont {Yeo}}, \bibinfo {author} {\bibfnamefont {M.}~\bibnamefont
  {Saffman}},\ and\ \bibinfo {author} {\bibfnamefont {T.}~\bibnamefont
  {Noel}},\ }\href {https://doi.org/10.1103/66s8-jj18} {\bibfield  {journal}
  {\bibinfo  {journal} {PRX Quantum}\ }\textbf {\bibinfo {volume} {6}},\
  \bibinfo {pages} {030334} (\bibinfo {year} {2025})}\BibitemShut {NoStop}%
\bibitem [{\citenamefont {Zhang}\ \emph {et~al.}(2026)\citenamefont {Zhang},
  \citenamefont {Liu}, \citenamefont {Bornet}, \citenamefont {Horvath},
  \citenamefont {Peng}, \citenamefont {Ma}, \citenamefont {Huang},
  \citenamefont {Puri},\ and\ \citenamefont {Thompson}}]{Zhang2026}%
  \BibitemOpen
  \bibfield  {author} {\bibinfo {author} {\bibfnamefont {B.}~\bibnamefont
  {Zhang}}, \bibinfo {author} {\bibfnamefont {G.}~\bibnamefont {Liu}}, \bibinfo
  {author} {\bibfnamefont {G.}~\bibnamefont {Bornet}}, \bibinfo {author}
  {\bibfnamefont {S.~P.}\ \bibnamefont {Horvath}}, \bibinfo {author}
  {\bibfnamefont {P.}~\bibnamefont {Peng}}, \bibinfo {author} {\bibfnamefont
  {S.}~\bibnamefont {Ma}}, \bibinfo {author} {\bibfnamefont {S.}~\bibnamefont
  {Huang}}, \bibinfo {author} {\bibfnamefont {S.}~\bibnamefont {Puri}},\ and\
  \bibinfo {author} {\bibfnamefont {J.~D.}\ \bibnamefont {Thompson}},\ }\href
  {https://doi.org/10.1038/s41567-026-03309-0} {\bibfield  {journal} {\bibinfo
  {journal} {Nature Physics}\ }\textbf {\bibinfo {volume} {22}},\ \bibinfo
  {pages} {910} (\bibinfo {year} {2026})}\BibitemShut {NoStop}%
\bibitem [{\citenamefont {Bluvstein}\ \emph {et~al.}(2025)\citenamefont
  {Bluvstein}, \citenamefont {Geim}, \citenamefont {Li}, \citenamefont
  {Evered}, \citenamefont {Ataides}, \citenamefont {Baranes}, \citenamefont
  {Gu}, \citenamefont {Manovitz}, \citenamefont {Xu}, \citenamefont
  {Kalinowski}, \citenamefont {Majidy}, \citenamefont {Kokail}, \citenamefont
  {Maskara}, \citenamefont {Trapp}, \citenamefont {Stewart}, \citenamefont
  {Hollerith}, \citenamefont {Zhou}, \citenamefont {Gullans}, \citenamefont
  {Yelin}, \citenamefont {Greiner}, \citenamefont {Vuletic}, \citenamefont
  {Cain},\ and\ \citenamefont {Lukin}}]{Bluvstein2025}%
  \BibitemOpen
  \bibfield  {author} {\bibinfo {author} {\bibfnamefont {D.}~\bibnamefont
  {Bluvstein}}, \bibinfo {author} {\bibfnamefont {A.~A.}\ \bibnamefont {Geim}},
  \bibinfo {author} {\bibfnamefont {S.~H.}\ \bibnamefont {Li}}, \bibinfo
  {author} {\bibfnamefont {S.~J.}\ \bibnamefont {Evered}}, \bibinfo {author}
  {\bibfnamefont {J.~P.~B.}\ \bibnamefont {Ataides}}, \bibinfo {author}
  {\bibfnamefont {G.}~\bibnamefont {Baranes}}, \bibinfo {author} {\bibfnamefont
  {A.}~\bibnamefont {Gu}}, \bibinfo {author} {\bibfnamefont {T.}~\bibnamefont
  {Manovitz}}, \bibinfo {author} {\bibfnamefont {M.}~\bibnamefont {Xu}},
  \bibinfo {author} {\bibfnamefont {M.}~\bibnamefont {Kalinowski}}, \bibinfo
  {author} {\bibfnamefont {S.}~\bibnamefont {Majidy}}, \bibinfo {author}
  {\bibfnamefont {C.}~\bibnamefont {Kokail}}, \bibinfo {author} {\bibfnamefont
  {N.}~\bibnamefont {Maskara}}, \bibinfo {author} {\bibfnamefont {E.~C.}\
  \bibnamefont {Trapp}}, \bibinfo {author} {\bibfnamefont {L.~M.}\ \bibnamefont
  {Stewart}}, \bibinfo {author} {\bibfnamefont {S.}~\bibnamefont {Hollerith}},
  \bibinfo {author} {\bibfnamefont {H.}~\bibnamefont {Zhou}}, \bibinfo {author}
  {\bibfnamefont {M.~J.}\ \bibnamefont {Gullans}}, \bibinfo {author}
  {\bibfnamefont {S.~F.}\ \bibnamefont {Yelin}}, \bibinfo {author}
  {\bibfnamefont {M.}~\bibnamefont {Greiner}}, \bibinfo {author} {\bibfnamefont
  {V.}~\bibnamefont {Vuletic}}, \bibinfo {author} {\bibfnamefont
  {M.}~\bibnamefont {Cain}},\ and\ \bibinfo {author} {\bibfnamefont {M.~D.}\
  \bibnamefont {Lukin}},\ }\bibfield  {journal} {\bibinfo  {journal}
  {arXiv:2506.20661}\ }\href {https://doi.org/10.48550/arXiv.2506.20661}
  {10.48550/arXiv.2506.20661} (\bibinfo {year} {2025})\BibitemShut {NoStop}%
\bibitem [{\citenamefont {Reichardt}\ \emph {et~al.}(2024)\citenamefont
  {Reichardt}, \citenamefont {Paetznick}, \citenamefont {Aasen}, \citenamefont
  {Basov}, \citenamefont {Bello-Rivas}, \citenamefont {Bonderson},
  \citenamefont {Chao}, \citenamefont {van Dam}, \citenamefont {Hastings},
  \citenamefont {Mishmash}, \citenamefont {Paz}, \citenamefont {da~Silva},
  \citenamefont {Sundaram}, \citenamefont {Svore}, \citenamefont {Vaschillo},
  \citenamefont {Wang}, \citenamefont {Zanner}, \citenamefont {Cairncross},
  \citenamefont {Chen}, \citenamefont {Crow}, \citenamefont {Kim},
  \citenamefont {Kindem}, \citenamefont {King}, \citenamefont {McDonald},
  \citenamefont {Norcia}, \citenamefont {Ryou}, \citenamefont {Stone},
  \citenamefont {Wadleigh}, \citenamefont {Barnes}, \citenamefont {Battaglino},
  \citenamefont {Bohdanowicz}, \citenamefont {Booth}, \citenamefont {Brown},
  \citenamefont {Brown}, \citenamefont {Cassella}, \citenamefont {Coxe},
  \citenamefont {Epstein}, \citenamefont {Feldkamp}, \citenamefont {Griger},
  \citenamefont {Halperin}, \citenamefont {Heinz}, \citenamefont {Hummel},
  \citenamefont {Jaffe}, \citenamefont {Jones}, \citenamefont {Kapit},
  \citenamefont {Kotru}, \citenamefont {Lauigan}, \citenamefont {Li},
  \citenamefont {Marjanovic}, \citenamefont {Megidish}, \citenamefont
  {Meredith}, \citenamefont {Morshead}, \citenamefont {Muniz}, \citenamefont
  {Narayanaswami}, \citenamefont {Nishiguchi}, \citenamefont {Paule},
  \citenamefont {Pawlak}, \citenamefont {Pudenz}, \citenamefont
  {Rodr?guez~P?rez}, \citenamefont {Simon}, \citenamefont {Smull},
  \citenamefont {Stack}, \citenamefont {Urbanek}, \citenamefont {van~de
  Veerdonk}, \citenamefont {Vendeiro}, \citenamefont {Weverka}, \citenamefont
  {Wilkason}, \citenamefont {Wu}, \citenamefont {Xie}, \citenamefont
  {Zalys-Geller}, \citenamefont {Zhang},\ and\ \citenamefont
  {Bloom}}]{Reichardt2024}%
  \BibitemOpen
  \bibfield  {author} {\bibinfo {author} {\bibfnamefont {B.~W.}\ \bibnamefont
  {Reichardt}}, \bibinfo {author} {\bibfnamefont {A.}~\bibnamefont
  {Paetznick}}, \bibinfo {author} {\bibfnamefont {D.}~\bibnamefont {Aasen}},
  \bibinfo {author} {\bibfnamefont {I.}~\bibnamefont {Basov}}, \bibinfo
  {author} {\bibfnamefont {J.~M.}\ \bibnamefont {Bello-Rivas}}, \bibinfo
  {author} {\bibfnamefont {P.}~\bibnamefont {Bonderson}}, \bibinfo {author}
  {\bibfnamefont {R.}~\bibnamefont {Chao}}, \bibinfo {author} {\bibfnamefont
  {W.}~\bibnamefont {van Dam}}, \bibinfo {author} {\bibfnamefont {M.~B.}\
  \bibnamefont {Hastings}}, \bibinfo {author} {\bibfnamefont {R.~V.}\
  \bibnamefont {Mishmash}}, \bibinfo {author} {\bibfnamefont {A.}~\bibnamefont
  {Paz}}, \bibinfo {author} {\bibfnamefont {M.~P.}\ \bibnamefont {da~Silva}},
  \bibinfo {author} {\bibfnamefont {A.}~\bibnamefont {Sundaram}}, \bibinfo
  {author} {\bibfnamefont {K.~M.}\ \bibnamefont {Svore}}, \bibinfo {author}
  {\bibfnamefont {A.}~\bibnamefont {Vaschillo}}, \bibinfo {author}
  {\bibfnamefont {Z.}~\bibnamefont {Wang}}, \bibinfo {author} {\bibfnamefont
  {M.}~\bibnamefont {Zanner}}, \bibinfo {author} {\bibfnamefont {W.~B.}\
  \bibnamefont {Cairncross}}, \bibinfo {author} {\bibfnamefont {C.-A.}\
  \bibnamefont {Chen}}, \bibinfo {author} {\bibfnamefont {D.}~\bibnamefont
  {Crow}}, \bibinfo {author} {\bibfnamefont {H.}~\bibnamefont {Kim}}, \bibinfo
  {author} {\bibfnamefont {J.~M.}\ \bibnamefont {Kindem}}, \bibinfo {author}
  {\bibfnamefont {J.}~\bibnamefont {King}}, \bibinfo {author} {\bibfnamefont
  {M.}~\bibnamefont {McDonald}}, \bibinfo {author} {\bibfnamefont {M.~A.}\
  \bibnamefont {Norcia}}, \bibinfo {author} {\bibfnamefont {A.}~\bibnamefont
  {Ryou}}, \bibinfo {author} {\bibfnamefont {M.}~\bibnamefont {Stone}},
  \bibinfo {author} {\bibfnamefont {L.}~\bibnamefont {Wadleigh}}, \bibinfo
  {author} {\bibfnamefont {K.}~\bibnamefont {Barnes}}, \bibinfo {author}
  {\bibfnamefont {P.}~\bibnamefont {Battaglino}}, \bibinfo {author}
  {\bibfnamefont {T.~C.}\ \bibnamefont {Bohdanowicz}}, \bibinfo {author}
  {\bibfnamefont {G.}~\bibnamefont {Booth}}, \bibinfo {author} {\bibfnamefont
  {A.}~\bibnamefont {Brown}}, \bibinfo {author} {\bibfnamefont {M.~O.}\
  \bibnamefont {Brown}}, \bibinfo {author} {\bibfnamefont {K.}~\bibnamefont
  {Cassella}}, \bibinfo {author} {\bibfnamefont {R.}~\bibnamefont {Coxe}},
  \bibinfo {author} {\bibfnamefont {J.~M.}\ \bibnamefont {Epstein}}, \bibinfo
  {author} {\bibfnamefont {M.}~\bibnamefont {Feldkamp}}, \bibinfo {author}
  {\bibfnamefont {C.}~\bibnamefont {Griger}}, \bibinfo {author} {\bibfnamefont
  {E.}~\bibnamefont {Halperin}}, \bibinfo {author} {\bibfnamefont
  {A.}~\bibnamefont {Heinz}}, \bibinfo {author} {\bibfnamefont
  {F.}~\bibnamefont {Hummel}}, \bibinfo {author} {\bibfnamefont
  {M.}~\bibnamefont {Jaffe}}, \bibinfo {author} {\bibfnamefont {A.~M.~W.}\
  \bibnamefont {Jones}}, \bibinfo {author} {\bibfnamefont {E.}~\bibnamefont
  {Kapit}}, \bibinfo {author} {\bibfnamefont {K.}~\bibnamefont {Kotru}},
  \bibinfo {author} {\bibfnamefont {J.}~\bibnamefont {Lauigan}}, \bibinfo
  {author} {\bibfnamefont {M.}~\bibnamefont {Li}}, \bibinfo {author}
  {\bibfnamefont {J.}~\bibnamefont {Marjanovic}}, \bibinfo {author}
  {\bibfnamefont {E.}~\bibnamefont {Megidish}}, \bibinfo {author}
  {\bibfnamefont {M.}~\bibnamefont {Meredith}}, \bibinfo {author}
  {\bibfnamefont {R.}~\bibnamefont {Morshead}}, \bibinfo {author}
  {\bibfnamefont {J.~A.}\ \bibnamefont {Muniz}}, \bibinfo {author}
  {\bibfnamefont {S.}~\bibnamefont {Narayanaswami}}, \bibinfo {author}
  {\bibfnamefont {C.}~\bibnamefont {Nishiguchi}}, \bibinfo {author}
  {\bibfnamefont {T.}~\bibnamefont {Paule}}, \bibinfo {author} {\bibfnamefont
  {K.~A.}\ \bibnamefont {Pawlak}}, \bibinfo {author} {\bibfnamefont {K.~L.}\
  \bibnamefont {Pudenz}}, \bibinfo {author} {\bibfnamefont {D.}~\bibnamefont
  {Rodr?guez~P?rez}}, \bibinfo {author} {\bibfnamefont {J.}~\bibnamefont
  {Simon}}, \bibinfo {author} {\bibfnamefont {A.}~\bibnamefont {Smull}},
  \bibinfo {author} {\bibfnamefont {D.}~\bibnamefont {Stack}}, \bibinfo
  {author} {\bibfnamefont {M.}~\bibnamefont {Urbanek}}, \bibinfo {author}
  {\bibfnamefont {R.~J.~M.}\ \bibnamefont {van~de Veerdonk}}, \bibinfo {author}
  {\bibfnamefont {Z.}~\bibnamefont {Vendeiro}}, \bibinfo {author}
  {\bibfnamefont {R.~T.}\ \bibnamefont {Weverka}}, \bibinfo {author}
  {\bibfnamefont {T.}~\bibnamefont {Wilkason}}, \bibinfo {author}
  {\bibfnamefont {T.-Y.}\ \bibnamefont {Wu}}, \bibinfo {author} {\bibfnamefont
  {X.}~\bibnamefont {Xie}}, \bibinfo {author} {\bibfnamefont {E.}~\bibnamefont
  {Zalys-Geller}}, \bibinfo {author} {\bibfnamefont {X.}~\bibnamefont
  {Zhang}},\ and\ \bibinfo {author} {\bibfnamefont {B.~J.}\ \bibnamefont
  {Bloom}}\ }\href {https://doi.org/10.48550/arXiv.2411.11822}
  {10.48550/arXiv.2411.11822} (\bibinfo {year} {2024})\BibitemShut {NoStop}%
\bibitem [{\citenamefont {Baranes}\ \emph {et~al.}(2026)\citenamefont
  {Baranes}, \citenamefont {Cain}, \citenamefont {Ataides}, \citenamefont
  {Bluvstein}, \citenamefont {Sinclair}, \citenamefont
  {Vuleti\ifmmode~\acute{c}\else \'{c}\fi{}}, \citenamefont {Zhou},\ and\
  \citenamefont {Lukin}}]{Baranes2026}%
  \BibitemOpen
  \bibfield  {author} {\bibinfo {author} {\bibfnamefont {G.}~\bibnamefont
  {Baranes}}, \bibinfo {author} {\bibfnamefont {M.}~\bibnamefont {Cain}},
  \bibinfo {author} {\bibfnamefont {J.~P.~B.}\ \bibnamefont {Ataides}},
  \bibinfo {author} {\bibfnamefont {D.}~\bibnamefont {Bluvstein}}, \bibinfo
  {author} {\bibfnamefont {J.}~\bibnamefont {Sinclair}}, \bibinfo {author}
  {\bibfnamefont {V.}~\bibnamefont {Vuleti\ifmmode~\acute{c}\else \'{c}\fi{}}},
  \bibinfo {author} {\bibfnamefont {H.}~\bibnamefont {Zhou}},\ and\ \bibinfo
  {author} {\bibfnamefont {M.~D.}\ \bibnamefont {Lukin}},\ }\href
  {https://doi.org/10.1103/ycwc-3myc} {\bibfield  {journal} {\bibinfo
  {journal} {Phys. Rev. X}\ }\textbf {\bibinfo {volume} {16}},\ \bibinfo
  {pages} {011002} (\bibinfo {year} {2026})}\BibitemShut {NoStop}%
\bibitem [{\citenamefont {Cain}\ \emph {et~al.}(2026)\citenamefont {Cain},
  \citenamefont {Xu}, \citenamefont {King}, \citenamefont {Picard},
  \citenamefont {Levine}, \citenamefont {Endres}, \citenamefont {Preskill},
  \citenamefont {Huang},\ and\ \citenamefont {Bluvstein}}]{Cain2026}%
  \BibitemOpen
  \bibfield  {author} {\bibinfo {author} {\bibfnamefont {M.}~\bibnamefont
  {Cain}}, \bibinfo {author} {\bibfnamefont {Q.}~\bibnamefont {Xu}}, \bibinfo
  {author} {\bibfnamefont {R.}~\bibnamefont {King}}, \bibinfo {author}
  {\bibfnamefont {L.~R.~B.}\ \bibnamefont {Picard}}, \bibinfo {author}
  {\bibfnamefont {H.}~\bibnamefont {Levine}}, \bibinfo {author} {\bibfnamefont
  {M.}~\bibnamefont {Endres}}, \bibinfo {author} {\bibfnamefont
  {J.}~\bibnamefont {Preskill}}, \bibinfo {author} {\bibfnamefont {H.-Y.}\
  \bibnamefont {Huang}},\ and\ \bibinfo {author} {\bibfnamefont
  {D.}~\bibnamefont {Bluvstein}},\ }\bibfield  {journal} {\bibinfo  {journal}
  {arXiv:2507.08439}\ }\href {https://doi.org/10.48550/arXiv.2603.28627}
  {10.48550/arXiv.2603.28627} (\bibinfo {year} {2026})\BibitemShut {NoStop}%
\bibitem [{\citenamefont {Jaksch}\ \emph {et~al.}(2000)\citenamefont {Jaksch},
  \citenamefont {Cirac}, \citenamefont {Zoller}, \citenamefont {Rolston},
  \citenamefont {C\^ot\'e},\ and\ \citenamefont {Lukin}}]{Jaksch2000}%
  \BibitemOpen
  \bibfield  {author} {\bibinfo {author} {\bibfnamefont {D.}~\bibnamefont
  {Jaksch}}, \bibinfo {author} {\bibfnamefont {J.~I.}\ \bibnamefont {Cirac}},
  \bibinfo {author} {\bibfnamefont {P.}~\bibnamefont {Zoller}}, \bibinfo
  {author} {\bibfnamefont {S.~L.}\ \bibnamefont {Rolston}}, \bibinfo {author}
  {\bibfnamefont {R.}~\bibnamefont {C\^ot\'e}},\ and\ \bibinfo {author}
  {\bibfnamefont {M.~D.}\ \bibnamefont {Lukin}},\ }\href
  {https://doi.org/10.1103/PhysRevLett.85.2208} {\bibfield  {journal} {\bibinfo
   {journal} {Phys. Rev. Lett}\ }\textbf {\bibinfo {volume} {85}},\ \bibinfo
  {pages} {2208} (\bibinfo {year} {2000})}\BibitemShut {NoStop}%
\bibitem [{\citenamefont {Lukin}\ \emph {et~al.}(2001)\citenamefont {Lukin},
  \citenamefont {Fleischhauer}, \citenamefont {C\^ot\'e}, \citenamefont {Duan},
  \citenamefont {Jaksch}, \citenamefont {Cirac},\ and\ \citenamefont
  {Zoller}}]{Lukin2001}%
  \BibitemOpen
  \bibfield  {author} {\bibinfo {author} {\bibfnamefont {M.~D.}\ \bibnamefont
  {Lukin}}, \bibinfo {author} {\bibfnamefont {M.}~\bibnamefont {Fleischhauer}},
  \bibinfo {author} {\bibfnamefont {R.}~\bibnamefont {C\^ot\'e}}, \bibinfo
  {author} {\bibfnamefont {L.~M.}\ \bibnamefont {Duan}}, \bibinfo {author}
  {\bibfnamefont {D.}~\bibnamefont {Jaksch}}, \bibinfo {author} {\bibfnamefont
  {J.~I.}\ \bibnamefont {Cirac}},\ and\ \bibinfo {author} {\bibfnamefont
  {P.}~\bibnamefont {Zoller}},\ }\href
  {https://doi.org/10.1103/PhysRevLett.87.037901} {\bibfield  {journal}
  {\bibinfo  {journal} {Phys. Rev. Lett.}\ }\textbf {\bibinfo {volume} {87}},\
  \bibinfo {pages} {037901} (\bibinfo {year} {2001})}\BibitemShut {NoStop}%
\bibitem [{\citenamefont {Ebadi}\ \emph {et~al.}(2021)\citenamefont {Ebadi},
  \citenamefont {Wang}, \citenamefont {Levine}, \citenamefont {Keesling},
  \citenamefont {Semeghini}, \citenamefont {Omran}, \citenamefont {Bluvstein},
  \citenamefont {Samajdar}, \citenamefont {Pichler}, \citenamefont {Ho} \emph
  {et~al.}}]{Ebadi2021}%
  \BibitemOpen
  \bibfield  {author} {\bibinfo {author} {\bibfnamefont {S.}~\bibnamefont
  {Ebadi}}, \bibinfo {author} {\bibfnamefont {T.~T.}\ \bibnamefont {Wang}},
  \bibinfo {author} {\bibfnamefont {H.}~\bibnamefont {Levine}}, \bibinfo
  {author} {\bibfnamefont {A.}~\bibnamefont {Keesling}}, \bibinfo {author}
  {\bibfnamefont {G.}~\bibnamefont {Semeghini}}, \bibinfo {author}
  {\bibfnamefont {A.}~\bibnamefont {Omran}}, \bibinfo {author} {\bibfnamefont
  {D.}~\bibnamefont {Bluvstein}}, \bibinfo {author} {\bibfnamefont
  {R.}~\bibnamefont {Samajdar}}, \bibinfo {author} {\bibfnamefont
  {H.}~\bibnamefont {Pichler}}, \bibinfo {author} {\bibfnamefont {W.~W.}\
  \bibnamefont {Ho}}, \emph {et~al.},\ }\href
  {https://doi.org/10.1038/s41586-021-03582-4} {\bibfield  {journal} {\bibinfo
  {journal} {Nature}\ }\textbf {\bibinfo {volume} {595}},\ \bibinfo {pages}
  {227} (\bibinfo {year} {2021})}\BibitemShut {NoStop}%
\bibitem [{\citenamefont {Bluvstein}\ \emph {et~al.}(2022)\citenamefont
  {Bluvstein}, \citenamefont {Levine}, \citenamefont {Semeghini}, \citenamefont
  {Wang}, \citenamefont {Ebadi}, \citenamefont {Kalinowski}, \citenamefont
  {Keesling}, \citenamefont {Maskara}, \citenamefont {Pichler}, \citenamefont
  {Greiner}, \citenamefont {Vuletic},\ and\ \citenamefont
  {Lukin}}]{Bluvstein2022}%
  \BibitemOpen
  \bibfield  {author} {\bibinfo {author} {\bibfnamefont {D.}~\bibnamefont
  {Bluvstein}}, \bibinfo {author} {\bibfnamefont {H.}~\bibnamefont {Levine}},
  \bibinfo {author} {\bibfnamefont {G.}~\bibnamefont {Semeghini}}, \bibinfo
  {author} {\bibfnamefont {T.~T.}\ \bibnamefont {Wang}}, \bibinfo {author}
  {\bibfnamefont {S.}~\bibnamefont {Ebadi}}, \bibinfo {author} {\bibfnamefont
  {M.}~\bibnamefont {Kalinowski}}, \bibinfo {author} {\bibfnamefont
  {A.}~\bibnamefont {Keesling}}, \bibinfo {author} {\bibfnamefont
  {N.}~\bibnamefont {Maskara}}, \bibinfo {author} {\bibfnamefont
  {H.}~\bibnamefont {Pichler}}, \bibinfo {author} {\bibfnamefont
  {M.}~\bibnamefont {Greiner}}, \bibinfo {author} {\bibfnamefont
  {V.}~\bibnamefont {Vuletic}},\ and\ \bibinfo {author} {\bibfnamefont {M.~D.}\
  \bibnamefont {Lukin}},\ }\href {https://doi.org/10.1038/s41586-022-04592-6}
  {\bibfield  {journal} {\bibinfo  {journal} {Nature}\ }\textbf {\bibinfo
  {volume} {604}},\ \bibinfo {pages} {451} (\bibinfo {year}
  {2022})}\BibitemShut {NoStop}%
\bibitem [{\citenamefont {Bluvstein}\ \emph {et~al.}(2024)\citenamefont
  {Bluvstein}, \citenamefont {Evered}, \citenamefont {Geim}, \citenamefont
  {Li}, \citenamefont {Zhou}, \citenamefont {Manovitz}, \citenamefont {Ebadi},
  \citenamefont {Cain}, \citenamefont {Kalinowski}, \citenamefont {Hangleiter},
  \citenamefont {Ataides}, \citenamefont {Maskara}, \citenamefont {Cong},
  \citenamefont {Gao}, \citenamefont {Rodriguez}, \citenamefont {Karolyshyn},
  \citenamefont {Semeghini}, \citenamefont {Gullans}, \citenamefont {Greiner},
  \citenamefont {Vuletic},\ and\ \citenamefont {Lukin}}]{Bluvstein2024}%
  \BibitemOpen
  \bibfield  {author} {\bibinfo {author} {\bibfnamefont {D.}~\bibnamefont
  {Bluvstein}}, \bibinfo {author} {\bibfnamefont {S.~J.}\ \bibnamefont
  {Evered}}, \bibinfo {author} {\bibfnamefont {A.~A.}\ \bibnamefont {Geim}},
  \bibinfo {author} {\bibfnamefont {S.~H.}\ \bibnamefont {Li}}, \bibinfo
  {author} {\bibfnamefont {H.}~\bibnamefont {Zhou}}, \bibinfo {author}
  {\bibfnamefont {T.}~\bibnamefont {Manovitz}}, \bibinfo {author}
  {\bibfnamefont {S.}~\bibnamefont {Ebadi}}, \bibinfo {author} {\bibfnamefont
  {M.}~\bibnamefont {Cain}}, \bibinfo {author} {\bibfnamefont {M.}~\bibnamefont
  {Kalinowski}}, \bibinfo {author} {\bibfnamefont {D.}~\bibnamefont
  {Hangleiter}}, \bibinfo {author} {\bibfnamefont {J.~P.~B.}\ \bibnamefont
  {Ataides}}, \bibinfo {author} {\bibfnamefont {N.}~\bibnamefont {Maskara}},
  \bibinfo {author} {\bibfnamefont {I.}~\bibnamefont {Cong}}, \bibinfo {author}
  {\bibfnamefont {X.}~\bibnamefont {Gao}}, \bibinfo {author} {\bibfnamefont
  {P.~S.}\ \bibnamefont {Rodriguez}}, \bibinfo {author} {\bibfnamefont
  {T.}~\bibnamefont {Karolyshyn}}, \bibinfo {author} {\bibfnamefont
  {G.}~\bibnamefont {Semeghini}}, \bibinfo {author} {\bibfnamefont {M.~J.}\
  \bibnamefont {Gullans}}, \bibinfo {author} {\bibfnamefont {M.}~\bibnamefont
  {Greiner}}, \bibinfo {author} {\bibfnamefont {V.}~\bibnamefont {Vuletic}},\
  and\ \bibinfo {author} {\bibfnamefont {M.~D.}\ \bibnamefont {Lukin}},\ }\href
  {https://doi.org/10.1038/s41586-023-06927-3} {\bibfield  {journal} {\bibinfo
  {journal} {Nature}\ }\textbf {\bibinfo {volume} {626}},\ \bibinfo {pages}
  {58} (\bibinfo {year} {2024})}\BibitemShut {NoStop}%
\bibitem [{\citenamefont {Dudinets}\ \emph {et~al.}(2026)\citenamefont
  {Dudinets}, \citenamefont {Straupe}, \citenamefont {Fedorov},\ and\
  \citenamefont {Lychkovskiy}}]{Dudinets2026}%
  \BibitemOpen
  \bibfield  {author} {\bibinfo {author} {\bibfnamefont {I.~V.}\ \bibnamefont
  {Dudinets}}, \bibinfo {author} {\bibfnamefont {S.~S.}\ \bibnamefont
  {Straupe}}, \bibinfo {author} {\bibfnamefont {A.~K.}\ \bibnamefont
  {Fedorov}},\ and\ \bibinfo {author} {\bibfnamefont {O.~V.}\ \bibnamefont
  {Lychkovskiy}},\ }\href {https://doi.org/10.1103/6bsx-k4m6} {\bibfield
  {journal} {\bibinfo  {journal} {Phys. Rev. A}\ }\textbf {\bibinfo {volume}
  {113}},\ \bibinfo {pages} {062602} (\bibinfo {year} {2026})}\BibitemShut
  {NoStop}%
\bibitem [{\citenamefont {Graham}\ \emph {et~al.}(2022)\citenamefont {Graham},
  \citenamefont {Song}, \citenamefont {Scott}, \citenamefont {Poole},
  \citenamefont {Phuttitarn}, \citenamefont {Jooya}, \citenamefont {Eichler},
  \citenamefont {Jiang}, \citenamefont {Marra}, \citenamefont {Grinkemeyer},
  \citenamefont {Kwon}, \citenamefont {Ebert}, \citenamefont {Cherek},
  \citenamefont {Lichtman}, \citenamefont {Gillette}, \citenamefont {Gilbert},
  \citenamefont {Bowman}, \citenamefont {Ballance}, \citenamefont {Campbell},
  \citenamefont {Dahl}, \citenamefont {Crawford}, \citenamefont {Noel},\ and\
  \citenamefont {Saffman}}]{Graham2022}%
  \BibitemOpen
  \bibfield  {author} {\bibinfo {author} {\bibfnamefont {T.~M.}\ \bibnamefont
  {Graham}}, \bibinfo {author} {\bibfnamefont {Y.}~\bibnamefont {Song}},
  \bibinfo {author} {\bibfnamefont {J.}~\bibnamefont {Scott}}, \bibinfo
  {author} {\bibfnamefont {C.}~\bibnamefont {Poole}}, \bibinfo {author}
  {\bibfnamefont {L.}~\bibnamefont {Phuttitarn}}, \bibinfo {author}
  {\bibfnamefont {K.}~\bibnamefont {Jooya}}, \bibinfo {author} {\bibfnamefont
  {P.}~\bibnamefont {Eichler}}, \bibinfo {author} {\bibfnamefont
  {X.}~\bibnamefont {Jiang}}, \bibinfo {author} {\bibfnamefont
  {A.}~\bibnamefont {Marra}}, \bibinfo {author} {\bibfnamefont
  {B.}~\bibnamefont {Grinkemeyer}}, \bibinfo {author} {\bibfnamefont
  {M.}~\bibnamefont {Kwon}}, \bibinfo {author} {\bibfnamefont {M.}~\bibnamefont
  {Ebert}}, \bibinfo {author} {\bibfnamefont {J.}~\bibnamefont {Cherek}},
  \bibinfo {author} {\bibfnamefont {M.~T.}\ \bibnamefont {Lichtman}}, \bibinfo
  {author} {\bibfnamefont {M.}~\bibnamefont {Gillette}}, \bibinfo {author}
  {\bibfnamefont {J.}~\bibnamefont {Gilbert}}, \bibinfo {author} {\bibfnamefont
  {D.}~\bibnamefont {Bowman}}, \bibinfo {author} {\bibfnamefont
  {T.}~\bibnamefont {Ballance}}, \bibinfo {author} {\bibfnamefont
  {C.}~\bibnamefont {Campbell}}, \bibinfo {author} {\bibfnamefont {E.~D.}\
  \bibnamefont {Dahl}}, \bibinfo {author} {\bibfnamefont {O.}~\bibnamefont
  {Crawford}}, \bibinfo {author} {\bibfnamefont {T.}~\bibnamefont {Noel}},\
  and\ \bibinfo {author} {\bibfnamefont {M.}~\bibnamefont {Saffman}},\ }\href
  {https://doi.org/10.1038/s41586-022-04603-6} {\bibfield  {journal} {\bibinfo
  {journal} {Nature}\ }\textbf {\bibinfo {volume} {604}},\ \bibinfo {pages}
  {457} (\bibinfo {year} {2022})}\BibitemShut {NoStop}%
\bibitem [{\citenamefont {Saffman}\ \emph {et~al.}(2010)\citenamefont
  {Saffman}, \citenamefont {Walker},\ and\ \citenamefont
  {M\o{}lmer}}]{Saffman2010}%
  \BibitemOpen
  \bibfield  {author} {\bibinfo {author} {\bibfnamefont {M.}~\bibnamefont
  {Saffman}}, \bibinfo {author} {\bibfnamefont {T.~G.}\ \bibnamefont
  {Walker}},\ and\ \bibinfo {author} {\bibfnamefont {K.}~\bibnamefont
  {M\o{}lmer}},\ }\href {https://doi.org/10.1103/RevModPhys.82.2313} {\bibfield
   {journal} {\bibinfo  {journal} {Rev. Mod. Phys.}\ }\textbf {\bibinfo
  {volume} {82}},\ \bibinfo {pages} {2313} (\bibinfo {year}
  {2010})}\BibitemShut {NoStop}%
\bibitem [{\citenamefont {Wilk}\ \emph {et~al.}(2010)\citenamefont {Wilk},
  \citenamefont {Ga\"etan}, \citenamefont {Evellin}, \citenamefont {Wolters},
  \citenamefont {Miroshnychenko}, \citenamefont {Grangier},\ and\ \citenamefont
  {Browaeys}}]{Wilk2010}%
  \BibitemOpen
  \bibfield  {author} {\bibinfo {author} {\bibfnamefont {T.}~\bibnamefont
  {Wilk}}, \bibinfo {author} {\bibfnamefont {A.}~\bibnamefont {Ga\"etan}},
  \bibinfo {author} {\bibfnamefont {C.}~\bibnamefont {Evellin}}, \bibinfo
  {author} {\bibfnamefont {J.}~\bibnamefont {Wolters}}, \bibinfo {author}
  {\bibfnamefont {Y.}~\bibnamefont {Miroshnychenko}}, \bibinfo {author}
  {\bibfnamefont {P.}~\bibnamefont {Grangier}},\ and\ \bibinfo {author}
  {\bibfnamefont {A.}~\bibnamefont {Browaeys}},\ }\href
  {https://doi.org/10.1103/PhysRevLett.104.010502} {\bibfield  {journal}
  {\bibinfo  {journal} {Phys. Rev. Lett.}\ }\textbf {\bibinfo {volume} {104}},\
  \bibinfo {pages} {010502} (\bibinfo {year} {2010})}\BibitemShut {NoStop}%
\bibitem [{\citenamefont {Shi}(2019)}]{Shi2019a}%
  \BibitemOpen
  \bibfield  {author} {\bibinfo {author} {\bibfnamefont {X.-F.}\ \bibnamefont
  {Shi}},\ }\href {https://doi.org/10.1103/PhysRevApplied.11.044035} {\bibfield
   {journal} {\bibinfo  {journal} {Phys. Rev. Applied}\ }\textbf {\bibinfo
  {volume} {11}},\ \bibinfo {pages} {044035} (\bibinfo {year}
  {2019})}\BibitemShut {NoStop}%
\bibitem [{\citenamefont {Levine}\ \emph {et~al.}(2019)\citenamefont {Levine},
  \citenamefont {Keesling}, \citenamefont {Semeghini}, \citenamefont {Omran},
  \citenamefont {Wang}, \citenamefont {Ebadi}, \citenamefont {Bernien},
  \citenamefont {Greiner}, \citenamefont {Vuleti\ifmmode~\acute{c}\else
  \'{c}\fi{}}, \citenamefont {Pichler},\ and\ \citenamefont
  {Lukin}}]{Levine2019}%
  \BibitemOpen
  \bibfield  {author} {\bibinfo {author} {\bibfnamefont {H.}~\bibnamefont
  {Levine}}, \bibinfo {author} {\bibfnamefont {A.}~\bibnamefont {Keesling}},
  \bibinfo {author} {\bibfnamefont {G.}~\bibnamefont {Semeghini}}, \bibinfo
  {author} {\bibfnamefont {A.}~\bibnamefont {Omran}}, \bibinfo {author}
  {\bibfnamefont {T.~T.}\ \bibnamefont {Wang}}, \bibinfo {author}
  {\bibfnamefont {S.}~\bibnamefont {Ebadi}}, \bibinfo {author} {\bibfnamefont
  {H.}~\bibnamefont {Bernien}}, \bibinfo {author} {\bibfnamefont
  {M.}~\bibnamefont {Greiner}}, \bibinfo {author} {\bibfnamefont
  {V.}~\bibnamefont {Vuleti\ifmmode~\acute{c}\else \'{c}\fi{}}}, \bibinfo
  {author} {\bibfnamefont {H.}~\bibnamefont {Pichler}},\ and\ \bibinfo {author}
  {\bibfnamefont {M.~D.}\ \bibnamefont {Lukin}},\ }\href
  {https://doi.org/10.1103/PhysRevLett.123.170503} {\bibfield  {journal}
  {\bibinfo  {journal} {Phys. Rev. Lett.}\ }\textbf {\bibinfo {volume} {123}},\
  \bibinfo {pages} {170503} (\bibinfo {year} {2019})}\BibitemShut {NoStop}%
\bibitem [{\citenamefont {Jandura}\ and\ \citenamefont
  {Pupillo}(2022)}]{Jandura2022}%
  \BibitemOpen
  \bibfield  {author} {\bibinfo {author} {\bibfnamefont {S.}~\bibnamefont
  {Jandura}}\ and\ \bibinfo {author} {\bibfnamefont {G.}~\bibnamefont
  {Pupillo}},\ }\href {https://doi.org/10.22331/q-2022-05-13-712} {\bibfield
  {journal} {\bibinfo  {journal} {Quantum}\ }\textbf {\bibinfo {volume} {6}},\
  \bibinfo {pages} {712} (\bibinfo {year} {2022})}\BibitemShut {NoStop}%
\bibitem [{\citenamefont {Jandura}\ \emph {et~al.}(2023)\citenamefont
  {Jandura}, \citenamefont {Thompson},\ and\ \citenamefont
  {Pupillo}}]{Jandura2023}%
  \BibitemOpen
  \bibfield  {author} {\bibinfo {author} {\bibfnamefont {S.}~\bibnamefont
  {Jandura}}, \bibinfo {author} {\bibfnamefont {J.~D.}\ \bibnamefont
  {Thompson}},\ and\ \bibinfo {author} {\bibfnamefont {G.}~\bibnamefont
  {Pupillo}},\ }\href {https://doi.org/10.1103/PRXQuantum.4.020336} {\bibfield
  {journal} {\bibinfo  {journal} {PRX Quantum}\ }\textbf {\bibinfo {volume}
  {4}},\ \bibinfo {pages} {020336} (\bibinfo {year} {2023})}\BibitemShut
  {NoStop}%
\bibitem [{\citenamefont {Fu}\ \emph {et~al.}(2022)\citenamefont {Fu},
  \citenamefont {Xu}, \citenamefont {Sun}, \citenamefont {Liu}, \citenamefont
  {He}, \citenamefont {Li}, \citenamefont {Liu}, \citenamefont {Li},
  \citenamefont {Wang}, \citenamefont {Liu},\ and\ \citenamefont
  {Zhan}}]{Fu2022}%
  \BibitemOpen
  \bibfield  {author} {\bibinfo {author} {\bibfnamefont {Z.}~\bibnamefont
  {Fu}}, \bibinfo {author} {\bibfnamefont {P.}~\bibnamefont {Xu}}, \bibinfo
  {author} {\bibfnamefont {Y.}~\bibnamefont {Sun}}, \bibinfo {author}
  {\bibfnamefont {Y.}~\bibnamefont {Liu}}, \bibinfo {author} {\bibfnamefont
  {X.}~\bibnamefont {He}}, \bibinfo {author} {\bibfnamefont {X.}~\bibnamefont
  {Li}}, \bibinfo {author} {\bibfnamefont {M.}~\bibnamefont {Liu}}, \bibinfo
  {author} {\bibfnamefont {R.}~\bibnamefont {Li}}, \bibinfo {author}
  {\bibfnamefont {J.}~\bibnamefont {Wang}}, \bibinfo {author} {\bibfnamefont
  {L.}~\bibnamefont {Liu}},\ and\ \bibinfo {author} {\bibfnamefont
  {M.}~\bibnamefont {Zhan}},\ }\href
  {https://doi.org/10.1103/PhysRevA.105.042430} {\bibfield  {journal} {\bibinfo
   {journal} {Physical Review A}\ }\textbf {\bibinfo {volume} {105}},\ \bibinfo
  {pages} {042430} (\bibinfo {year} {2022})}\BibitemShut {NoStop}%
\bibitem [{\citenamefont {Vybornyi}\ \emph {et~al.}(2023)\citenamefont
  {Vybornyi}, \citenamefont {Gerasimov}, \citenamefont {Kupriyanov},
  \citenamefont {Straupe},\ and\ \citenamefont {Tikhonov}}]{Vybornyi2023}%
  \BibitemOpen
  \bibfield  {author} {\bibinfo {author} {\bibfnamefont {I.}~\bibnamefont
  {Vybornyi}}, \bibinfo {author} {\bibfnamefont {L.}~\bibnamefont {Gerasimov}},
  \bibinfo {author} {\bibfnamefont {D.}~\bibnamefont {Kupriyanov}}, \bibinfo
  {author} {\bibfnamefont {S.}~\bibnamefont {Straupe}},\ and\ \bibinfo {author}
  {\bibfnamefont {K.}~\bibnamefont {Tikhonov}},\ }\href
  {https://arxiv.org/pdf/2206.12171.pdf} {\bibfield  {journal} {\bibinfo
  {journal} {Journal of the Optical Society of America B}\ }\textbf {\bibinfo
  {volume} {41}},\ \bibinfo {pages} {134} (\bibinfo {year} {2023})}\BibitemShut
  {NoStop}%
\bibitem [{\citenamefont {Saffman}(2016)}]{Saffman2016}%
  \BibitemOpen
  \bibfield  {author} {\bibinfo {author} {\bibfnamefont {M.}~\bibnamefont
  {Saffman}},\ }\href {https://doi.org/10.1088/0953-4075/49/20/202001}
  {\bibfield  {journal} {\bibinfo  {journal} {J. Phys. B: At. Mol. Phys.}\
  }\textbf {\bibinfo {volume} {49}},\ \bibinfo {pages} {202001} (\bibinfo
  {year} {2016})}\BibitemShut {NoStop}%
\bibitem [{\citenamefont {Henriet}\ \emph {et~al.}(2020)\citenamefont
  {Henriet}, \citenamefont {Beguin}, \citenamefont {Signoles}, \citenamefont
  {Lahaye}, \citenamefont {Browaeys}, \citenamefont {Reymond},\ and\
  \citenamefont {Jurczak}}]{Henriet2020}%
  \BibitemOpen
  \bibfield  {author} {\bibinfo {author} {\bibfnamefont {L.}~\bibnamefont
  {Henriet}}, \bibinfo {author} {\bibfnamefont {L.}~\bibnamefont {Beguin}},
  \bibinfo {author} {\bibfnamefont {A.}~\bibnamefont {Signoles}}, \bibinfo
  {author} {\bibfnamefont {T.}~\bibnamefont {Lahaye}}, \bibinfo {author}
  {\bibfnamefont {A.}~\bibnamefont {Browaeys}}, \bibinfo {author}
  {\bibfnamefont {G.-O.}\ \bibnamefont {Reymond}},\ and\ \bibinfo {author}
  {\bibfnamefont {C.}~\bibnamefont {Jurczak}},\ }\href
  {https://doi.org/10.22331/q-2020-09-21-327} {\bibfield  {journal} {\bibinfo
  {journal} {Quantum}\ }\textbf {\bibinfo {volume} {4}},\ \bibinfo {pages}
  {327} (\bibinfo {year} {2020})}\BibitemShut {NoStop}%
\bibitem [{\citenamefont {Li}\ \emph {et~al.}(2024)\citenamefont {Li},
  \citenamefont {Hou}, \citenamefont {Wang}, \citenamefont {Wang},
  \citenamefont {He}, \citenamefont {Zhou}, \citenamefont {Wang}, \citenamefont
  {Liu}, \citenamefont {Wang}, \citenamefont {Xu},\ and\ \citenamefont
  {Zhan}}]{Li2024}%
  \BibitemOpen
  \bibfield  {author} {\bibinfo {author} {\bibfnamefont {X.}~\bibnamefont
  {Li}}, \bibinfo {author} {\bibfnamefont {J.-Y.}\ \bibnamefont {Hou}},
  \bibinfo {author} {\bibfnamefont {J.-C.}\ \bibnamefont {Wang}}, \bibinfo
  {author} {\bibfnamefont {G.-W.}\ \bibnamefont {Wang}}, \bibinfo {author}
  {\bibfnamefont {X.-D.}\ \bibnamefont {He}}, \bibinfo {author} {\bibfnamefont
  {F.}~\bibnamefont {Zhou}}, \bibinfo {author} {\bibfnamefont {Y.-B.}\
  \bibnamefont {Wang}}, \bibinfo {author} {\bibfnamefont {M.}~\bibnamefont
  {Liu}}, \bibinfo {author} {\bibfnamefont {J.}~\bibnamefont {Wang}}, \bibinfo
  {author} {\bibfnamefont {P.}~\bibnamefont {Xu}},\ and\ \bibinfo {author}
  {\bibfnamefont {M.-S.}\ \bibnamefont {Zhan}},\ }\bibfield  {journal}
  {\bibinfo  {journal} {arXiv:2411.08502}\ }\href
  {https://doi.org/10.48550/arXiv.2411.08502} {10.48550/arXiv.2411.08502}
  (\bibinfo {year} {2024})\BibitemShut {NoStop}%
\bibitem [{\citenamefont {Ma}\ \emph {et~al.}(2023)\citenamefont {Ma},
  \citenamefont {Liu}, \citenamefont {Peng}, \citenamefont {Zhang},
  \citenamefont {Jandura}, \citenamefont {Claes}, \citenamefont {Burgers},
  \citenamefont {Pupillo}, \citenamefont {Puri},\ and\ \citenamefont
  {Thompson}}]{Ma2023}%
  \BibitemOpen
  \bibfield  {author} {\bibinfo {author} {\bibfnamefont {S.}~\bibnamefont
  {Ma}}, \bibinfo {author} {\bibfnamefont {G.}~\bibnamefont {Liu}}, \bibinfo
  {author} {\bibfnamefont {P.}~\bibnamefont {Peng}}, \bibinfo {author}
  {\bibfnamefont {B.}~\bibnamefont {Zhang}}, \bibinfo {author} {\bibfnamefont
  {S.}~\bibnamefont {Jandura}}, \bibinfo {author} {\bibfnamefont
  {J.}~\bibnamefont {Claes}}, \bibinfo {author} {\bibfnamefont {A.~P.}\
  \bibnamefont {Burgers}}, \bibinfo {author} {\bibfnamefont {G.}~\bibnamefont
  {Pupillo}}, \bibinfo {author} {\bibfnamefont {S.}~\bibnamefont {Puri}},\ and\
  \bibinfo {author} {\bibfnamefont {J.~D.}\ \bibnamefont {Thompson}},\ }\href
  {https://doi.org/10.1038/s41586-023-06438-1} {\bibfield  {journal} {\bibinfo
  {journal} {Nature}\ }\textbf {\bibinfo {volume} {622}},\ \bibinfo {pages}
  {279} (\bibinfo {year} {2023})}\BibitemShut {NoStop}%
\bibitem [{\citenamefont {Bezuglov}\ \emph {et~al.}(2025)\citenamefont
  {Bezuglov}, \citenamefont {Beterov}, \citenamefont {Cinins}, \citenamefont
  {Miculis}, \citenamefont {Entin}, \citenamefont {Betleni}, \citenamefont
  {Suliman}, \citenamefont {Gromyko}, \citenamefont {Tretyakov}, \citenamefont
  {Yakshina},\ and\ \citenamefont {Ryabtsev}}]{Bezuglov2025}%
  \BibitemOpen
  \bibfield  {author} {\bibinfo {author} {\bibfnamefont {N.~N.}\ \bibnamefont
  {Bezuglov}}, \bibinfo {author} {\bibfnamefont {I.~I.}\ \bibnamefont
  {Beterov}}, \bibinfo {author} {\bibfnamefont {A.}~\bibnamefont {Cinins}},
  \bibinfo {author} {\bibfnamefont {K.}~\bibnamefont {Miculis}}, \bibinfo
  {author} {\bibfnamefont {V.~M.}\ \bibnamefont {Entin}}, \bibinfo {author}
  {\bibfnamefont {P.~I.}\ \bibnamefont {Betleni}}, \bibinfo {author}
  {\bibfnamefont {G.}~\bibnamefont {Suliman}}, \bibinfo {author} {\bibfnamefont
  {V.~V.}\ \bibnamefont {Gromyko}}, \bibinfo {author} {\bibfnamefont {D.~B.}\
  \bibnamefont {Tretyakov}}, \bibinfo {author} {\bibfnamefont {E.~A.}\
  \bibnamefont {Yakshina}},\ and\ \bibinfo {author} {\bibfnamefont {I.~I.}\
  \bibnamefont {Ryabtsev}},\ }\href {https://doi.org/10.1103/s918-nqsp}
  {\bibfield  {journal} {\bibinfo  {journal} {Phys. Rev. A}\ }\textbf {\bibinfo
  {volume} {112}},\ \bibinfo {pages} {063103} (\bibinfo {year}
  {2025})}\BibitemShut {NoStop}%
\bibitem [{\citenamefont {Spirin}\ \emph {et~al.}(2026)\citenamefont {Spirin},
  \citenamefont {Yakshina}, \citenamefont {Sologub}, \citenamefont {Suliman},
  \citenamefont {Minnigaliev}, \citenamefont {Gromyko}, \citenamefont
  {Betleni}, \citenamefont {Tretyakov}, \citenamefont {Entin}, \citenamefont
  {Bezuglov}, \citenamefont {Beterov},\ and\ \citenamefont
  {Ryabtsev}}]{Spirin2026}%
  \BibitemOpen
  \bibfield  {author} {\bibinfo {author} {\bibfnamefont {S.~A.}\ \bibnamefont
  {Spirin}}, \bibinfo {author} {\bibfnamefont {E.~A.}\ \bibnamefont
  {Yakshina}}, \bibinfo {author} {\bibfnamefont {A.~K.}\ \bibnamefont
  {Sologub}}, \bibinfo {author} {\bibfnamefont {G.}~\bibnamefont {Suliman}},
  \bibinfo {author} {\bibfnamefont {A.~M.}\ \bibnamefont {Minnigaliev}},
  \bibinfo {author} {\bibfnamefont {V.~V.}\ \bibnamefont {Gromyko}}, \bibinfo
  {author} {\bibfnamefont {P.}~\bibnamefont {Betleni}}, \bibinfo {author}
  {\bibfnamefont {D.~B.}\ \bibnamefont {Tretyakov}}, \bibinfo {author}
  {\bibfnamefont {V.~M.}\ \bibnamefont {Entin}}, \bibinfo {author}
  {\bibfnamefont {N.~N.}\ \bibnamefont {Bezuglov}}, \bibinfo {author}
  {\bibfnamefont {I.~I.}\ \bibnamefont {Beterov}},\ and\ \bibinfo {author}
  {\bibfnamefont {I.~I.}\ \bibnamefont {Ryabtsev}}\ }\href
  {https://doi.org/10.48550/arXiv.2607.15764} {10.48550/arXiv.2607.15764}
  (\bibinfo {year} {2026})\BibitemShut {NoStop}%
\bibitem [{\citenamefont {Finkelstein}\ \emph {et~al.}(2024)\citenamefont
  {Finkelstein}, \citenamefont {Tsai}, \citenamefont {Sun}, \citenamefont
  {Scholl}, \citenamefont {Direkci}, \citenamefont {Gefen}, \citenamefont
  {Choi}, \citenamefont {Shaw},\ and\ \citenamefont {Endres}}]{Endres2024}%
  \BibitemOpen
  \bibfield  {author} {\bibinfo {author} {\bibfnamefont {T.}~\bibnamefont
  {Finkelstein}}, \bibinfo {author} {\bibfnamefont {R.~B.-S.}\ \bibnamefont
  {Tsai}}, \bibinfo {author} {\bibfnamefont {X.}~\bibnamefont {Sun}}, \bibinfo
  {author} {\bibfnamefont {P.}~\bibnamefont {Scholl}}, \bibinfo {author}
  {\bibfnamefont {S.}~\bibnamefont {Direkci}}, \bibinfo {author} {\bibfnamefont
  {T.}~\bibnamefont {Gefen}}, \bibinfo {author} {\bibfnamefont
  {J.}~\bibnamefont {Choi}}, \bibinfo {author} {\bibfnamefont {A.}~\bibnamefont
  {Shaw}},\ and\ \bibinfo {author} {\bibfnamefont {M.}~\bibnamefont {Endres}},\
  }\href {https://doi.org/10.1038/s41586-024-08005-8} {\bibfield  {journal}
  {\bibinfo  {journal} {Nature}\ }\textbf {\bibinfo {volume} {634}},\ \bibinfo
  {pages} {321} (\bibinfo {year} {2024})}\BibitemShut {NoStop}%
\bibitem [{\citenamefont {Cole}\ \emph {et~al.}(2025)\citenamefont {Cole},
  \citenamefont {Buchemmavari},\ and\ \citenamefont {Saffman}}]{Cole2025}%
  \BibitemOpen
  \bibfield  {author} {\bibinfo {author} {\bibfnamefont {D.~C.}\ \bibnamefont
  {Cole}}, \bibinfo {author} {\bibfnamefont {V.}~\bibnamefont {Buchemmavari}},\
  and\ \bibinfo {author} {\bibfnamefont {M.}~\bibnamefont {Saffman}}\ }\href
  {https://doi.org/10.48550/arXiv.2512.22767} {10.48550/arXiv.2512.22767}
  (\bibinfo {year} {2025})\BibitemShut {NoStop}%
\bibitem [{\citenamefont {Isenhower}\ \emph {et~al.}(2010)\citenamefont
  {Isenhower}, \citenamefont {Urban}, \citenamefont {Zhang}, \citenamefont
  {Gill}, \citenamefont {Henage}, \citenamefont {Johnson}, \citenamefont
  {Walker},\ and\ \citenamefont {Saffman}}]{Isenhower2010}%
  \BibitemOpen
  \bibfield  {author} {\bibinfo {author} {\bibfnamefont {L.}~\bibnamefont
  {Isenhower}}, \bibinfo {author} {\bibfnamefont {E.}~\bibnamefont {Urban}},
  \bibinfo {author} {\bibfnamefont {X.~L.}\ \bibnamefont {Zhang}}, \bibinfo
  {author} {\bibfnamefont {A.~T.}\ \bibnamefont {Gill}}, \bibinfo {author}
  {\bibfnamefont {T.}~\bibnamefont {Henage}}, \bibinfo {author} {\bibfnamefont
  {T.~A.}\ \bibnamefont {Johnson}}, \bibinfo {author} {\bibfnamefont {T.~G.}\
  \bibnamefont {Walker}},\ and\ \bibinfo {author} {\bibfnamefont
  {M.}~\bibnamefont {Saffman}},\ }\href
  {https://doi.org/10.1103/PhysRevLett.104.010503} {\bibfield  {journal}
  {\bibinfo  {journal} {Phys. Rev. Lett.}\ }\textbf {\bibinfo {volume} {104}},\
  \bibinfo {pages} {010503} (\bibinfo {year} {2010})}\BibitemShut {NoStop}%
\bibitem [{\citenamefont {Ebadi}\ \emph {et~al.}(2022)\citenamefont {Ebadi},
  \citenamefont {Keesling}, \citenamefont {Cain}, \citenamefont {Wang},
  \citenamefont {Levine}, \citenamefont {Bluvstein}, \citenamefont {Semeghini},
  \citenamefont {Omran}, \citenamefont {Liu}, \citenamefont {Samajdar} \emph
  {et~al.}}]{Ebadi2022}%
  \BibitemOpen
  \bibfield  {author} {\bibinfo {author} {\bibfnamefont {S.}~\bibnamefont
  {Ebadi}}, \bibinfo {author} {\bibfnamefont {A.}~\bibnamefont {Keesling}},
  \bibinfo {author} {\bibfnamefont {M.}~\bibnamefont {Cain}}, \bibinfo {author}
  {\bibfnamefont {T.~T.}\ \bibnamefont {Wang}}, \bibinfo {author}
  {\bibfnamefont {H.}~\bibnamefont {Levine}}, \bibinfo {author} {\bibfnamefont
  {D.}~\bibnamefont {Bluvstein}}, \bibinfo {author} {\bibfnamefont
  {G.}~\bibnamefont {Semeghini}}, \bibinfo {author} {\bibfnamefont
  {A.}~\bibnamefont {Omran}}, \bibinfo {author} {\bibfnamefont {J.-G.}\
  \bibnamefont {Liu}}, \bibinfo {author} {\bibfnamefont {R.}~\bibnamefont
  {Samajdar}}, \emph {et~al.},\ }\href
  {https://doi.org/10.1126/science.abo6587} {\bibfield  {journal} {\bibinfo
  {journal} {Science}\ }\textbf {\bibinfo {volume} {376}},\ \bibinfo {pages}
  {1209} (\bibinfo {year} {2022})}\BibitemShut {NoStop}%
\bibitem [{\citenamefont {Chinnarasu}\ \emph {et~al.}(2025)\citenamefont
  {Chinnarasu}, \citenamefont {Poole}, \citenamefont {Phuttitarn},
  \citenamefont {Noori}, \citenamefont {Graham}, \citenamefont {Coppersmith},
  \citenamefont {Balantekin},\ and\ \citenamefont {Saffman}}]{Chinnarasu2025}%
  \BibitemOpen
  \bibfield  {author} {\bibinfo {author} {\bibfnamefont {R.}~\bibnamefont
  {Chinnarasu}}, \bibinfo {author} {\bibfnamefont {C.}~\bibnamefont {Poole}},
  \bibinfo {author} {\bibfnamefont {L.}~\bibnamefont {Phuttitarn}}, \bibinfo
  {author} {\bibfnamefont {A.}~\bibnamefont {Noori}}, \bibinfo {author}
  {\bibfnamefont {T.~M.}\ \bibnamefont {Graham}}, \bibinfo {author}
  {\bibfnamefont {S.~N.}\ \bibnamefont {Coppersmith}}, \bibinfo {author}
  {\bibfnamefont {A.~B.}\ \bibnamefont {Balantekin}},\ and\ \bibinfo {author}
  {\bibfnamefont {M.}~\bibnamefont {Saffman}},\ }\href
  {https://doi.org/10.1103/PRXQuantum.6.020350} {\bibfield  {journal} {\bibinfo
   {journal} {PRX Quantum}\ }\textbf {\bibinfo {volume} {6}},\ \bibinfo {pages}
  {020350} (\bibinfo {year} {2025})}\BibitemShut {NoStop}%
\bibitem [{\citenamefont {Chow}\ \emph {et~al.}(2024)\citenamefont {Chow},
  \citenamefont {Buchemmavari}, \citenamefont {Omanakuttan}, \citenamefont
  {Little}, \citenamefont {Pandey}, \citenamefont {Deutsch},\ and\
  \citenamefont {Jau}}]{Chow2024}%
  \BibitemOpen
  \bibfield  {author} {\bibinfo {author} {\bibfnamefont {M.~N.~H.}\
  \bibnamefont {Chow}}, \bibinfo {author} {\bibfnamefont {V.}~\bibnamefont
  {Buchemmavari}}, \bibinfo {author} {\bibfnamefont {S.}~\bibnamefont
  {Omanakuttan}}, \bibinfo {author} {\bibfnamefont {B.~J.}\ \bibnamefont
  {Little}}, \bibinfo {author} {\bibfnamefont {S.}~\bibnamefont {Pandey}},
  \bibinfo {author} {\bibfnamefont {I.~H.}\ \bibnamefont {Deutsch}},\ and\
  \bibinfo {author} {\bibfnamefont {Y.-Y.}\ \bibnamefont {Jau}},\ }\href
  {https://doi.org/10.1103/PRXQuantum.5.040343} {\bibfield  {journal} {\bibinfo
   {journal} {PRX Quantum}\ }\textbf {\bibinfo {volume} {5}},\ \bibinfo {pages}
  {040343} (\bibinfo {year} {2024})}\BibitemShut {NoStop}%
\bibitem [{\citenamefont {Beterov}\ \emph {et~al.}(2024)\citenamefont
  {Beterov}, \citenamefont {Yakshina}, \citenamefont {Suliman}, \citenamefont
  {Betleni}, \citenamefont {Prilutskaya}, \citenamefont {Skvortsova},
  \citenamefont {Zagirov}, \citenamefont {Tretyakov}, \citenamefont {Entin},
  \citenamefont {Bezuglov},\ and\ \citenamefont {Ryabtsev}}]{Beterov2024}%
  \BibitemOpen
  \bibfield  {author} {\bibinfo {author} {\bibfnamefont {I.}~\bibnamefont
  {Beterov}}, \bibinfo {author} {\bibfnamefont {E.}~\bibnamefont {Yakshina}},
  \bibinfo {author} {\bibfnamefont {G.}~\bibnamefont {Suliman}}, \bibinfo
  {author} {\bibfnamefont {P.}~\bibnamefont {Betleni}}, \bibinfo {author}
  {\bibfnamefont {A.}~\bibnamefont {Prilutskaya}}, \bibinfo {author}
  {\bibfnamefont {D.}~\bibnamefont {Skvortsova}}, \bibinfo {author}
  {\bibfnamefont {T.}~\bibnamefont {Zagirov}}, \bibinfo {author} {\bibfnamefont
  {D.}~\bibnamefont {Tretyakov}}, \bibinfo {author} {\bibfnamefont
  {V.}~\bibnamefont {Entin}}, \bibinfo {author} {\bibfnamefont
  {N.}~\bibnamefont {Bezuglov}},\ and\ \bibinfo {author} {\bibfnamefont
  {I.}~\bibnamefont {Ryabtsev}},\ }\href
  {https://doi.org/https://doi.org/10.31857/S0044451024100109} {\bibfield
  {journal} {\bibinfo  {journal} {Journal of Experimental and Theoretical
  Physics}\ }\textbf {\bibinfo {volume} {166}},\ \bibinfo {pages} {516}
  (\bibinfo {year} {2024})}\BibitemShut {NoStop}%
\bibitem [{\citenamefont {Locher}\ \emph {et~al.}(2025)\citenamefont {Locher},
  \citenamefont {Old}, \citenamefont {Brechtelsbauer}, \citenamefont
  {Holschbach}, \citenamefont {B\"{u}chler}, \citenamefont {Weber},\ and\
  \citenamefont {M\"{u}ller}}]{Rydopt2025}%
  \BibitemOpen
  \bibfield  {author} {\bibinfo {author} {\bibfnamefont {D.~F.}\ \bibnamefont
  {Locher}}, \bibinfo {author} {\bibfnamefont {J.}~\bibnamefont {Old}},
  \bibinfo {author} {\bibfnamefont {K.}~\bibnamefont {Brechtelsbauer}},
  \bibinfo {author} {\bibfnamefont {J.}~\bibnamefont {Holschbach}}, \bibinfo
  {author} {\bibfnamefont {H.~P.}\ \bibnamefont {B\"{u}chler}}, \bibinfo
  {author} {\bibfnamefont {S.}~\bibnamefont {Weber}},\ and\ \bibinfo {author}
  {\bibfnamefont {M.}~\bibnamefont {M\"{u}ller}},\ }\bibfield  {journal}
  {\bibinfo  {journal} {arxiv:2512.00843}\ }\href
  {https://doi.org/10.48550/arXiv.2512.00843} {10.48550/arXiv.2512.00843}
  (\bibinfo {year} {2025})\BibitemShut {NoStop}%
\bibitem [{\citenamefont {Beterov}\ \emph {et~al.}(2026)\citenamefont
  {Beterov}, \citenamefont {Kozenko}, \citenamefont {Ryabtsev},\ and\
  \citenamefont {Xu}}]{Beterov2026}%
  \BibitemOpen
  \bibfield  {author} {\bibinfo {author} {\bibfnamefont {I.~I.}\ \bibnamefont
  {Beterov}}, \bibinfo {author} {\bibfnamefont {K.~V.}\ \bibnamefont
  {Kozenko}}, \bibinfo {author} {\bibfnamefont {I.~I.}\ \bibnamefont
  {Ryabtsev}},\ and\ \bibinfo {author} {\bibfnamefont {P.}~\bibnamefont {Xu}},\
  }\href {https://doi.org/10.1103/g6sw-w48t} {\bibfield  {journal} {\bibinfo
  {journal} {Phys. Rev. A}\ }\textbf {\bibinfo {volume} {113}},\ \bibinfo
  {pages} {042614} (\bibinfo {year} {2026})}\BibitemShut {NoStop}%
\bibitem [{\citenamefont {Maller}\ \emph {et~al.}(2015)\citenamefont {Maller},
  \citenamefont {Lichtman}, \citenamefont {Xia}, \citenamefont {Sun},
  \citenamefont {Piotrowicz}, \citenamefont {Carr}, \citenamefont {Isenhower},\
  and\ \citenamefont {Saffman}}]{Maller2015}%
  \BibitemOpen
  \bibfield  {author} {\bibinfo {author} {\bibfnamefont {K.~M.}\ \bibnamefont
  {Maller}}, \bibinfo {author} {\bibfnamefont {M.~T.}\ \bibnamefont
  {Lichtman}}, \bibinfo {author} {\bibfnamefont {T.}~\bibnamefont {Xia}},
  \bibinfo {author} {\bibfnamefont {Y.}~\bibnamefont {Sun}}, \bibinfo {author}
  {\bibfnamefont {M.~J.}\ \bibnamefont {Piotrowicz}}, \bibinfo {author}
  {\bibfnamefont {A.~W.}\ \bibnamefont {Carr}}, \bibinfo {author}
  {\bibfnamefont {L.}~\bibnamefont {Isenhower}},\ and\ \bibinfo {author}
  {\bibfnamefont {M.}~\bibnamefont {Saffman}},\ }\href
  {https://doi.org/10.1103/PhysRevA.92.022336} {\bibfield  {journal} {\bibinfo
  {journal} {Phys. Rev. A}\ }\textbf {\bibinfo {volume} {92}},\ \bibinfo
  {pages} {022336} (\bibinfo {year} {2015})}\BibitemShut {NoStop}%
\bibitem [{\citenamefont {{\v{S}}ibali{\'c}}\ \emph {et~al.}(2017)\citenamefont
  {{\v{S}}ibali{\'c}}, \citenamefont {Pritchard}, \citenamefont {Adams},\ and\
  \citenamefont {Weatherill}}]{Vsibalic2017ARC}%
  \BibitemOpen
  \bibfield  {author} {\bibinfo {author} {\bibfnamefont {N.}~\bibnamefont
  {{\v{S}}ibali{\'c}}}, \bibinfo {author} {\bibfnamefont {J.~D.}\ \bibnamefont
  {Pritchard}}, \bibinfo {author} {\bibfnamefont {C.~S.}\ \bibnamefont
  {Adams}},\ and\ \bibinfo {author} {\bibfnamefont {K.~J.}\ \bibnamefont
  {Weatherill}},\ }\href {https://doi.org/10.1016/j.cpc.2017.06.015} {\bibfield
   {journal} {\bibinfo  {journal} {Computer Physics Communications}\ }\textbf
  {\bibinfo {volume} {220}},\ \bibinfo {pages} {319} (\bibinfo {year}
  {2017})}\BibitemShut {NoStop}%
\bibitem [{\citenamefont {Saffman}\ \emph {et~al.}(2020)\citenamefont
  {Saffman}, \citenamefont {Beterov}, \citenamefont {Dalal}, \citenamefont
  {P\'aez},\ and\ \citenamefont {Sanders}}]{Saffman2020}%
  \BibitemOpen
  \bibfield  {author} {\bibinfo {author} {\bibfnamefont {M.}~\bibnamefont
  {Saffman}}, \bibinfo {author} {\bibfnamefont {I.~I.}\ \bibnamefont
  {Beterov}}, \bibinfo {author} {\bibfnamefont {A.}~\bibnamefont {Dalal}},
  \bibinfo {author} {\bibfnamefont {E.~J.}\ \bibnamefont {P\'aez}},\ and\
  \bibinfo {author} {\bibfnamefont {B.~C.}\ \bibnamefont {Sanders}},\ }\href
  {https://doi.org/10.1103/PhysRevA.101.062309} {\bibfield  {journal} {\bibinfo
   {journal} {Phys. Rev. A}\ }\textbf {\bibinfo {volume} {101}},\ \bibinfo
  {pages} {062309} (\bibinfo {year} {2020})}\BibitemShut {NoStop}%
\bibitem [{\citenamefont {Grimm}\ \emph {et~al.}(2000)\citenamefont {Grimm},
  \citenamefont {Weidem\"uller},\ and\ \citenamefont {Ovchnnikov}}]{Grimm2000}%
  \BibitemOpen
  \bibfield  {author} {\bibinfo {author} {\bibfnamefont {R.}~\bibnamefont
  {Grimm}}, \bibinfo {author} {\bibfnamefont {M.}~\bibnamefont
  {Weidem\"uller}},\ and\ \bibinfo {author} {\bibfnamefont {Y.~B.}\
  \bibnamefont {Ovchnnikov}},\ }\href
  {https://doi.org/10.1016/S1049-250X(08)60186-X} {\bibfield  {journal}
  {\bibinfo  {journal} {Adv. At., Mol., Opt. Phys.}\ }\textbf {\bibinfo
  {volume} {42}},\ \bibinfo {pages} {95} (\bibinfo {year} {2000})}\BibitemShut
  {NoStop}%
\bibitem [{\citenamefont {Demtr\"{o}der}(2015)}]{Demtroder2015}%
  \BibitemOpen
  \bibfield  {author} {\bibinfo {author} {\bibfnamefont {W.}~\bibnamefont
  {Demtr\"{o}der}},\ }\href {https://doi.org/10.1007/978-3-662-44641-6} {\emph
  {\bibinfo {title} {Laser spectroscopy 2: Experimental techniques, fifth
  edition}}}\ (\bibinfo  {publisher} {Springer},\ \bibinfo {year} {2015})\ pp.\
  \bibinfo {pages} {1--757}\BibitemShut {NoStop}%
\bibitem [{\citenamefont {Ryabtsev}\ \emph {et~al.}(2011)\citenamefont
  {Ryabtsev}, \citenamefont {Beterov}, \citenamefont {Tretyakov}, \citenamefont
  {Entin},\ and\ \citenamefont {Yakshina}}]{Ryabtsev2011}%
  \BibitemOpen
  \bibfield  {author} {\bibinfo {author} {\bibfnamefont {I.~I.}\ \bibnamefont
  {Ryabtsev}}, \bibinfo {author} {\bibfnamefont {I.~I.}\ \bibnamefont
  {Beterov}}, \bibinfo {author} {\bibfnamefont {D.~B.}\ \bibnamefont
  {Tretyakov}}, \bibinfo {author} {\bibfnamefont {V.~M.}\ \bibnamefont
  {Entin}},\ and\ \bibinfo {author} {\bibfnamefont {E.~A.}\ \bibnamefont
  {Yakshina}},\ }\href {https://doi.org/10.1103/PhysRevA.84.053409} {\bibfield
  {journal} {\bibinfo  {journal} {Phys. Rev. A}\ }\textbf {\bibinfo {volume}
  {84}},\ \bibinfo {pages} {053409} (\bibinfo {year} {2011})}\BibitemShut
  {NoStop}%
\bibitem [{\citenamefont {Beterov}\ \emph {et~al.}(2009)\citenamefont
  {Beterov}, \citenamefont {Ryabtsev}, \citenamefont {Tretyakov},\ and\
  \citenamefont {Entin}}]{Beterov2009}%
  \BibitemOpen
  \bibfield  {author} {\bibinfo {author} {\bibfnamefont {I.~I.}\ \bibnamefont
  {Beterov}}, \bibinfo {author} {\bibfnamefont {I.~I.}\ \bibnamefont
  {Ryabtsev}}, \bibinfo {author} {\bibfnamefont {D.~B.}\ \bibnamefont
  {Tretyakov}},\ and\ \bibinfo {author} {\bibfnamefont {V.~M.}\ \bibnamefont
  {Entin}},\ }\href {https://doi.org/10.1103/PhysRevA.79.052504} {\bibfield
  {journal} {\bibinfo  {journal} {Phys. Rev. A}\ }\textbf {\bibinfo {volume}
  {79}},\ \bibinfo {pages} {052504} (\bibinfo {year} {2009})}\BibitemShut
  {NoStop}%
\bibitem [{\citenamefont {Gillen-Christandl}\ \emph {et~al.}(2016)\citenamefont
  {Gillen-Christandl}, \citenamefont {Gillen}, \citenamefont {Piotrowicz},\
  and\ \citenamefont {Saffman}}]{Gillen-Christandl2016}%
  \BibitemOpen
  \bibfield  {author} {\bibinfo {author} {\bibfnamefont {K.}~\bibnamefont
  {Gillen-Christandl}}, \bibinfo {author} {\bibfnamefont {G.~D.}\ \bibnamefont
  {Gillen}}, \bibinfo {author} {\bibfnamefont {M.~J.}\ \bibnamefont
  {Piotrowicz}},\ and\ \bibinfo {author} {\bibfnamefont {M.}~\bibnamefont
  {Saffman}},\ }\href {https://doi.org/10.1007/s00340-016-6407-y} {\bibfield
  {journal} {\bibinfo  {journal} {Applied Physics B}\ }\textbf {\bibinfo
  {volume} {122}},\ \bibinfo {pages} {131} (\bibinfo {year}
  {2016})}\BibitemShut {NoStop}%
\end{thebibliography}
